\documentclass[prd,12pt]{article}

\usepackage{tikz}
\usepackage{amsmath,amssymb,graphicx,multirow,xspace,slashed}
\usepackage[colorlinks=true,urlcolor=blue,anchorcolor=blue,citecolor=red,filecolor=blue,linkcolor=blue,menucolor=blue,pagecolor=blue]{hyperref}
\usepackage[compress,numbers]{natbib}
\usepackage{subfigure, placeins}
\usepackage{booktabs}
\usepackage{blindtext, rotating}
\usepackage{afterpage}
\usepackage{enumitem}
\usepackage{ marvosym }
\usepackage{authblk} 
\usepackage{verbatim}
\usepackage{soul} 
\usepackage[normalem]{ulem}
\usepackage{pifont}
\usepackage[usestackEOL]{stackengine}
\usepackage{subfiles}
\newcommand\sq{\framebox(10,10){}\kern\fboxrule}

\setstackgap{S}{\fboxrule}

\usepackage[utf8]{inputenc}

\usepackage{floatrow}
\newfloatcommand{capbtabbox}{table}[][\FBwidth]

\usepackage[font=footnotesize,labelfont=bf]{caption}

\usepackage{lineno}

\usepackage{ulem}

\allowdisplaybreaks

\long\def\symbolfootnote[#1]#2{\begingroup%
\def\thefootnote{\fnsymbol{footnote}}\footnote[#1]{#2}\endgroup}

\newcommand{\newc}{\newcommand}
\newc{\gsim}{\lower.7ex\hbox{$\;\stackrel{\textstyle>}{\sim}\;$}}
\newc{\lsim}{\lower.7ex\hbox{$\;\stackrel{\textstyle<}{\sim}\;$}}
\newc{\gev}{\,{\rm GeV}}
\newc{\mev}{\,{\rm MeV}}
\newc{\ev}{\,{\rm eV}}
\newc{\kev}{\,{\rm keV}}
\newc{\tev}{\,{\rm TeV}}
\newc{\MHT}{$H_T^{\text{miss}}$}
\newc{\MET}{$\slashed{E}_T$}
\newc{\MTT}{$M_{T2}$}

\newc{\mz}{M_Z}
\newc{\mpl}{M_*}
\newc{\mw}{m_{\rm weak}}
\newc{\nr}[1]{N^c_R{}_{#1}}

\newcommand{\bq}{\bar{q}}

\newcommand{\Nqin}{N_{q}^{\mathrm{initial}}}

\newcommand\Sfactor{{\mathcal{S}}}

\def\beq{\begin{equation}}
\def\eeq{\end{equation}}
\newcommand{\bea}{\begin{eqnarray}\begin{aligned}}
\newcommand{\eea}{\end{aligned}\end{eqnarray}}
\def\bitem{\begin{itemize}}
\def\eitem{\end{itemize}}
\newcommand{\nn}{\nonumber}

\definecolor{darkgreen}{rgb}{0,0.5,0}
\definecolor{goodyellow}{rgb}{0.9,0.7,0}

\numberwithin{equation}{section}

\newcommand\fverb{\setbox\fverbbox=\hbox\bgroup\verb}

\newbox\fverbbox

\renewcommand{\arraystretch}{1.3}

\begin{document}
\baselineskip 0.6cm

\begin{titlepage}

\thispagestyle{empty}

\begin{center}

\vskip 0.1cm

\hspace{4.8in}
MIT-CTP/5285

\vspace{1.3cm}

{\Huge \bf Thermal Squeezeout of Dark Matter}

\vskip 0.5cm

\vskip 0.5cm

\vskip 1.0cm
{\large Pouya Asadi$^{1}$, Eric David Kramer$^2$, Eric Kuflik$^2$, Gregory W. Ridgway$^{1}$, Tracy R. Slatyer$^{1}$, Juri Smirnov$^{3,4}$}
\vskip 1.0cm
{\it $^1$ Center for Theoretical Physics, Massachusetts Institute of Technology, \\ Cambridge, MA 02139, USA.\\}
{\it $^2$ Racah Institute of Physics, Hebrew University of Jerusalem, Jerusalem 91904, Israel. \\}
{\it $^3$ Center for Cosmology and AstroParticle Physics (CCAPP), The Ohio State University, Columbus, OH 43210, USA\\}
{\it $^4$ Department of Physics, The Ohio State University, Columbus, OH 43210, USA\\}

\vskip 0.3cm

\end{center}

\vskip 0.6cm

\begin{abstract}

We carry out a detailed study of the confinement phase transition in a dark sector with a $SU(N)$ gauge group and a single generation of dark heavy quark. We focus on heavy enough quarks such that their abundance freezes out before the phase transition and the phase transition is of first-order. We find that during this phase transition the quarks are trapped inside contracting pockets of the deconfined phase and are compressed enough to interact at a significant rate, giving rise to a second stage of annihilation that can dramatically change the resulting dark matter abundance. 
As a result, the dark matter can be heavier than the often-quoted unitarity bound of $\sim100$~TeV. Our findings are almost completely independent of the details of the portal between the dark sector and the Standard Model. 
We comment briefly on possible signals of such a sector. Our main findings are summarized in a companion letter, while here we provide further details on different parts of the calculation.

\end{abstract}

\flushbottom

\end{titlepage}

\setcounter{page}{1}

\tableofcontents

\vskip 1cm

\newpage

\section{Introduction}
\label{sec:intro}

The cosmic abundance of dark matter (DM) is comparable to the abundance of Standard Model (SM) particles up to an $\mathcal{O}(1)$ factor \cite{Aghanim:2018eyx}. The similarity of these two ostensibly unrelated abundances raises the suspicion that the two sectors may have been in chemical equilibrium at some point in their history, implying some sort of interaction portal between the SM and DM.
Numerous experimental efforts to look for such a portal are under way. Nonetheless, the particle nature of DM and any potential portals to the SM remain unknown at present.

In addition to probing interactions between DM and the SM, many experimental and theoretical efforts aim to probe possible dynamics within the dark sector itself. Often simplified dark sectors with only a single DM particle are considered. 
Yet, the rich gauge structure of the SM offers no particular reason to believe that the dark sector will be significantly simpler. A wide range of more involved dark sectors have been studied, especially scenarios with a new confining force in the dark sector; see for instance Refs.~\cite{Gudnason:2006yj,Alves:2009nf,Kilic:2009mi,Hambye:2009fg,Kribs:2009fy,Alves:2010dd,Bai:2010qg,Feng:2011ik,Fok:2011yc,Lewis:2011zb,Frigerio:2012uc,Buckley:2012ky,Appelquist:2013ms,Bai:2013xga,Bhattacharya:2013kma,Cline:2013zca,Boddy:2014yra,Boddy:2014qxa,Hochberg:2014kqa,Appelquist:2015yfa,Antipin:2015xia,GarciaGarcia:2015fol,Soni:2016gzf,Kribs:2016cew,Harigaya:2016nlg,Mitridate:2017oky,DeLuca:2018mzn,Contino:2018crt,Gross:2018zha,Beylin:2019gtw,Dondi:2019olm,Buttazzo:2019mvl,Landini:2020daq,Brower:2020mab,Contino:2020god,Beylin:2020bsz}. 
Depending on the details of the sector, different hadronic states can be the DM candidate in different theories and the stabilizing symmetry and the DM mass scale can vary widely \cite{Dondi:2019olm}. Such a sector can further give rise to rich dynamics that can potentially solve other problems in the SM as well, e.g. see Refs.~\cite{Barr:1991,Kribs:2009fy,Huang:2017kzu} where the observed baryon asymmetry is tied to the DM abundance. 

An interesting class of confining dark sector models is the scenario where all the dark quarks are substantially heavier than the dark confinement scale, $\Lambda$. 
These models and their experimental signals have been studied extensively, see for instance Refs.~\cite{Kribs:2009fy,Harigaya:2016nlg,Mitridate:2017oky,DeLuca:2018mzn,Gross:2018zha}. 
For sufficiently heavy quarks, lattice calculations have shown that the phase transition in such a sector is of first-order for $SU(N)$ with $N=3$~\cite{Ben_PT, Kaczmarek:1999mm,PhysRevD.60.034504,Aoki:2006we,Saito:2011fs} or $N > 3$ \cite{Lucini:2003zr,largeN_lattice}. 
There has been a recent surge of interest in the study of the potential effects of first-order phase transitions in other DM models, e.g. see Refs.~\cite{Baker:2019ndr,Chway:2019kft,Davoudiasl:2019xeb,Chao:2020adk,Kang:2021epo,Azatov:2021ifm}\footnote{See also Refs.~\cite{Konstandin:2011dr,Hambye:2018qjv,Baratella:2018pxi} for another mechanism affecting DM abundance in the presence of significant supercooling during a phase transition.}, but the effects of the phase transition on the relic abundance of dark matter in confining dark sector models are mostly unexplored, with the exception of a recent study of dark sectors with only light quarks ($m_q \leqslant \Lambda$) \cite{Baldes:2020kam}.

In this work, we consider the simplest such confining model -- an $SU(3)$ gauge theory with one heavy quark in the fundamental representation -- and focus on the effects of the first-order phase transition on the DM relic abundance calculation. 
Similar to the arguments put forward in Ref.~\cite{Witten:1984rs}, we will argue that toward the end of the phase transition we will be left with pockets of the high temperature, i.e. deconfined, phase submerged in a sea of low temperature, i.e. confined, phase. We will argue that the heavy quarks are all initially trapped inside these contracting pockets.\footnote{This depends on the representation of the quarks under the dark confining gauge group. For instance, if the quarks were in the adjoint representation (similar to the model in Ref.~\cite{Contino:2018crt}), they could combine with the surrounding gluons, form color-neutral hadrons, and move into the confined phase. Similarly, if in addition to the heavy quarks the spectrum were to include light quarks in the fundamental representation as well, the heavy quarks would not remain trapped within the pocket as they could make a color-neutral bound state with one of many light quarks surrounding them to escape the pocket. } 
To determine important properties of these pockets such as their initial characteristic size and contraction rate, we will develop a simplified model to numerically simulate the phase transition.

As a pocket contracts, the dark quarks within it are compressed, allowing them to recouple and go through a second stage of annihilation. 
We calculate the fraction of the quarks that survive this new annihilation stage and escape the pockets in the form of dark baryons. We refer to this process as ``thermal squeezeout'', as the dark quarks are squeezed within the pockets and eventually leak out, in contrast to the standard ``thermal freezeout''. 
We find a dramatic suppression in the final abundance thanks to this phenomenon, which points to much heavier dark matter parameter space than was previously thought.
In particular, the fact that the local DM density is much larger than the globally averaged DM density during this second stage of annihilation invalidates the homogeneity assumption made in the unitarity bound argument of Ref.~\cite{Griest:1989wd}.\footnote{See also Refs.~\cite{vonHarling:2014kha,Smirnov:2019ngs} for more recent studies of the unitarity bound on thermal DM models.} 
As a result, this model allows the DM candidate to be heavier than the perturbative unitarity bound on Weakly-Interacting Massive Particles (WIMPs), despite being thermal.

Since this second stage of annihilation is controlled by the dynamics within the dark sector and not the interaction between the dark sector and SM, our results are largely independent of the portal to the SM. In fact, we do not constrain ourselves to any specific portal in this paper. The only assumptions we make about such a portal is that (i) it exists, (ii) it keeps the SM and the dark sector in thermal contact during the phase transition, and (iii) it respects the dark baryon number that stabilizes the dark baryons. These assumptions streamline our calculations significantly. 
However, it is worth considering the possibility of models in which we can relax one or more of these assumptions; we leave this for future work.

The current work is merely the start of a broader program of studying such models in more details. The phenomenology of all such models should be revisited in light of the dramatic change in the relic abundance calculation. 
Depending on the gauge group under study, the quarks' representation, and the portal, different models (with vastly different phenomenology) can be constructed. 

Our study indicates a natural window of DM masses between 1-100~PeV for such a setup. While conventional searches may lose sensitivity for such a high DM mass, the stochastic gravitational wave background due to the first-order phase transition in this scenario can be detected by planned future facilities. 
(See Ref.~\cite{Moore:2014lga} for the projected reach of such facilities.)
While this signal depends on the UV parameters that control the thermodynamics of the phase transition, it does not depend on the nature of the portal.

The rest of this paper is organized as follows. In Sec.~\ref{sec:overview} we provide an overview of the cosmology of our dark sector.  In Sec.~\ref{sec:boltzeqs} we write down and solve the Boltzmann equations that determine the relic abundance of the dark matter. In Sec.~\ref{sec:pheno} we provide an overview of the possible phenomenological implications of our dark sector before concluding in Sec.~\ref{sec:conclusion}. 
We also provide three appendices to supply more details. 
In App.~\ref{appx:thermo} we provide more details on the thermodynamics of first-order phase transitions in an expanding universe and detail a simulation we performed to fix the phase transition parameters that enter the relic abundance calculation. In Apps.~\ref{appx:xsec} and~\ref{appx:binding} we review some results in the literature for the cross sections and binding energies of heavy quarks and their bound states.

\section{A Qualitative Overview of the Cosmology}
\label{sec:overview}

We consider a dark sector with a non-abelian $SU(3)$ gauge group and a single flavor of heavy quarks $q$ in the fundamental representation
\begin{align}
\label{eq:Lag}
\mathcal{L} \supset -\frac{1}{4} G^{\mu \nu} G_{\mu \nu} +  \bar{q} \left( i \gamma_\mu D^\mu -  m_{q} \right)  q\,,
\end{align}
where $G^{\mu\nu}$ is the dark gluon field strength and $m_q$ is the dark quark mass with $m_q \gg \Lambda$ (in practice we consider $m_q \geqslant 100\Lambda$), 
where $\Lambda$ is the dark confinement scale at which a phase transition takes place. 

Given that $m_q \gg \Lambda$ we expect that such heavy quarks decouple from the thermal bath well before the phase transition, so that the phase diagram of this model is very close to that of pure Yang-Mills for $T \ll m_q$. Since the heavy quark regime can be well-approximated by the pure-gauge regime, the phase transition behavior is almost independent of the number of heavy quark flavors~\cite{Bonati:2012pe}. 
The only constraining condition on the number of quark flavors is then asymptotic freedom, or in special cases asymptotic safety~\cite{Pelaggi:2017abg}. 
Various lattice gauge theory studies have established that the $SU(3)$ phase transition takes place at a critical temperature very near the confinement scale, $T_c \approx \Lambda$, and is first-order~\cite{Kaczmarek:1999mm,PhysRevD.60.034504,Aoki:2006we,Saito:2011fs}.
We will therefore assume that this dark sector features a first-order phase transition exactly at $T_c = \Lambda$.

The effects of this phase transition on the relic abundance of DM have been relatively unexplored in previous studies of such a confining gauge sector, e.g. \cite{Mitridate:2017oky}.
We will discuss in this section how this phase transition dramatically changes the relic abundance calculation by causing a second stage of significant DM annihilation.

We remain agnostic about how the dark quark mass is generated as it will not affect our study. We also do not commit to any specific portal between the dark sector and the SM. We merely assume such a portal exists and keeps the dark sector in kinetic equilibrium with the SM. The portal enables the decay of the dark glueballs and mesons to the SM while respecting the dark baryon number symmetry, thus stabilizing the dark baryons. 
These baryons, which are three-quark bound states, are the DM candidate in this setup.  We will also assume a symmetric initial condition, $n_q = n_{\bq}$.

In this section we provide an overview of the cosmology of such a sector, focusing primarily on the effect that the phase transition has on the DM relic abundance. The goal is to provide the reader with a broad picture of the various moving parts in this study, while leaving some of the more detailed calculations for later sections and App.~\ref{appx:thermo}.

\subsection{Pre-confinement epoch}
\label{subsec:preconf}

For high enough temperatures, $T > T_c$, the dark sector exists in a deconfined thermal state in which quarks move freely within a gluon bath.
Naively this setup seems at odds with confinement, which requires that colored objects not propagate freely over distances greater than the confinement length, $\Lambda^{-1}$.
However, qualitatively, these colored quarks can move freely because they are connected to a network of thermal gluons~\cite{Patel:1983sc}. 
These gluons screen a quark's color charge so that the quark effectively behaves like a color neutral object, in a process analogous to Debye shielding in  plasmas. 
More quantitatively, lattice simulations have shown that when $T \geq T_c$, the potential between two heavy quarks flattens when they are separated by a distance of roughly more than $\Lambda^{-1}$~\cite{Kaczmarek:1999mm}. In other words, distant quarks in a gluon bath do not influence one another.

In the deconfined phase, the quark relic abundance calculation proceeds analogously to a standard WIMP relic abundance calculation. 
For large enough dark quark masses, $m_q \gtrsim 20 \Lambda$, which will be satisfied for all the parameter space we consider in our analysis, 
the dominant number changing process $q\bar{q} \leftrightarrow gg$ freezes out before confinement. The Boltzmann equation governing this freeze-out is simply
\begin{equation}
	\dot{n}_q + 3 H(T) n_q = - \langle \sigma v \rangle  \left(n_q^2 - \left( n_q^{\text{eq}} \right)^2\right),
\label{eq:pre-conf-freeze}
\end{equation}
where $H(T)$ is the Hubble constant at temperature $T$, $\langle \sigma v\rangle$ is the thermally averaged annihilation rate for $q\bar{q} \leftrightarrow gg$, $n_q$ is the quark number density, and $n_q^\text{eq}$ is its value in thermal equilibrium with a thermal bath of temperature $T$ and with zero chemical potential. 
Since $T \gg \Lambda$ in this epoch, $\langle \sigma v\rangle$ can be calculated perturbatively~\cite{Mitridate:2017izz}, 
\begin{equation}
    \langle 	\sigma v \rangle = \zeta \, \pi \frac{\alpha^2 (m_q)}{m_q^2} \, ,
\label{eq:pre-conf-xsec}
\end{equation}
where $\alpha (m_q)$ is the dark, strong coupling constant evaluated at the dark quark mass scale and $\zeta$ is a prefactor encapsulating plasma effects and non-relativistic enhancements (with numerical values presented in Fig.~\ref{fig:crosssections}). 
We will find that the exact size of this cross section does not qualitatively change the main results of this paper. 
For further elaboration about this cross section and others, see App.~\ref{appx:xsec}.\footnote{In addition to $\bq q$ annihilation, quarks are able to bind into diquarks via the attractive anti-triplet channel~\cite{Mitridate:2017izz} and diquarks can bind with quarks to form baryons. 
We have checked that in the pre-confinement epoch and for the parameter space we are considering, this bound state production is negligible (see Sec.~\ref{sec:boltzeqs}).
}
For the running coupling constant we use~\cite{PDG}
\begin{equation}
    \alpha (m_q) = \frac{12\pi}{(11 N_c - 2 N_f(m_q)) \log \frac{m_q^2}{\Lambda^2}} \, ,
\label{eq:alpha}
\end{equation}
where $N_f(\mu)$ is the number of light flavors contributing to the beta function at mass scale $\mu$.
We set $N_c=3$ and $N_f(m_q) = 1$.

\begin{figure}
\resizebox{\columnwidth}{!}{
\includegraphics[scale=1]{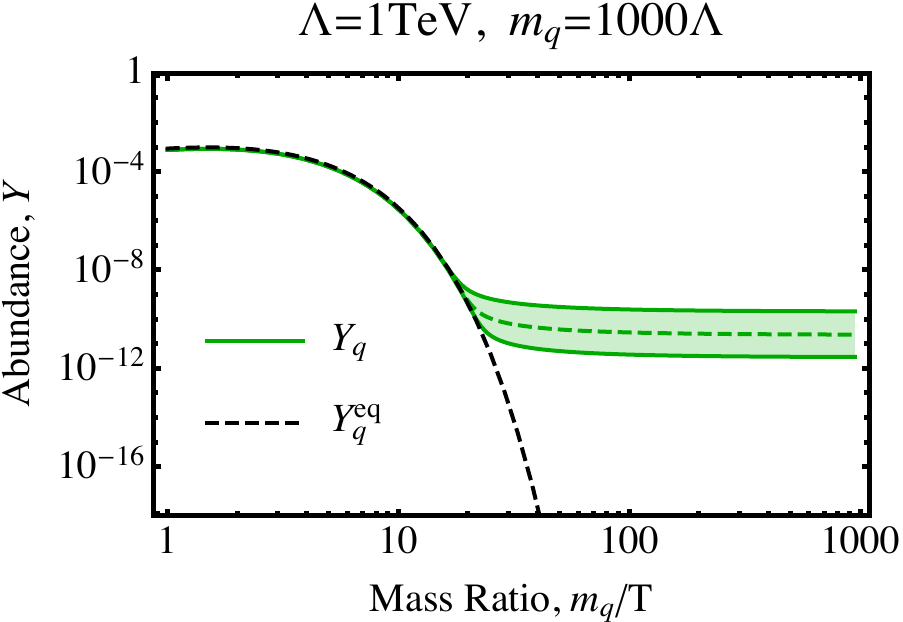}
\includegraphics[scale=1]{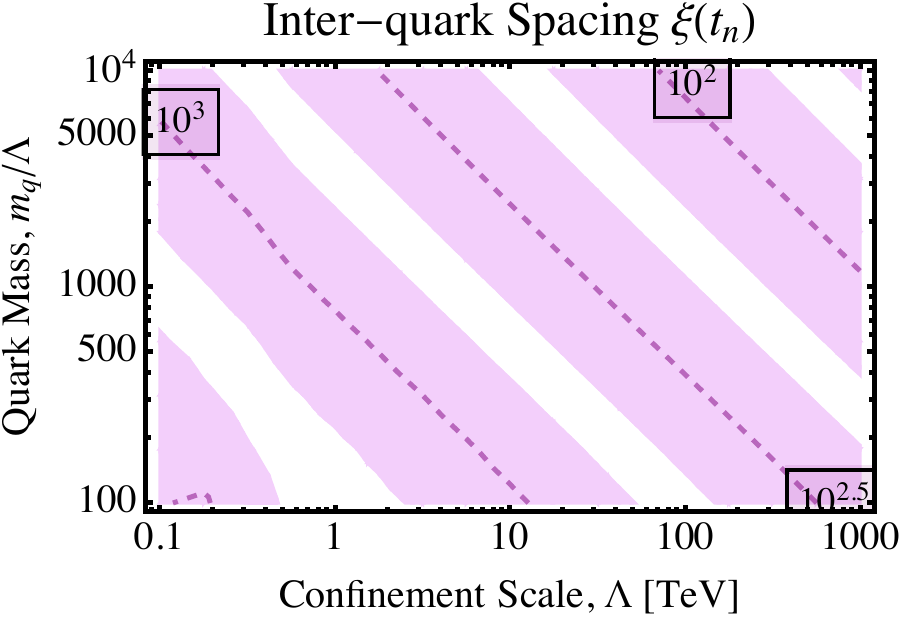}
}
\caption{ \textbf{Left:} The yield (number density normalized by the entropy density of the universe), $Y = \frac{n_q}{s}$, of the quarks for a quark mass of $10^3\,$TeV and confinement scale of $1\,$TeV. 
\textbf{Right:} The average separation of quarks as defined in Eq.~\eqref{eq:x-def} at the onset of the phase transition for various values of $\Lambda$ and $m_q/\Lambda$. 
In each plot we vary the cross section by a factor of $10$ above or below the central value (dashed lines) in Eq.~\eqref{eq:pre-conf-xsec} to produce the shaded regions.
}
\label{fig:pre-plot}
\end{figure}

In the left panel of Fig.~\ref{fig:pre-plot} we show the resulting quark number density evolution for specific choices of quark mass and confinement scale. 
A generic obstacle in the study of strong sectors is the uncertainty in determining cross sections. 
To characterize this uncertainty, we vary the cross section within an order of magnitude around the central value in Eq.~\eqref{eq:pre-conf-xsec}, which produces the green bands.

Importantly, we find that heavy quarks are well-separated just before the phase transition begins. 
To characterize their separation, we define the typical inter-quark spacing in units of the confinement length,
\begin{equation}
    \xi(t_n)\equiv \frac{\Lambda}{ \left(n_q(t_n) \right)^{1/3}} \, ,
\label{eq:x-def}
\end{equation}
where $t_n$ is the time at which extensive bubble nucleation starts, i.e. the onset of the phase transition, and $n_q(t_n)$ is the number density of the quarks at this time.  
This quantity measures, in units of $\Lambda^{-1}$, the typical distance between quarks at the onset of the phase transition. When $\xi(t)$ is large, quarks are separated by much more than a confinement length. In the right panel of Fig.~\ref{fig:pre-plot} we show $\xi(t_n)$ as a function of $m_q$ and $\Lambda$. 
Indeed, quarks are generically further from each other than a confinement length just as the phase transition begins. 
Were quarks not so well separated, the details of the phase transition would have a less dramatic effect on the DM relic abundance calculation and we would be able to use the combinatorial method of Ref.~\cite{Mitridate:2017oky}.

\subsection{Bubble dynamics}
\label{subsec:bubbles}

Once the universe cools down to the critical temperature, $T_c = \Lambda$, a first-order phase transition begins. 
Phase conversion cannot occur right at the critical temperature as both phases have the same free energy, so the temperature of the deconfined phase initially cools slightly below $T_c$.
As the deconfined phase supercools further into a metastable state, bubbles of the confined phase begin to nucleate and expand at a non-negligible rate.

As deconfined phase is converted to confined phase, latent heat is released. In contrast to weakly coupled phase transitions, there is no perturbative parameter suppressing the latent heat, meaning that phase conversion will serve as a significant heat source in the temperature evolution of the universe.
As a result, the plasma heats back up to a temperature very close to $T_c$ quickly after bubble nucleation becomes efficient. 
Since the nucleation rate is exponentially sensitive to the degree of supercooling, $(T-T_c)/T_c$, 
subsequent nucleation of bubbles is completely suppressed.

For the phase transition to continue, at least some of the bubbles from the brief period of efficient nucleation must continue to grow. 
To determine the bubble growth rate, we borrow an argument from \cite{Witten:1984rs}.
As bubbles grow, the local temperature at the bubble walls increases towards $T_c$, diminishing the free energy difference between the two phases that drives the expansion. 
The expansion rate is then limited by the rate at which the wall can cool. The cooling rate is controlled by the temperature gradient between the wall and surrounding fluid -- if there were no temperature difference, 
heat would not flow. 
Since the wall temperature cannot exceed $T_c$ without reversing direction, we assume that this temperature difference is bounded above by the small degree of supercooling $(T_c-T)/T_c$. By modeling the heat dynamics near the wall in App.~\ref{appx:thermo}, we estimate that the wall velocity is also bounded above by the degree of supercooling, $v_w \leq (T_c-T)/T_c$. 
For simplicity, we assume that $v_w$ saturates this bound.

\begin{figure}
    \caption{The degree of supercooling prior to percolation. 
    The temperature supercools until bubbles nucleate efficiently. 
    The nucleated bubbles quickly expand, depositing latent heat that drives $T$ back up to $T_c$ and eliminates further bubble nucleation.
    The temperature then stays roughly constant as latent heat deposited from bubble growth nearly cancels Hubble cooling. 
    This plot ends at percolation, when half of the universe is in each phase. 
    Notice that the timescale of the phase transition is much shorter than the Hubble timescale.
    }
    \resizebox{0.55\columnwidth}{!}{
    \includegraphics[scale=0.35]{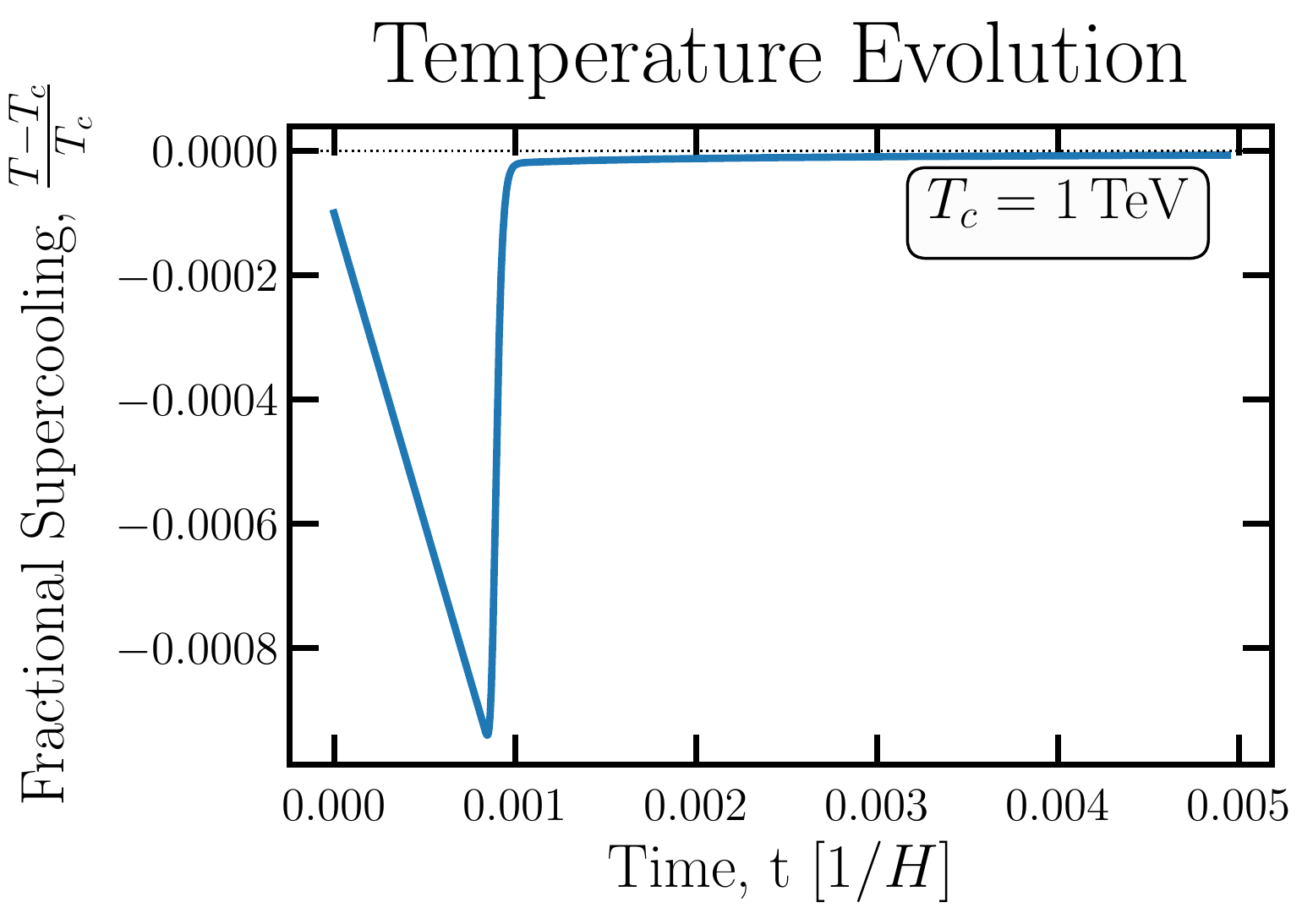}
     }
     \label{fig:supercool}
\end{figure}

In Fig.~\ref{fig:supercool} we plot the degree of supercooling as a function of time during the bubble expansion stage of the phase transition. 
This result comes from a simple simulation that we develop to track the nucleation and growth of bubbles during this epoch. Further details about this simulation can be found in App.~\ref{appx:thermo}. The stages of the phase transition discussed above are visible in this plot. 
The universe initially supercools through Hubble expansion until bubble nucleation becomes efficient, leading to quick bubble growth and latent heat injection that reheats the universe. 
The heating and cooling rates then roughly balance one another, leaving the temperature at a value very near $T_c$, which suppresses further bubble nucleation as explained above.

Eventually, half of the universe converts to the confined phase. 
At around this so-called percolation time, most bubbles are in contact with one another and start coalescing.
Soon after we are left with isolated ``pockets'' of the deconfined phase submerged in a sea of the confined phase. 
To properly compute the spectrum of shapes and sizes of these pockets would require a full numerical 3D bubble simulation.
Instead, to simplify our analysis, we will assume that soon after percolation there is a characteristic size of a typical pocket, that pockets can be approximated as spherical, and that the details of the spectrum of pocket shapes and sizes will give only sub-dominant corrections to our results.

To determine this characteristic pocket size, we first determine the characteristic size of bubbles just before percolation. Using our simulation from App.~\ref{appx:thermo},
we find that at percolation the spectrum of bubble radii peaks strongly at 
\begin{alignat}{1}
    R_0  \approx 10^{-6} \times \left(\frac{\Lambda}{M_{pl}}\right)^{-0.9} \frac{1}{\Lambda} \, ,
    \label{eq:R0-main-text}
\end{alignat}
where $M_{pl}=2.4 \times 10^{18}$~GeV is the reduced Planck mass. We now borrow another argument from \cite{Witten:1984rs} to argue that for most values of $\Lambda$, these bubbles coalesce quickly until they reach a larger characteristic size, denoted as $R_1$.

\begin{figure}
    \caption{\textbf{Left:} The typical radius of bubbles just before percolation (blue line) and the characteristic coalescence radius $R_1$ (orange line).
    For any $\Lambda$ with $R_0 \leq R_1$, we assume that bubbles quickly coalesce and grow to radius $R_1$ at percolation.
    \textbf{Right:} The asymptotic velocity of the pocket wall during its contraction as a function of the confinement scale when quark pressure is ignored. In the more realistic case where internal quark pressure is allowed to resist the contraction of the pocket, we expect $v_w$ to be much smaller, and to not necessarily asymptote to a constant value at small radii.
    }
     \label{fig:R0-R1}
     \resizebox{0.9\columnwidth}{!}{
     \includegraphics[scale=0.45]{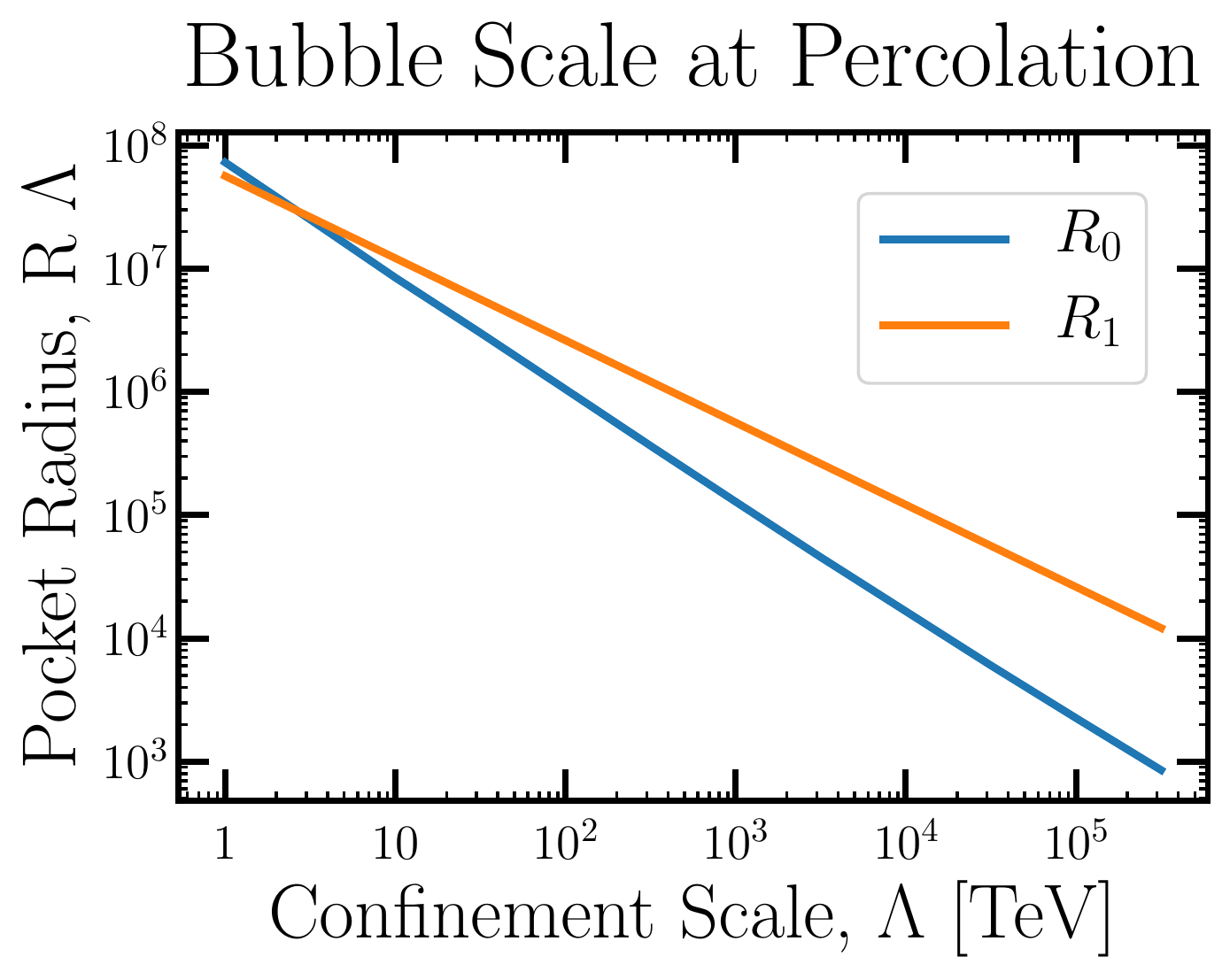}
     \includegraphics[scale=0.45]{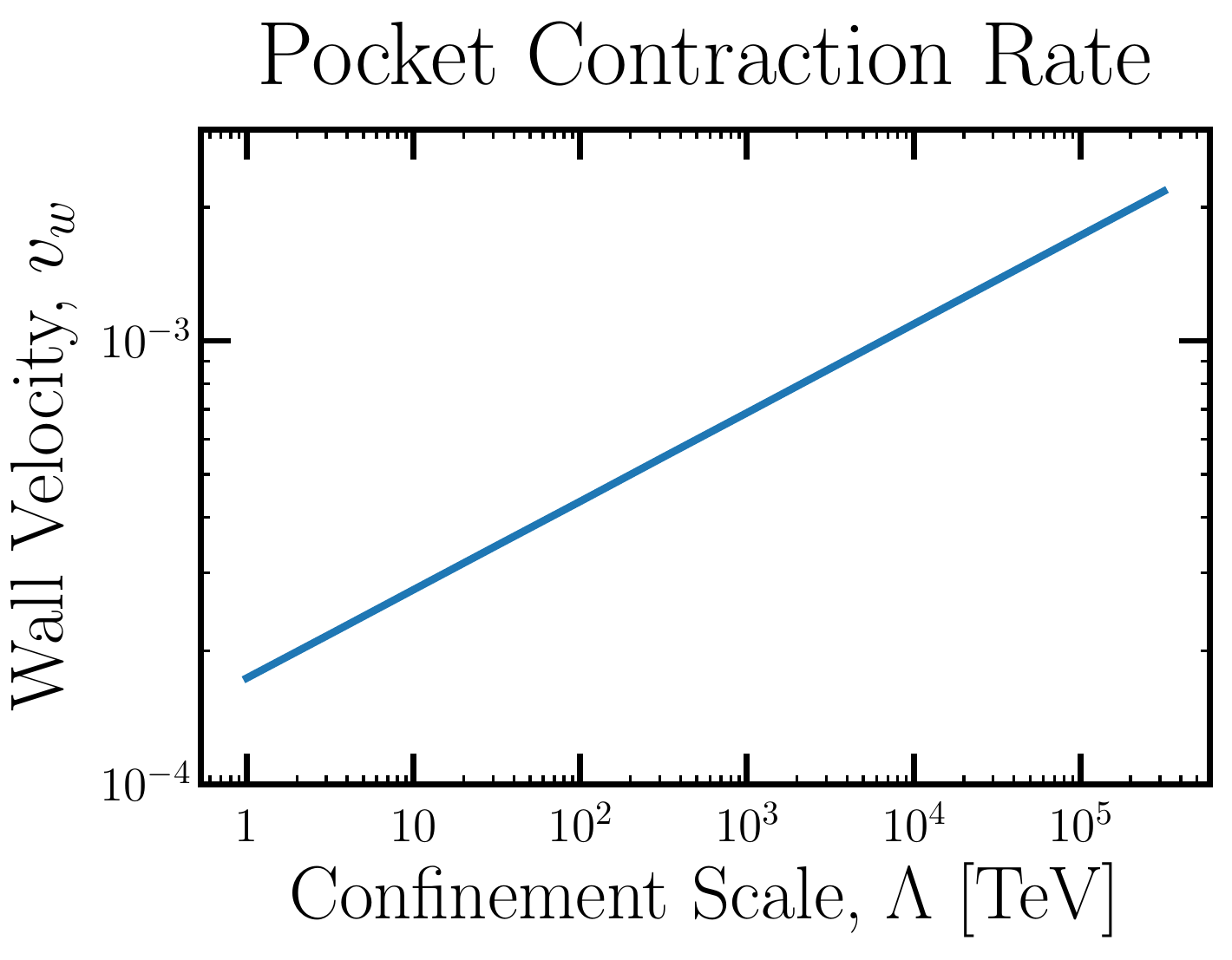}
     }
\end{figure}

The central idea is that small bubbles coalesce and merge quickly into bigger bubbles, and that the time scale for two bubbles in contact to merge becomes longer as bubble sizes grow. 
Intuitively, the larger the coalescing bubbles, the more matter has to be moved via the bubbles' surface tension, which takes more time. Thus there is a special bubble size, $R_1$, above which bubbles merge slower than the timescale over which the phase transition takes place. 
We find this critical size to be 
\begin{equation}
    R_{1} \approx \left(	\frac{M_\text{pl}}{10^4 \, \Lambda}	\right)^{2/3} \frac{1}{\Lambda} \, .
\label{eq:Rinitial}
\end{equation}
Figure \ref{fig:R0-R1} shows that for $\Lambda \gsim 1$~TeV the typical size of bubbles just before percolation ($R_0$) is always smaller than $R_1$. Thus, we assume that, for this range of $\Lambda$, at percolation all bubbles quickly coalesce until they reach a size of $R_1$. For smaller $\Lambda$s we assume that all bubbles will have radius $R_0$ instead. 
We then make the simplifying assumption that the characteristic size of pockets just after percolation is the same as the characteristic size of bubbles just before percolation, i.e.
\begin{equation}
    R_i = \max \left(R_0, R_1 \right)\, ,
    \label{eq:Ri}
\end{equation}
where $R_i$ is the characteristic initial pocket radius after percolation.

It is more complicated to determine the wall velocity of the contracting pocket. 
The main complication is that quarks are trapped within pockets, which we will show in the next section, and will generically slow down the wall. For now we will assume that we can neglect the effect of the enclosed quarks, which will lead to an overestimate of $v_w$.
In Sec.~\ref{subsec:QP_cosmo} we will revisit the effect these quarks have on $v_w$.

In App.~\ref{appx:thermo} we find that at radii much smaller than $R_i$, the pocket contraction rate asymptotes to a constant value, which is shown in the right panel of Fig.~\ref{fig:R0-R1} and can be fit by
\begin{alignat}{1}
    v_w (\Lambda) \approx 0.2 \times \left(\frac{\Lambda}{M_{pl}}\right)^{0.2}.
    \label{eq:vw}
\end{alignat}
It will turn out in Sec.~\ref{sec:boltzeqs} that the relic abundance of DM is set while $R \ll R_i$, so we can neglect the initial stages when $v_w$ varies and treat it as a constant. 
The pockets' radii therefore shrink as a function of time according to
\begin{equation}
    R(t') = R_i - v_w t'
\label{eq:Rpocket}
\end{equation}
where $t'$ is the time after percolation.

To the best of our knowledge, the problem of characteristic bubble properties, e.g. the wall velocity and charactristic size at percolation, is not completely settled for first-order phase transitions even in weakly interacting theories (see \cite{Balaji:2020yrx} and the references therein for recent discussions on calculating the wall velocity). 
In our numerical calculations in Sec.~\ref{sec:boltzeqs}, we characterize these uncertainties by varying both $v_w$ and $R_i$ within one order of magnitude of the results shown in Fig.~\ref{fig:R0-R1}.

We note here that since the quark temperature is fixed near $T_c$ throughout the phase transition, the typical quark velocity is
\begin{equation}
    v_q \sim \sqrt{\Lambda/m_q} \, .
    \label{eq:vq}
\end{equation}
For the range of parameters that we will be interested in, we find $v_q \gg v_w$. This inequality will become important in the next section when we analyze the effects that the walls have on the quarks.

To summarize, the phase transition begins with an initial, complicated stage of bubble nucleation and growth until bubbles come into contact with one another. 
It then enters an even more complicated bubble coalescence stage. The space between bubbles is made of pockets of the deconfined phase with the same characteristic size, $R_i$. 
These pockets become isolated and eventually spherical, then contract initially with a velocity that is determined by the local heat diffusion rate. The contraction rate gradually slows down due to the pressure of the enclosed quarks, and the pockets eventually vanish. 
The phase transition has completed at this point, and the universe can proceed with its standard expansion history.

Further details of this phase transition, as well as an overview of the relevant thermodynamics can be found in App.~\ref{appx:thermo}. 
Our study of the phase transition's effect on the DM relic abundance is insensitive to many of the details of the phase transition; 
we merely need an expression for the characteristic initial radius of pockets and their wall velocity, which are respectively provided in Eqs.~\eqref{eq:Ri} and \eqref{eq:vw}. 
We emphasize that this latter expression for $v_w$, which neglects the effect of quark pressure, overestimates the wall velocity during the contraction phase .

\subsection{Heavy quarks during the phase transition}
\label{subsec:quarkhistory}

During the entire process of bubble nucleation and expansion described in the previous section, bubble walls run into quarks and anti-quarks. In this section we study these encounters in detail, and argue that the walls are impermeable to quarks, but permeable to color-neutral bound states. While we will focus on the interaction between walls and quarks, our conclusions hold for anti-quarks as well.

\begin{figure}
\resizebox{\columnwidth}{!}{
\includegraphics[scale=1]{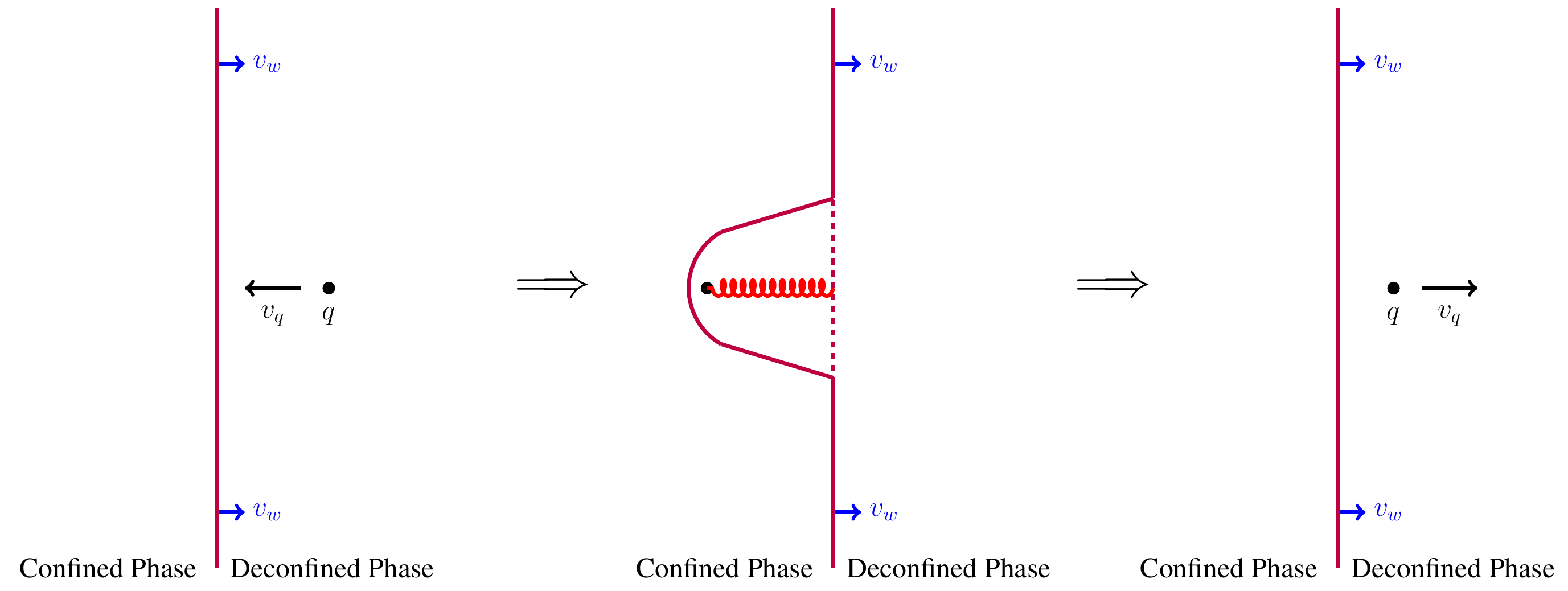}
}
\caption{
A depiction of a quark interacting with a phase boundary.
As bubbles of confined phase grow (the wall moves to the right in the figure), their walls run in to quarks that move with typical velocity, $v_q$, which is much larger than the wall velocity, $v_w$ (\textbf{left} configuration). 
The quark locally deforms the wall (\textbf{center} configuration), introducing an opposing force via surface tension. 
One can think of the quark as connected to the deconfined phase through a gluon string \cite{Patel:1983sc}. Since string breaking is shut off, i.e. the quark-gluon configuration does not have sufficient energy to pull a heavy $\bar{q}q$ pair out of the thermal background, eventually the quark comes to a halt and then rebounds back into the deconfined phase with its initial speed (the \textbf{right} configuration).  }
\label{fig:wall}
\end{figure}

When a wall encounters a quark, the quark can push against it and deform it locally. 
Whereas in electroweak-like phase transitions a particle is able to penetrate through the wall at the cost of only a finite mass difference~\cite{Baker:2019ndr,Chway:2019kft}, 
the energy cost for an isolated quark to enter the confined phase is unbounded~\cite{Kaczmarek:1999mm},
preventing it from traveling far into the confined phase. Therefore, a quark can enter a bubble only if it either forms a color-neutral bound state before it enters the bubble, 
or deforms the wall so that it remains immersed in the color-screening gluon bath (see Fig.~\ref{fig:wall}).

There are two ways in which the quark could form a bound state. 
First, $\bq q$ pairs could be spontaneously created, binding with the quark as it passes through the wall. 
We can imagine a scenario as in Fig.~\ref{fig:wall} in which the quark pushes into the bubble 
and is connected to a gluon string \cite{Patel:1983sc} starting from its initial point of contact with the wall.
If the quark were light enough, at some point this stretched string could break into a $\bq q$ pair 
and the $\bq$ could bind with the incoming quark to form a color-singlet bound state that enters the bubble (see \cite{Baldes:2020kam} for an example in which this process is efficient). 
However, for a heavy quark, the string breaking rate is extremely suppressed;
this rate can be approximated \cite{Kang:2008ea} using the Schwinger mechanism~\cite{PhysRev.82.664},
\begin{equation}
    (t_{\mathrm{string}})^{-1} \sim \frac{m_q}{4 \pi^3} e^{-m_q^2/\Lambda^2} \, .
\label{eq:t-string-lifetime}
\end{equation}
The exponential of the square of the large ratio $m_q/\Lambda$ makes this string breaking timescale much larger than all other timescales, completely shutting off this process. 
The inefficiency of string breaking and the quark's inability to pass through bubble walls is the main feature distinguishing our model from those that involve light quarks.

The second way a quark could form a bound state is by encountering an anti-quark or two quarks somewhere within the deconfined phase, binding, 
then escaping into the confined phase before the bound state dissociates. 
These processes are important and will be analyzed in the next section via Boltzmann equations.

If a quark has not managed to bind into a color singlet state by the time it reaches a bubble wall, it deforms the wall to avoid entering the confined phase. 
As the wall deforms, its surface area increases, which increases the energy of the system. 
The surface tension therefore creates a force that opposes this deformation.
If we estimate this force to be of order $\Lambda^2$ on dimensional grounds, then we find that the timescale for the surface tension to restore the shape of the wall and reverse the quark's velocity is
\begin{equation}
    t_{\mathrm{rebound}} \sim \frac{v_q}{\dot{v}_q} \sim \frac{v_q}{\Lambda^2/m_q} =  \sqrt{\frac{m_q}{\Lambda}}\frac{1}{\Lambda} \, .
\label{eq:t-rebound}
\end{equation}
This rebound timescale is much shorter than the string breaking timescale. 
It is also orders of magnitude smaller than the phase transition timescale, which we find in App.~\ref{appx:thermo} to be  $t_{\mathrm{PT}} \sim 10^{-2} H^{-1} \sim 10^{-2} M_\text{pl}/\Lambda^2$ (also, see Fig.~\ref{fig:supercool}). 
Finally, the pocket contraction timescale is of order $t_\text{contract} \sim R/v_w$. 
Since $R > \Lambda^{-1}$ and we find that $v_w \lesssim 10^{-3}$, we have $t_\text{contract} > 10^3/\Lambda$. 
Since we only consider quark masses that satisfy $\sqrt{m_q/\Lambda} \leq 10^2$ in this paper, we have $t_\text{rebound} \ll t_\text{contract}$. 
Since this rebound timescale is the shortest timescale in the problem, quarks rebound off walls very quickly before any other process can take place. 
Therefore, the bubble walls act like very stiff surfaces that quickly reflect quarks that come into contact with them.

\begin{figure}
\resizebox{1\columnwidth}{!}{
\includegraphics[scale=1]{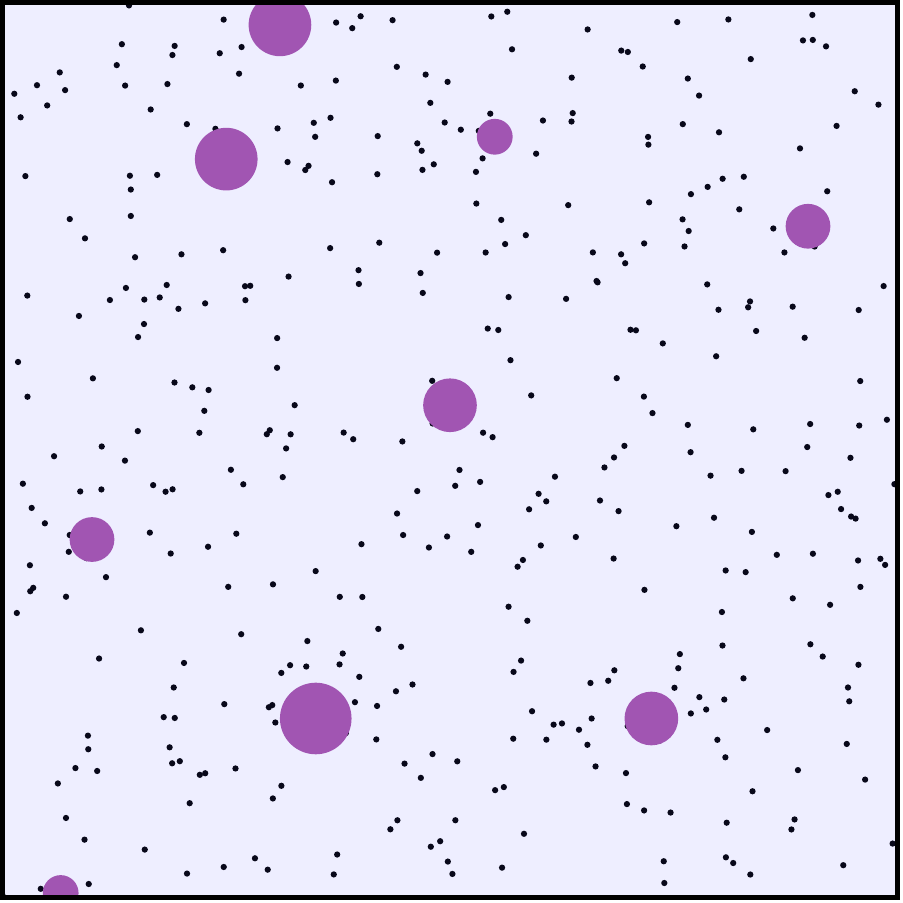}
\includegraphics[scale=1]{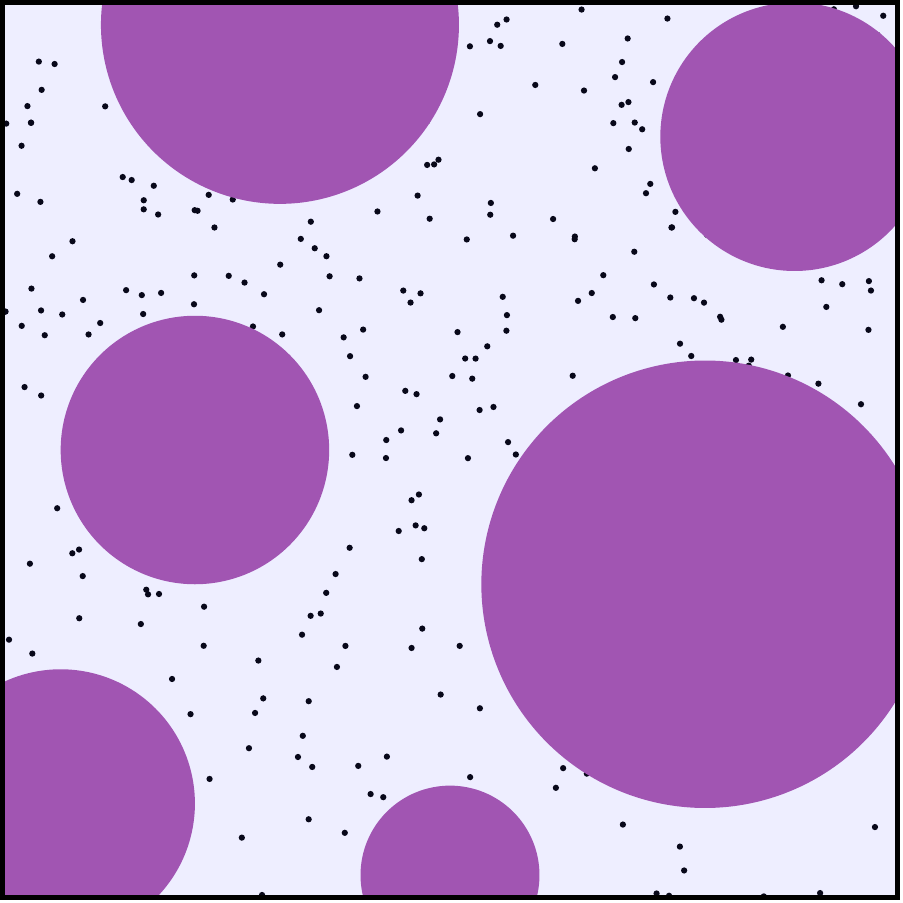}
\includegraphics[scale=1]{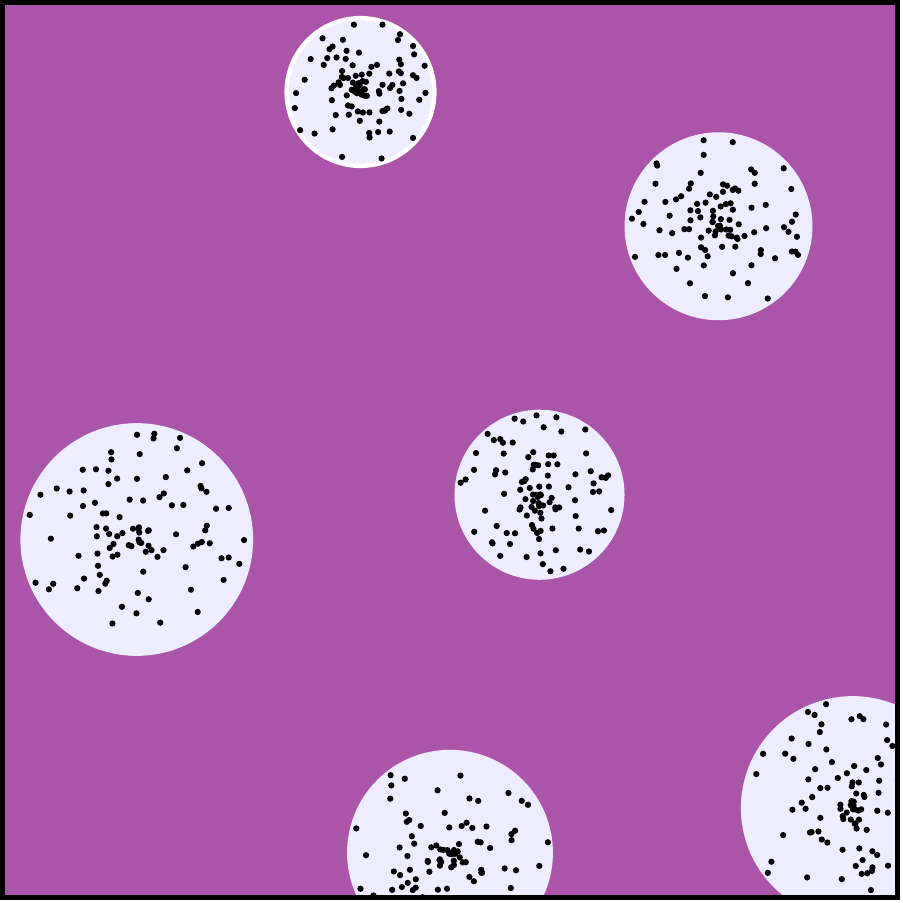}
}\\
\resizebox{1\columnwidth}{!}{
\includegraphics[scale=1]{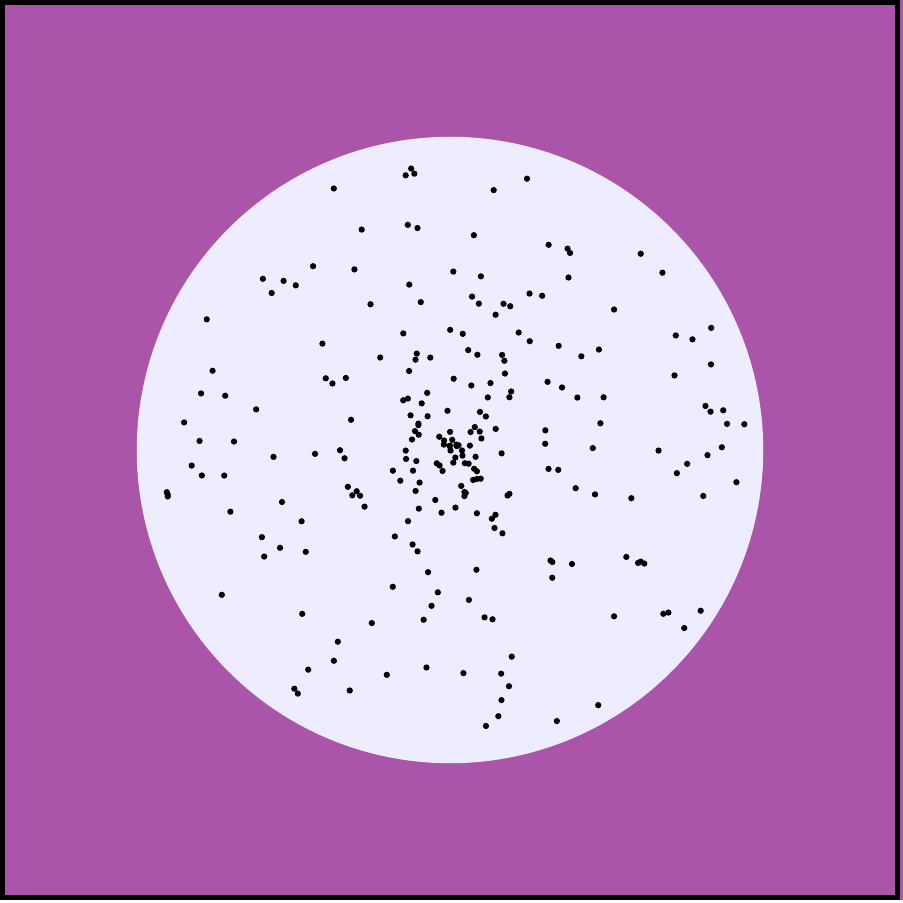}
\includegraphics[scale=1]{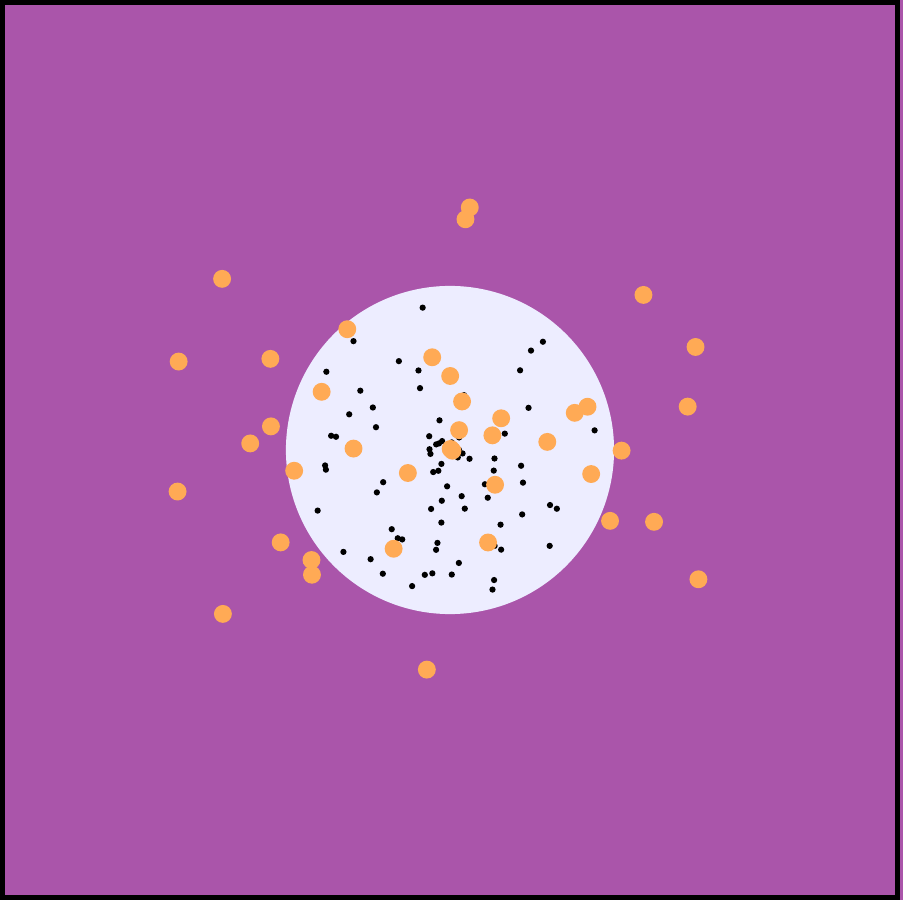}
\includegraphics[scale=1]{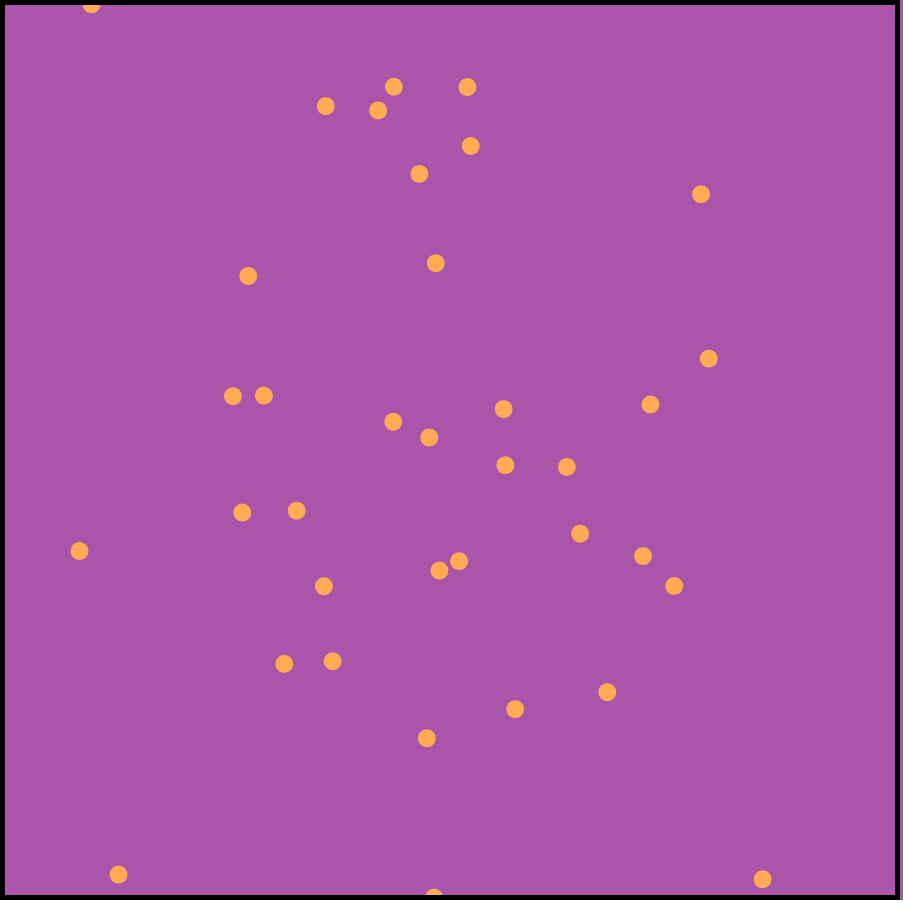}
}
\caption{ 
Schematic illustration of different stages of the phase transition and its effect on the DM abundance. 
Violet indicates the confined phase and light blue the deconfined phase. 
\textbf{Top Left:} Once the temperature drops slightly below $T_c=\Lambda$, bubbles of confined phase begin nucleating everywhere. The nucleated bubbles start growing and push quarks (black dots) around. 
\textbf{Top Middle:} The bubbles grow to a point that a $\mathcal{O}(1)$ fraction of the universe has converted into the confined phase. At this point bubbles start coalescing and quickly grow larger. 
\textbf{Top Right:} As the bubbles keep growing and combining, eventually we are left with isolated pockets of the deconfined phase submerged in a sea of the confined phase. \textbf{Bottom Left:} A single isolated pocket with quarks trapped inside it. Each pocket contracts as the phase transition continues. 
\textbf{Bottom Middle:} The particles in the pocket are compressed and their interactions recouple. During this phase the particles can either annihilate or bind with one another. Non-singlet states can not enter the confined phase, but once they form color-neutral bound states (orange dots) they can escape into the confined phase. 
\textbf{Bottom Right:} In the end, the pockets vanish and only the fraction of quarks that ended up inside color-neutral baryons survive. These particles diffuse away from the original pocket's position due to their local overdensity.}
\label{fig:phases}
\end{figure}

As these bubbles grow, the walls sweep quarks and anti-quarks into the ever-shrinking deconfined regions, increasing the quark density over time. 
Moreover, since $v_q \gg v_w$, quarks that are swept in can quickly travel through the shrinking deconfined region and maintain homogeneity, 
meaning that $n_q$ is independent of position in the pocket throughout the phase transition.
Eventually, these particles end up inside the isolated pockets formed toward the final stage of the phase transition (see Fig.~\ref{fig:phases}). 

We note that in models with additional light (though not massless) quarks, there would be no first-order phase transition~\cite{Karsch:2001vs}, and even if such a transition did exist, quarks would easily pass through bubble walls and would most likely be unaffected by the phase transition.

Within any fixed volume of the universe, including the isolated pockets, the baryon number is a fluctuating random variable. 
Although the baryon number averaged over all pockets must be zero due to our symmetric initial condition, any given pocket is expected to have an overabundance or underabundance of quarks relative to anti-quarks, which we call the pocket asymmetry, $\eta$. We find that the initial total number of quarks in a pocket, $\Nqin$, is large, 
so that by the central limit theorem the standard deviation of fluctuations above and below the mean is $\sqrt{\Nqin}$. 
Therefore, no matter how efficient $\bq q$ annihilation processes are in these contracting pockets, on average, at least a fraction 
\begin{equation} 
	\eta_\text{rms} \equiv \sqrt{\left<\eta^2\right>} = \sqrt{\Nqin}/\Nqin = 1/\sqrt{\Nqin}
\label{eq:eta} 
\end{equation} 
of the initial quarks (or antiquarks) in a pocket will survive.
This observation will have important consequences for our relic abundance calculation in the next section. 

Once isolated pockets have formed and their asymmetries have been set, 
they will contract and compress quarks and anti-quarks until formerly frozen-out interactions turn back on.
These interactions include $\bq q$ annihilation as well as $q + q$ binding via the attractive anti-triplet channel~\cite{Mitridate:2017izz}.
As these diquarks build up their occupation number, they can eventually bind with quarks to form color-singlet baryons that can 
quickly fly out of the pocket.\footnote{
Notice that the formation of baryons through an intermediate diquark is more efficient than the formation of baryons via direct $3$-body recombination, which we ignore.
}
These escaping stable baryons constitute the DM candidate of our model, while the rest of the particles eventually dump their energy into the SM sector through an unspecified portal interaction.

We define a survival factor as the fraction of quarks and antiquarks that escape the pocket within baryons and antibaryons,
\begin{equation} 
	\Sfactor \equiv \frac{N_{q}^{\mathrm{survived}}}{\Nqin} .
\label{eq:Sdef} 
\end{equation} 
In the next section we write down the Boltzmann equations governing the quark dynamics within contracting pockets and calculate this survival factor. 
As remarked above, $\Sfactor$ is bounded below by the asymmetry of a given pocket, and the expectation value of this lower bound is 
\begin{equation}
     \Sfactor \geqslant 
     \eta_\text{rms} = \frac{1}{\sqrt{\Nqin}} \, .
 \label{eq:Slimit} 
\end{equation} 
After these surviving baryons escape the pockets and after the phase transition eventually completes, these baryons continue to diffuse away until they re-establish homogeneity in the universe. If the asymmetry bound is not saturated, baryons and anti-baryons can continue to annihilate as they diffuse outside of the pocket. A more detailed study of this final annihilation stage requires integrating inhomogeneous Boltzmann equations, which we leave for future work.

In summary, the dark matter undergoes a short squeeze, where collapsing bubbles during the phase transition induce a second stage of rapid annihilation that drastically depletes the universe's pre-existing stock of dark matter. 
This extra annihilation after freeze-out opens up parameter space that had previously been ruled out due to overproduction of DM.
We will show that this effect allows for this thermal DM to be heavier than the conventional unitarity bound of $\sim 300$TeV \cite{Griest:1989wd}.

\subsection{The quark pressure on the wall}
\label{subsec:QP_cosmo}

Although we have considered the effect that the wall has on the trapped quarks, we have ignored the effect that the trapped quarks have on the wall. 
In this section, we will argue that the trapped quarks generically slow down the contraction rate of the wall.\footnote{We thank Filippo Sala for pointing out this effect.}

Much like a piston, a pocket wall can contract only if it does work on the enclosed gas of heavy quarks. 
Since we have assumed that this gas is thermally coupled to the rest of the SM bath, which has a much larger heat capacity than that of the dark sector, the quarks contract at constant temperature.
Using an ideal gas equation of state, we can write the pressure of this quark gas as $p_q = n_q T$.
The work that the wall does on the gas when it contracts by an amount $dR$ is therefore $p_q dV = 4\pi R^2 n_q T dR$. 

The forces that are responsible for this change of pocket radius are the surface tension and net gluonic pressure, the latter of which is directed inward whenever $T<T_c$.
We will show in the next section that during the earliest stages of pocket contraction, quark interactions are inefficient and the total number of quarks in the pocket is initially conserved. 
As a result, when the pocket shrinks the work required to contract the pocket grows like $R^{-1}$.
At the same time, the work that the surface tension and net gluonic pressure do when contracting the pocket by $dR$ is proportional to the change in area and volume, respectively, and so shrink like $R$ and $R^2$.
Altogether, as $R$ contracts, the forces pushing out grow while the forces pushing in shrink. We therefore expect that $v_w$ decreases with decreasing $R$. 

As this physics involves non-equilibrium, strong dynamics, we cannot reliably compute $v_w$ as a function of $R$. 
Instead, in the remainder of this section we will argue that the effect of quark pressure is to slow down $v_w$ by orders of magnitude relative to the upper bound of Eq.~\eqref{eq:vw} we computed when we neglected the quark pressure. For more details relevant to the following discussion, we refer the reader to App.~\ref{appx:pq}.

When we simulate pocket contraction while keeping track of the quark density within a pocket (see next section), we find that there always comes a point when the quark pressure has grown to such an extent that, were we to suddenly include it, the quark pressure would exactly oppose all inward-pointing forces. 
This point of mechanical equilibrium is defined by $\sigma \, dA + \left(\sum p \right) dV = 0$, where $\sigma$ is the surface tension, $dA$ the change in surface area, $dV$ the change in volume, and $\sum p$ the sum of pressures acting on the wall.
(Inward-pointing pressures are defined to be positive while outward facing pressures are negative.)
If we were to suddenly include the effects of quark pressure, the motion of the wall would suddenly become calculable, since the state of the wall would be determined by equilibrium physics. 
The pocket would suddenly slow down and proceed to adiabatically shrink while maintaining mechanical equilibrium. 
Number changing processes would deplete $n_q$, diminishing the quarks' outward-pointing pressure, and the universe would supercool further, increasing the net gluonic inward-pointing pressure. 
We find that in this scenario, $v_w$ suddenly drops by orders of magnitude when mechanical equilibrium is achieved, and $v_w$ steadily decreases many more orders of magnitude as the pocket contracts.

The discontinuous drop in $v_w$ signals a breakdown of our assumption that quark pressure was negligible before mechanical equilibrium was achieved.
This simulation merely demonstrates that it is inconsistent to neglect quark pressure, and that it can potentially slow down the pocket contraction rate by orders of magnitude.
We therefore expect that a more realistic simulation that correctly includes the effects of quark pressure from the very beginning will lead to a $v_w$ that gradually decreases from our upper bound of Eq.~\eqref{eq:vw}, which eventually overestimates $v_w$ by orders of magnitude.

We will use the results of our pocket evolution simulations to calculate a few parameters that enter the Boltzmann Equations that govern the abundances of various bound states in the pocket. While the expression for the pocket radius, Eq.~\eqref{eq:Ri}, is robust to the uncertainties introduced by quark pressure, we argued that the wall velocity $v_w$ is sensitive to this uncertainty. In the next section, we will study the evolution of the bound state abundances in the pocket in two extreme cases: (i) when the effect of quark pressure on $v_w$ is completely ignored, or (ii) when its effect dramatically reduces $v_w$.

\section{Boltzmann Equations During Compression}
\label{sec:boltzeqs}

As described above, toward the end of the phase transition, the deconfined regions form isolated pockets that contain all of the dark quarks. 
In this section, we describe the dynamics of the dark quarks and their bound states as the contracting pockets compress them. The Boltzmann equations that we solve keep track of the many processes by which quarks either ultimately annihilate into gluons or 
form baryons that escape the pockets and become dark matter.
We will solve the Boltzmann equations for a typical pocket with initial characteristic radius $R_i$ and pocket asymmetry set to its root-mean-square value, $\sqrt{\Nqin}$. We assume that the $\Sfactor$ of this typical pocket is approximately equal to $\Sfactor$ averaged over the full distribution of initial pocket radii and pocket asymmetries. 
The total number of DM particles that survive until today will then equal the total number of DM particles entering the phase transition times $\Sfactor$.

\subsection{Ingredients of the Boltzmann equations}
\label{subsec:ingredients}

We begin by listing the degrees of freedom that we will include in our Boltzmann equations, which have been tabulated in Tab.~\ref{tab:relics}. 
We have neglected a host of exotic hadronic bound states like tetra- and penta-quark states because we assume that they are unstable and promptly decay to the states listed in Tab.~\ref{tab:relics}. 
We also do not consider excited states of any of the bound states. 
To simplify the notation, we label states by their quark number throughout the text (for example, a baryon state is a $3$ state while an anti-diquark is a $-2$ state).

\begin{table}
\begin{tabular}{|c|c|c|}
\hline 
State & Dark Quark Number & Color Rep. \\ 
\hline 
Gluons & 0 & \textbf{8} \\ 
\hline 
Quark & 1 & \textbf{3} \\ 
\hline 
Diquark & 2 & $\bar{\textbf{3}}$ \\ 
\hline 
Baryon & 3 & \textbf{1} \\ 
\hline 
\end{tabular} 
\caption{Different degrees of freedom entering the Boltzmann equations of the contracting pockets. We use the quark number of each state to refer them throughout the text. The existence of anti-particles, with negative dark quark numbers and conjugate representations under $SU(3)$, are implied. }
\label{tab:relics}
\end{table}

We also neglect the mesons $\bar{q}q$ in our analysis. 
This can be justified by comparing their decay rate to the fastest annihilation rate that we will encounter (see App.~\ref{appx:xsec})
\begin{eqnarray}
\label{eq:mesonrates}
    \Gamma_{\bar{q}q} & \sim & \alpha^5 m_q, \\
    \langle \sigma v \rangle_{\mathrm{max}} n_q & \sim & \frac{1}{\alpha^3} \frac{\alpha^2}{m_q^2} \left( \frac{\Lambda}{\xi(t)}	\right)^3, \nonumber \\
    \Longrightarrow \frac{\langle \sigma v \rangle_{\mathrm{max}} n_q}{\Gamma_{\bar{q}q}} & \sim & \left( \frac{1}{\xi(t)\alpha^2} \frac{\Lambda}{m_q} \right)^3 \ll 1 ,  \nonumber 
\end{eqnarray}
where the last inequality is obtained because we have heavy quarks ($m_q/\Lambda \geqslant 100$) and the interquark spacing in units of $\Lambda$ satisfies $\xi(t) \geq 1$. 
Such a fast meson decay rate ensures that these states are kept in equilibrium so that their number density is negligibly small. We also have verified numerically that including the mesons in our Boltzmann equations below has a negligible effect on our results.

Let us now look into the Boltzmann equation for the particles in Tab.~\ref{tab:relics} as they are compressed by the contracting pockets. 
We start with the Liouville operators. 
For the colored particles, i.e. $\pm1$ (quarks/anti-quarks) and $\pm2$ (diquarks/anti-diquarks), we have
\begin{equation}
L[i] = \dot{n}_i - 3 \frac{v_w}{R} n_i, ~~~ i=1,2, 
\label{eq:L12}
\end{equation}
where the second term captures the effect of pocket compression. Notice that we have not included the usual factor of $+3Hn_i$ for the dilution of space due to Hubble expansion. 
As argued above, $t_\text{PT} \ll H^{-1}$. 
Therefore, the Hubble dilution rate is negligible during the phase transition and can be ignored.

For the color-neutral particles, i.e. baryons and anti-baryons, the compression term will be absent.
Unlike the colored particles, the baryons are not constrained by confinement to remain in the deconfined pockets. The baryons formed in the pocket can then be thought of as a gas created in a container without walls.
The gas of baryons will thus escape with a rate governed by its internal pressure, or equivalently by the 
thermal velocities of the baryons.\footnote{Notice that the justification for why baryons in the pocket are homogeneously distributed is different than that of the quarks and diquarks. 
Gradients in the baryon density naturally arise as the baryons flow from their high density points of creation to the low density exterior of the pockets. 
However, a homogeneous component of baryons is constantly being produced within a pocket due to the binding of (homogeneously distributed) quarks and diquarks. We find that the rate of production is faster than the escape rate, so the baryon density in the pocket remains homogeneous to a good approximation.
}

Once the baryons escape the pocket they are no longer tracked by the Boltzmann equations.  
So we must include baryon escape as a sink term in our Boltzmann equations, which we do by modifying the Liouville operator,
\begin{equation}
    L[3] = \dot{n}_3 + 3 \frac{v_q}{R} n_3.
\label{eq:L3}
\end{equation}
To derive this escape rate, consider a small time step, $dt$. 
In each time step the pocket radius contracts by $v_w dt$. The typical baryon moves a distance of about $v_q dt$, where we ignore the distinction between the baryon and quark velocities. We then overestimate the escape rate by an $\mathcal{O}(1)$ factor by assuming that all baryons at the edge of the bubble move radially outward, giving a total number of escaped baryons of
\begin{equation}
    dN_3^{\mathrm{esc}} = 4 \pi R^2 n_3 (R)(v_q+v_w) dt \, .
\label{eq:baryonesc}
\end{equation}
Combining this with the rate of change for pockets volume gives the density loss rate due to baryon escape used in Eq.~\eqref{eq:L3}.

It will be convenient to track the evolution of the total number of particles in a pocket as opposed to number densities. Define the pocket volume, 
\begin{equation}
    V(R)=\frac{4\pi}3 R^3 \, .
\label{eq:volume}
\end{equation}
Multiplying the number density of species $i$ by the volume of a pocket then gives the total number of species $i$ in the pocket,
\begin{equation}
N_i \equiv V n_i\,.
\label{eq:changevariables}
\end{equation}
It will also be convenient to replace the time coordinate with $R$ using Eq.~\eqref{eq:Rpocket}. We can then rewrite the Liouville operators as 
\begin{eqnarray}
\label{eq:Kliouville}
    L[i] &=& -\frac{v_w}{V} N_i',~~~i=1,2,\\
    L[3] &=& -\frac{v_w}{V} \left( N_3' - \frac{3}{R} \frac{v_q+v_w
    }{v_w} N_3  \right),
\end{eqnarray}
where $N' \equiv dN/dR$ and we have used $\dot{R} = -v_w$.

Now that we have dealt with the Liouville operators we write down the collision operators. 
We will only be concerned with $2$-to-$2$ processes since $n$-to-$2$ processes are Boltzmann suppressed while $2$-to-$n$ processes are suppressed by extra factors of $\alpha(m_q)$ and phase space factors. We denote each of these terms with the following notation,
\begin{alignat}{1}
    \Big\langle (a,b) \rightarrow (\alpha,\beta)  \Big\rangle &= \langle \sigma v \rangle_{ab \rightarrow \alpha\beta} \left(	n_a n_b - n_\alpha n_\beta \frac{n_a^{eq} n_b^{eq}}{n_\alpha^{eq} n_\beta^{eq}}	\right) \nn \\
    &=\frac{\langle \sigma v \rangle_{ab \rightarrow \alpha\beta}}{V^2} \left(	N_a N_b - N_\alpha N_\beta f_{ab,\alpha \beta}	\right) \, ,
\label{eq:intterms}
\end{alignat}
with $a,b,\alpha,\beta=0,\pm1, \pm2,\pm3$, and $f_{ab,\alpha \beta} \equiv \frac{N_a^{eq} N_b^{eq}}{N_\alpha^{eq} N_\beta^{eq}}$. 
For gluons we have $n_0 = n_0^{(eq)}$, i.e. the gluons are always in equilibrium.

Once we have identified all the important interactions to be included in our Boltzmann equations, we can write down the complete system of differential equations for $N_i(R)$. 
We supply these equations with the initial conditions, which were derived in Sec.~\ref{sec:overview}.
The initial pocket radius is $R_i$ while the initial quark number in the pocket, $N_1$, is found by multiplying the number density result of the pre-confinement freeze-out calculation in Eq.~\eqref{eq:pre-conf-freeze} by $\frac{4\pi}3 R_i^3$. 
We find that the initial conditions for $N_2$ and $N_3$ are irrelevant, as they quickly approach an equilibrium value independent of whatever values we initially choose (so long as $N_2, N_3 \ll N_1$ initially). 
All that is left is to write down these equations and solve them.

\subsection{Complete set of Boltzmann equations}
\label{subsec:numerical}

The complete set of Boltzmann equations is:
\begin{eqnarray}
\label{eq:fullboltz}
    L[i] &=& C[i], ~~~ i=1,2,3. \\
    C[1] &=& - \Big\langle  (-3,1)	  \rightarrow	(-1,-1) 	\Big\rangle - \Big\langle  (-3,1)	  \rightarrow	(-2,0) 	\Big\rangle + 2 \Big\langle  (3,-1)	  \rightarrow	(1,1) 	\Big\rangle  \nonumber \\
    &+& \Big\langle  (3,-2)	  \rightarrow	(1,0) 	\Big\rangle - \Big\langle  (1,-1)	  \rightarrow	(0,0) 	\Big\rangle + \Big\langle  (2,2)	  \rightarrow	(3,1) 	\Big\rangle - 2 \Big\langle  (1,1)	  \rightarrow	(2,0) 	\Big\rangle \nonumber \\
    &+& \Big\langle  (-3,2)	  \rightarrow	(-2,1) 	\Big\rangle + \Big\langle  (2,-2)	  \rightarrow	(1,-1) 	\Big\rangle + \Big\langle  (2,-1)	  \rightarrow	(1,0) 	\Big\rangle  \nonumber \\
    &-& \Big\langle  (2,1)	  \rightarrow	(3,0) 	\Big\rangle - \Big\langle  (-2,1)	  \rightarrow	(-1,0) \Big\rangle + \Big\langle  (3,-3)	  \rightarrow	(1,-1) 	\Big\rangle \quad, \nonumber \\
    C[2] &=&  \Big\langle  (1,1)	  \rightarrow	(2,0) 	\Big\rangle - \Big\langle  (-3,2)	  \rightarrow	(-1,0) 	\Big\rangle +  \Big\langle  (3,-1)	  \rightarrow	(2,0) 	\Big\rangle  \nonumber \\
    &-& \Big\langle  (2,-2)	  \rightarrow	(0,0) 	\Big\rangle + \Big\langle  (3,-2)	  \rightarrow	(2,-1) 	\Big\rangle + \Big\langle  (3,-3)	  \rightarrow	(2,-2) 	\Big\rangle \nonumber \\
    &-& \Big\langle  (2,-1)	  \rightarrow	(1,0) 	\Big\rangle - 2\Big\langle  (2,2)	  \rightarrow	(3,1) 	\Big\rangle - \Big\langle  (2,1)	  \rightarrow	(3,0) 	\Big\rangle   \nonumber \\
    &-& \Big\langle  (-3,2)	  \rightarrow	(-2,1) 	\Big\rangle -  \Big\langle  (2,-2)	  \rightarrow	(1,-1) 	\Big\rangle \quad , \nonumber \\
    C[3] &=&  \Big\langle  (2,1)	  \rightarrow	(3,0) 	\Big\rangle + \Big\langle  (2,2)	  \rightarrow	(3,1) 	\Big\rangle -  \Big\langle  (3,-3)	  \rightarrow	(0,0) 	\Big\rangle - \Big\langle  (3,-1)	  \rightarrow	(2,0) 	\Big\rangle \nonumber \\
    &-& \Big\langle  (3,-1)	  \rightarrow	(1,1) 	\Big\rangle - \Big\langle  (3,-3)	  \rightarrow	(1,-1) 	\Big\rangle - \Big\langle  (3,-3)	  \rightarrow	(2,-2) 	\Big\rangle \nonumber \\
    &-& \Big\langle  (3,-2)	  \rightarrow	(2,-1) \Big\rangle  -  \Big\langle  (3,-2)	  \rightarrow	(1,0) 		\Big\rangle \nonumber 
\end{eqnarray}
where $\Big\langle (\cdot \,,\cdot) \rightarrow (\cdot\,,\cdot)  \Big\rangle$ is defined in Eq.~\eqref{eq:intterms}.
The right-hand side consists of all interactions that are consistent with quark number conservation. 
We also make the approximation that 
\begin{alignat}{1}
	N_{i} = N_{-i}.
\end{alignat}
While this equality is not strictly satisfied due to the pocket asymmetry, we are able to make it because only one of three scenarios can occur:
either (i) the symmetric component is never depleted to the point that the asymmetry is important, (ii) it is completely depleted and the accidental asymmetric abundance is all that survives, or (iii) the symmetric and the asymmetric components are comparable and our answer is off by an $\mathcal{O}(1)$ factor.
As argued before, this asymmetry introduces a lower bound on $\Sfactor$ (Eq.~\eqref{eq:Slimit}).

Despite Eq.~\eqref{eq:fullboltz} having numerous terms, solving these equations numerically is rather straightforward.
For convenience we list the important parameters entering into these equations and their reference values in Tab.~\ref{tab:quantities}.
We remind the reader that Eq.~\eqref{eq:vw} overestimates $v_w$ since it neglects the quark pressure's ability to oppose pocket contraction. As we will discuss below, we can bracket the effect that a slower $v_w$ would have on the final DM relic abundance quite robustly, see Sec.~\ref{subsec:discussion} for further details. We also reemphasize that we have used simple approximations for some of other quantities -- particularly the bubble radius -- and a rigorous determination of them is only possible through more extensive numerical calculations.

\begin{table}
\resizebox{\columnwidth}{!}{
\begin{tabular}{|c|c|c|c|c|c|c|}
\hline 
Quantity & $v_w$ & $R_i (\Lambda)$ & $\xi(t_n)$ & $v_q$ & $\sigma v$ & Binding energies \\ 
\hline 
Central Value & See main text. & Eq.~\eqref{eq:Ri} & Eq.~\eqref{eq:x-def} & Eq.~\eqref{eq:vq} & See App.~\ref{appx:xsec} & See App.~\ref{appx:binding} \\ 
\hline 
\end{tabular} 
}
\caption{The relevant quantities in the Boltzmann equations and our expression for each. More discussion on how we treat $v_w$ is included in the main text.}
\label{tab:quantities}
\end{table}

\begin{figure}
\resizebox{0.7\columnwidth}{!}{
\includegraphics[scale=2]{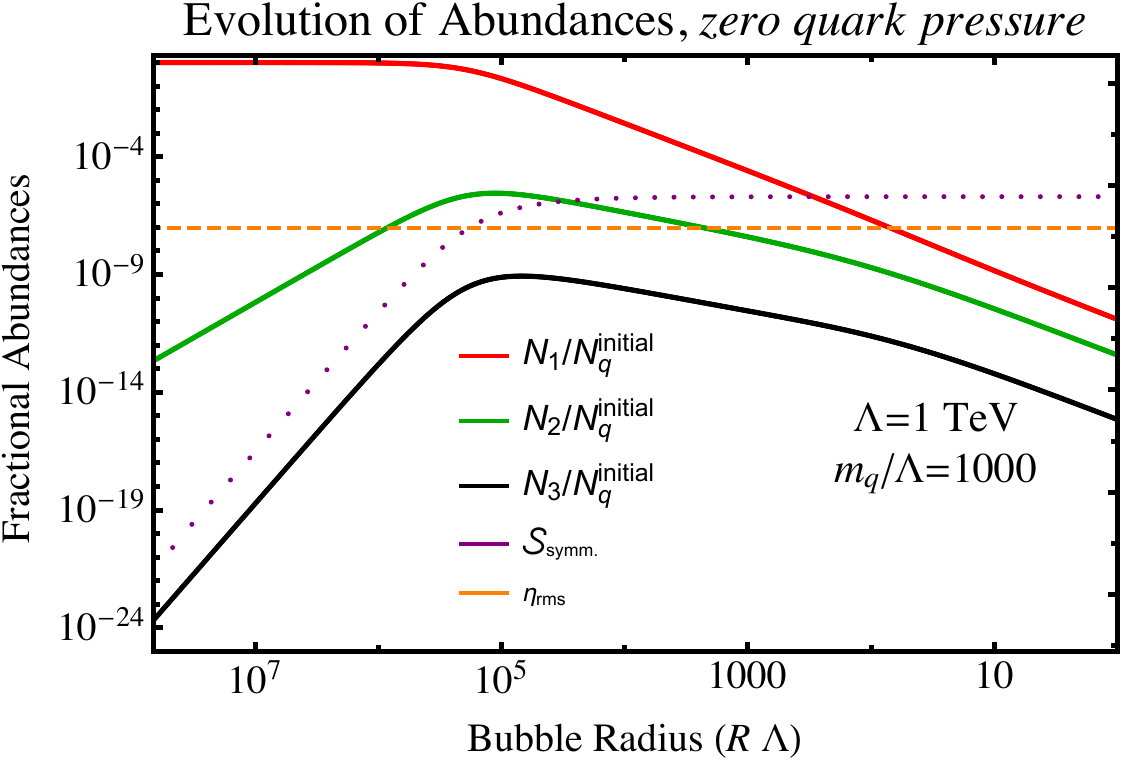}
}\\
\resizebox{0.7\columnwidth}{!}{
\includegraphics[scale=2]{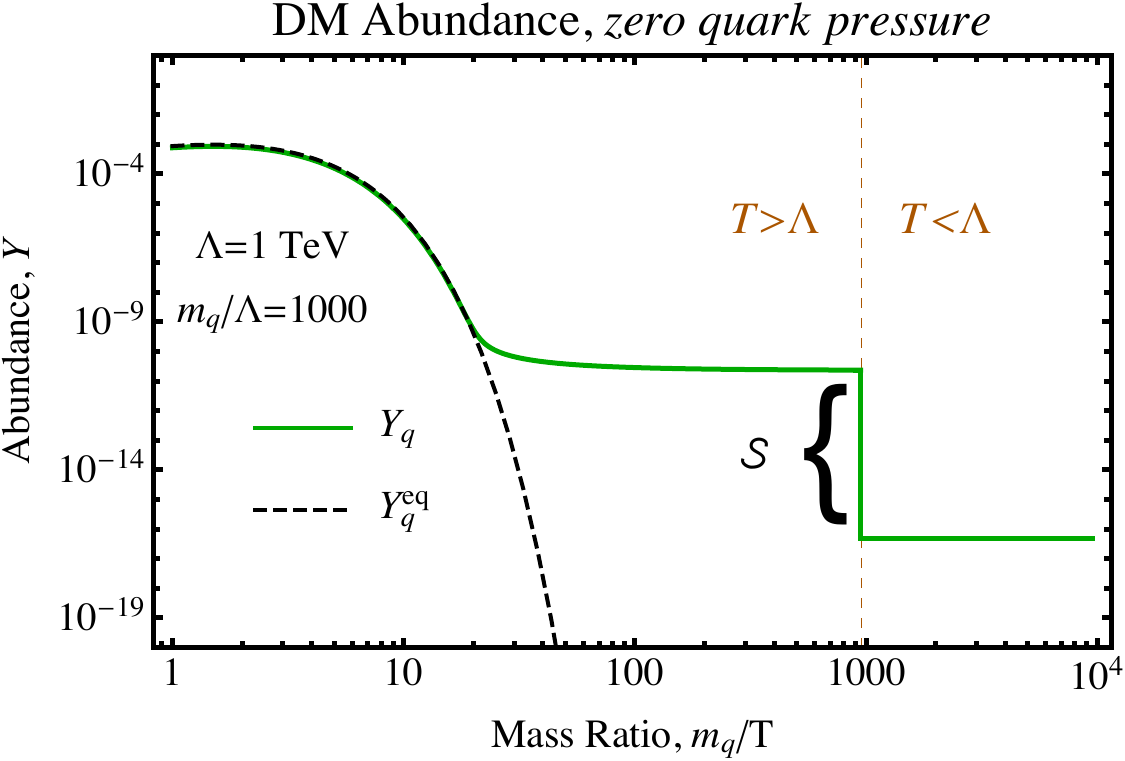}
}
\caption{
\textbf{Top:} Evolution of the fractional number of free quarks (red), diquarks (green), and baryons (black) inside the pocket, normalized to the initial quark number. 
We neglect the effect of quark pressure on $v_w$ in solving the Boltzmann equations for this plot, meaning that we use Eq.~\eqref{eq:vw} for $v_w$.
As the phase transition proceeds, the pocket radius decreases. Initially, almost all the quarks are unbound. 
As the pocket contracts, more bound states are formed and fewer quarks are found as free particles. 
As their numbers increase, the various states' annihilation rates increase as well. 
At some point their production and annihilation rates are comparable and the number of bound states inside the pocket reaches its maximum. 
The accumulative surviving fraction assuming zero pocket asymmetry predicted by Eq.~\eqref{eq:final-Sfactor-symm} is denoted by the dotted purple line. The asymptotic value of this line is equal to $\Sfactor{_\mathrm{symm.}}$. We denote the asymmetry lower bound on $\Sfactor$ from Eq.~\eqref{eq:Slimit} by the orange dashed line. \textbf{Bottom:} The DM abundance evolution for this mass and confinement scale. The $T > \Lambda$ region is similar to Fig.~\ref{fig:pre-plot}; the confinement takes place at $T=\Lambda$ and gives rise to the abundance suppression predicted by Eq.~\eqref{eq:final-Sfactor}. 
}
\label{fig:abundances-R}
\end{figure}

In Fig.~\ref{fig:abundances-R}, we show the solution of the Boltzmann equations in Eq.~\eqref{eq:fullboltz} for a specific quark mass and confinement scale when we neglect quark pressure and use Eq.~\eqref{eq:vw} for the pocket wall velocity.
There are a number of important observations to be made about this figure. 
First, the fractions of diquarks and baryons are initially very small, justifying why we did not include them in our calculations prior to pocket formation. 
Next, as the pockets contract, the number of bound states initially grows while the number of free quarks decreases due to binding or annihilation to gluons. 
As the number of free quarks decreases, the annihilation or escape of bound states become more important than their production, so their occupation numbers reach a maximum and monotonically decrease from there. 
Finally, we see that each step in the chain of bound state formation ($q + q + q \to qq + q \to qqq$) results in a suppression, 
i.e. the total number of diquarks is suppressed compared to the total number of free quarks, while the total number of baryons is suppressed compared to the diquarks. 
We anticipate that had we started with a larger $SU(N)$ gauge group ($N \geqslant 4$), 
the bound states with higher quark numbers would have been further suppressed and the final DM survival factor would be lower. 
We leave a more detailed analysis of this scenario to future work.

Finally, to calculate the survival factor, we simply integrate Eq.~\eqref{eq:baryonesc} to calculate the total number of baryons that escaped during the contraction of the pocket and normalize to the initial quark number in the pocket.
Rewriting Eq.~\eqref{eq:baryonesc} in terms of $N$ and $R$, we find 
\begin{equation}
    \Sfactor_{\mathrm{symm.}} = \frac{3\int dN_3^{\mathrm{esc}} }{N_q^{\mathrm{initial}}} = \frac{9}{N_1(R_i)}   \int dR \frac{v_q+v_w}{v_w R} N_3(R),
\label{eq:final-Sfactor-symm}
\end{equation}
where we have used Eqs.~\eqref{eq:Rpocket} and \eqref{eq:changevariables} to change variables, and the subscript in $\Sfactor_{\mathrm{symm.}}$ is to indicate that this is the survival factor of the symmetric component of the dark quarks. 
The factor of $3$ in the first equality accounts for the fact that three quarks exists within every one baryon that escapes. 

Note that in deriving this result we assumed no asymmetry exists in the pocket. Combining this result with the lower bound on $\Sfactor$ from the asymmetry component, Eq.~\eqref{eq:Slimit}, we have 
\begin{equation}
     \Sfactor = \max \left\lbrace\Sfactor_{\mathrm{symm.}}, \eta_{\mathrm{rms}}\right\rbrace.
\label{eq:final-Sfactor}
\end{equation}
In Fig.~\ref{fig:abundances-R} we show $\Sfactor_{\mathrm{symm.}}$ and $\eta_{\mathrm{rms}}$ as well. We find that for the chosen $\Lambda$ and $m_q$, while neglecting the effect of quark pressure on the wall velocity, $\Sfactor_{\mathrm{symm.}} \geqslant \eta_{\mathrm{rms}}$, i.e. the local pocket asymmetry is not saturated during the contraction, but $\Sfactor$ is within $\sim 1$ order of magnitude of this asymmetry bound. In fact, we find this is true for all the points in the parameter space that we study. 
In the upcoming section we will describe how we can leverage this observation to bracket the range of parameter space that gives rise to the correct DM relic abundance.

\subsection{Analytic approximation}
\label{subsec:analytic}

While the Boltzmann equations in Eq.~\eqref{eq:fullboltz} can be solved numerically, the large number of terms involved can muddle one's intuition. In this section, we develop a simple analytic approximation for solving these equations and determining $\Sfactor_{\mathrm{symm.}}$.

From the full set of interactions included in Eq.~\eqref{eq:fullboltz}, we identify and neglect all but the most relevant processes that provide a closed set of equations with an analytic, asymptotic solution that shows good qualitative agreement with the numerical treatment. 
The subset of processes that we include are the formation of diquarks, and the subsequent capture of quarks that lead to the formation of baryons. The reduced set of Boltzmann equations is then 
\begin{align}
\label{eq:boltz}
-\frac{v_w}{V}\,N_1' &= -\Big\langle (1,-1) \rightarrow (0,0) \Big\rangle -2 \Big\langle  (1,1)	  \rightarrow	(2,0) 	\Big\rangle   - \Big\langle (2,1) \rightarrow (3,0) \Big\rangle \nn \\ 
& \;\;\;\;\; + \Big\langle (2,-1) \rightarrow (1,0) \Big\rangle  \, ,  \nonumber \\
-\frac{v_w}{V}\,N_2' &= - \Big\langle (2,-2) \rightarrow (0,0) \Big\rangle - \Big\langle (2,1) \rightarrow (3,0) \Big\rangle + 2 \Big\langle (1,1) \rightarrow (2,0) \Big\rangle  \, , \\ \nonumber
-\frac{v_w}{V}\,N_3' &= - \Big\langle (3,-3) \rightarrow (0,0) \Big\rangle-\Big\langle (3,-1) \rightarrow (2,0) \Big\rangle + \Big\langle (2,1) \rightarrow (3,0) \Big\rangle - \frac{dN^\text{esc}_3}{dR}\,.
\end{align}

The analytic solution for this set of equations is obtained relying on several assumptions. 
\begin{itemize}
\item The initial dark quark abundance $\Nqin$ is determined by the pre-confinement freezeout of the elementary constituents. 
\item As long as the annihilation rate and the baryon escape rate in the contracting pocket is slower than the pocket contraction rate $v_w/R$, 
the total quark number is conserved. Once those rates are of the same order, the annihilation process ``recouples", and the free quark abundance drops toward zero. 
The condition $\Gamma_{\rm ann} \approx N_q^{\mathrm{initial}} \langle \sigma v \rangle_{1(-1)\rightarrow 0 0}/V = v_w/R$ defines the recoupling pocket radius
\begin{equation}
    R_{\rm rec} = \sqrt{ \frac{3 N_q^{\mathrm{initial}} \langle \sigma v \rangle_{1(-1)\rightarrow 0 0}}{(4 \pi v_w ) } }.
    \label{eq:Rrec}
\end{equation} 
\item 
The initial number of bound states, $N_X$ ($X=2,3$), is negligible. As the pocket contracts, bound states start forming. Thus, we can write $N_X \sim R^{-n}$, with $n>0$, which implies $N_X' \sim N_X/R$. Inserting this into the Boltzmann equation shows that there is a small parameter controlling the rate of change, which is proportional to $\delta = N_X \langle \sigma v \rangle/(R^2 v_w) \propto \langle \sigma v \rangle/R_1^2 \sim \langle \sigma v \rangle \Lambda^{10/3}/M_{\rm pl}^{4/3} \ll 1$. Thus expanding in $\delta$ the leading order result is obtained by setting $N'_X \approx 0$, which is the equilibrium condition before the recoupling due to pocket contraction. 
\end{itemize}

Given the above assumptions, before the annihilation process recouples, we have the quark number conservation $N_1 = \Nqin - 2 N_2 - 3 N_3$.  
Applying the equilibrium condition before recoupling and neglecting the escape and annihilation terms for the bound states at that point gives
\begin{align}
 & 2 \, \langle \sigma v  \rangle_{11\rightarrow 20}   \left( N_1^2 - \tilde{f}_1 N_2 V \right) = \langle \sigma v  \rangle_{21\rightarrow 30}  \left( N_2 N_1  - \tilde{f}_2 N_3 V \right)    \, , \nn \\ 
 &  \langle \sigma v  \rangle_{21\rightarrow 30}   \left( N_1 N_2- \tilde{f}_2 N_3 V \right)  =   \langle \sigma v  \rangle_{3(-1)\rightarrow 20}  \left( N_3 N_1  - \tilde{f}_3 N_2 V \right)    \,,
\end{align}
where
\begin{alignat}{1}
    \tilde{f}_1 &= \frac{\left( n_1^{eq}\right)^2}{n_2^{eq}} \propto \exp{(- \Delta E_1/T_c)} \, , \nn \\
    \tilde{f}_2 &= \frac{n_2^{eq} n_1^{eq}}{n_3^{eq}} \propto \exp{(- \Delta E_2/T_c)} \, ,\nn \\
    \tilde{f}_3 &= \frac{n_3^{eq} n_1^{eq}}{n_2^{eq}} \propto \exp{(- \Delta E_3/T_c)}\, ,
\end{alignat}
and $\Delta E_i$ denote the heat released during the above processes. 
The solution to the above algebraic set of equations provide the abundances of quarks and bound states in the contracting pocket before recoupling as a function of the pocket radius $R$. 
The total abundance of produced color-singlet baryons is given by the total baryon abundance $N_3$ evaluated at the recoupling radius $R_{\rm rec}$. 
Notice that $e^{-\Delta E_{1,2}/T_c} \sim e^{-\alpha^2 m_q/T_c} \gg e^{-\Delta E_3/T_c} \sim e^{-m_q/T_c}$ , where $\alpha$ is evaluated at the bound state's Bohr radius. Thus, we identify a strong hierarchy  $\tilde{f}_{1} ~,~ \tilde{f}_{2} \gg \tilde{f}_{3}$.

As a result, a simple analytic expression for the baryon fraction that survives the phase transition relative to the initial quark abundance $N_q^{\mathrm{initial}}$ can be found. In the limit of inefficient bound state breaking reactions $\tilde{f}_{1,2} \, V \ll 1$ it is 
\begin{align}
\frac{N_3}{N_q^{\mathrm{initial}}} = \frac{2 \langle \sigma v  \rangle_{21\rightarrow 30} \langle \sigma v  \rangle_{11\rightarrow 20} }{\langle \sigma v  \rangle_{3(-1)\rightarrow 20}  \langle \sigma v  \rangle_{21\rightarrow 30}  + 4 \langle \sigma v  \rangle_{3(-1)\rightarrow 20}  \langle \sigma v  \rangle_{11\rightarrow 20} } \,.
\end{align} 
Thus, assuming all the cross sections are of the same order of magnitude, we see that the baryon survival factor is of order one, if deeply bound states dominate the system. This is the case if the scale hierarchy $m_q \gg \Lambda$ is taken to be extremely large.

In the regime of efficient bound state breaking $\tilde{f}_{1,2} V \gg 1$, we find stronger DM abundance suppression. To simplify things even further, we assume the terms with $\tilde{f}$ dominate and that $\tilde{f}_1 \sim \tilde{f}_{2}$.\footnote{Both $\tilde{f}_1$ and $\tilde{f}_{2}$ depend on the ratio $m_q/\Lambda$. For $m_q/\Lambda \lesssim 1000$ they are within an orders of magnitude of each other, justifying our assumption. 
Neglecting this difference allows us to find a simple analytic formula that sheds light on the effect of various quantities on the survival factor. 
} With these assumptions, we find that at the recoupling radius we have 
\begin{align}
\label{eq:Nanalytic}
\frac{N_3}{N_q^{\mathrm{initial}}} = \frac{4\pi  v_w^3}{3 \tilde{f}_1^2 N_q^{\mathrm{initial}} \langle \sigma v  \rangle^3_{1(-1)\rightarrow 00}}\, .
\end{align} 
Now if we assume in Eq.~\eqref{eq:final-Sfactor-symm} the integral is dominated by the contribution around the recoupling point where the bound states total numbers peak, we find 
\begin{align}
\label{eq:Sanalitic}
\Sfactor_{\mathrm{symm.}}\approx 9 \frac{v_q}{v_w} \frac{4\pi  v_w^3  }{3 \tilde{f}_1^2 N_q^{\mathrm{initial}} \langle \sigma v  \rangle^3_{1(-1)\rightarrow 00}}\, .
\end{align} 

We can better understand from this equation the effects that various parameters have on the survival factor. Increasing the quark velocity $v_q$ enhances their escape rate (see Eq.~\eqref{eq:baryonesc}) thus increasing $\Sfactor_{\mathrm{symm.}}$. 
We also see that by increasing $\langle \sigma v  \rangle_{1(-1) \rightarrow 00}$, the survival factor decreases, which was expected since by increasing this cross section quarks annihilate more against each other instead of binding in bound states. For shallower bound states the binding processes are less favored, thus we expect the survival factor should decrease. This is exactly what Eq.~\eqref{eq:Sanalitic} suggests: for shallower bound states, the Boltzmann suppression in $\tilde{f}_1$ becomes less severe and $\tilde{f}_1$ increases, thus $\Sfactor_{\mathrm{symm.}}$ decreases. 

The initial density of quarks in a pocket is determined via a pre-confinement, perturbative freezeout calculation. Yet, $N_q^{\mathrm{initial}}$ in Eq.~\eqref{eq:Sanalitic} depends on the initial pocket radius too. Thus, through $N_q^{\mathrm{initial}}$ we find that $\Sfactor_{\mathrm{symm.}} \sim R_i^{-3}$.

Finally, by decreasing $v_w$, according to Eq.~\eqref{eq:Rrec}, the recoupling radius increases, which gives less time for the baryon abundance in the pocket to build up before the interactions become efficient again, see Fig.~\ref{fig:abundances-R}. 
A larger recoupling radius means a smaller peak value for the $N_3$ abundance, like the one seen at $R \, \Lambda \sim 10^5$ in Fig.~\ref{fig:abundances-R}, which in turn decreases the survival factor $\Sfactor_{\rm symm.}$. This behavior is exactly what we see in Eq.~\eqref{eq:Sanalitic}.

\subsection{The effect of quark pressure and summary of assumptions}
\label{subsec:QP_boltz}

The $v_w$ scaling of Eq.~\eqref{eq:Sanalitic} helps us better understand how our determination of $\Sfactor$ would change had we included the effect of quark pressure on $v_w$. 
This equation suggests that by using Eq.~\eqref{eq:vw} for $v_w$ and ignoring the fact that quark pressure can oppose pocket contraction, we are actually calculating an upper bound on the survival factor, since we are certainly overestimating $v_w$. 
Also, this $v_w$ scaling, combined with the proximity of $\Sfactor_{\rm symm.}$ to the asymmetry bound $\eta_{\rm rms}$ across our parameter space when we use Eq.~\eqref{eq:vw}, motivates us to believe that when the quark pressure is properly taken into account, we should expect that we saturate the the asymmetry bound $\Sfactor = \eta_{\mathrm{rms}}$ for every point in the parameter space that we study (see App.~\ref{appx:pq} for more empirical evidence of this claim). 
In the upcoming section we use these two limits to bracket the parameter space of the model that reproduces the observed DM relic abundance. 
We refer to these two limiting scenarios as the \textit{zero quark pressure} and the \textit{asymmetry} scenarios.

Before jumping to the result of solving the Boltzmann equations, it is useful to review all the parameters affecting our calculation of $\Sfactor$ and the final DM abundance. The UV model has a very limited set of parameters: the confinement scale, $\Lambda$, and the quark mass, $m_q$. These parameters feed into the calculation of a few secondary quantities that directly affect the calculation of $\Sfactor$ and are listed in Tab.~\ref{tab:quantities}. A precise calculation of these secondary quantities requires various non-perturbative studies. These quantities can be divided into two broad categories: \textit{macroscopic} and \textit{microscopic}. 

The \textit{macroscopic} quantities are those concerning the dynamics of the bubbles and pockets, i.e. their initial radius $R_i$ and their wall velocity $v_w$. While our expressions for these quantities in Eqs.~\eqref{eq:Ri} and \eqref{eq:vw} were based on a simplified simulation of the phase transition (see App.~\ref{appx:thermo}), there is extensive literature concerned with the detailed calculation of these quantities. Unfortunately, this literature has not yet settled on a single, definitive calculation of these quantities, which is why we content ourselves with simple order of magnitude estimates. (See, for example, Refs.~\cite{Ignatius:1993qn,Huber:2011aa,Ellis:2018mja,Balaji:2020yrx} and references within for various calculations of the wall velocity.)

The \textit{microscopic} quantities include various cross sections and binding energies. They also determine the dimensionless inter-quark spacing, $\xi$, which directly affects our final results as well. 
We use the results from \cite{Mitridate:2017izz} for the cross sections and the binding energies. We summarize the relevant quantities in Apps.~\ref{appx:xsec}-\ref{appx:binding}. 

It is also worth reiterating a few important assumptions that significantly streamlined our analysis. Recall that in Sec.~\ref{subsec:bubbles} we argued that the wall velocity is controlled by the amount of supercooling and quark pressure during the phase transition. 
Following that assumption, we found that the typical velocity of quarks in Eq.~\eqref{eq:vq} is much faster than the wall velocity even when the quark pressure effect is neglected in Eq.~\eqref{eq:vw}. 
Therefore, any density gradient within a pocket caused by the compression of the walls can be quickly smoothed out by the thermal motions of the quarks.
As a result, we assume that the particles within the pockets are homogeneously distributed, which simplifies our analysis significantly.

We also neglect the abundance of the bound states before the phase transition. Furthermore, as suggested in Fig.~\ref{fig:pre-plot}, we assume the quarks are initially well-separated inside the pockets, and that they rebound off the wall surface promptly. As we will argue later, all of our assumptions determine the parts of the parameter space where our analysis is valid.


\subsection{Results and discussion}
\label{subsec:discussion}

We now turn to the central results of this paper.  We scan over a range of $\Lambda$ and $m_q/\Lambda$ values, solving the Boltzmann equations at each point to calculate the survival factor $\Sfactor$. As mentioned above, we use Eq.~\eqref{eq:vw} for the wall velocity when solving the equations and finding the viable part of the model's parameter space that produces the correct present-day abundance of DM. We argued that this \textit{zero quark pressure} scenario and the \textit{asymmetry} scenario, in which we assume $\Sfactor=\eta_{\rm rms}$, are the two limiting cases that bracket the uncertainties in our DM relic abundance calculation. We will find that these two scenarios only give rise to an $\mathcal{O}(1)$ difference in the DM mass range that can explain the observed relic abundance. 

\begin{figure}
\resizebox{1\columnwidth}{!}{
\includegraphics[scale=1]{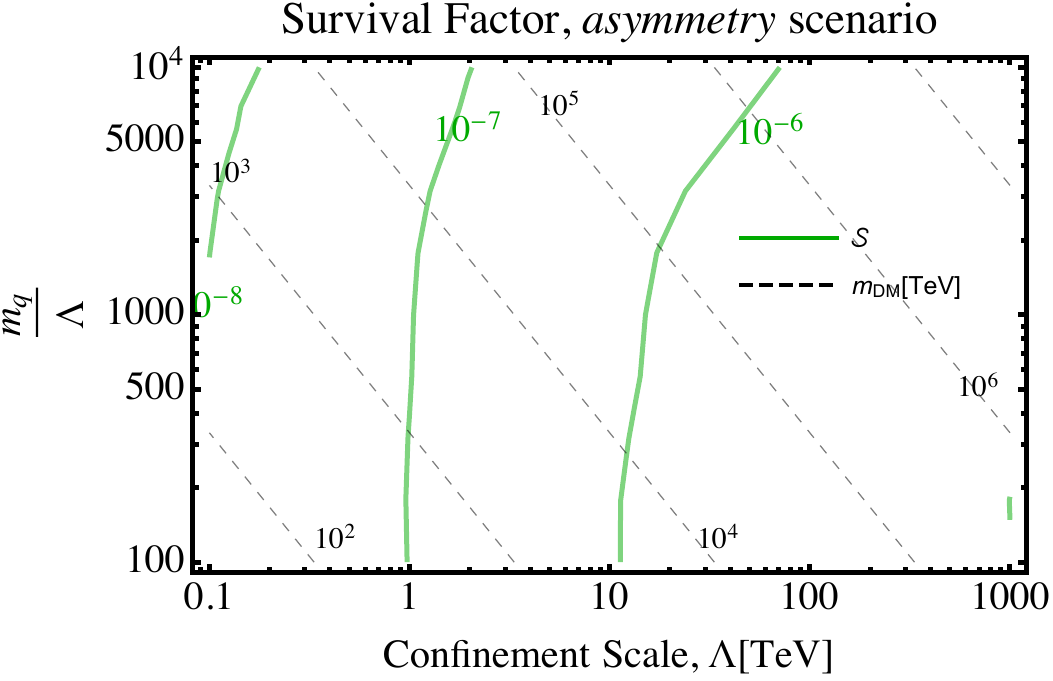}
\includegraphics[scale=1]{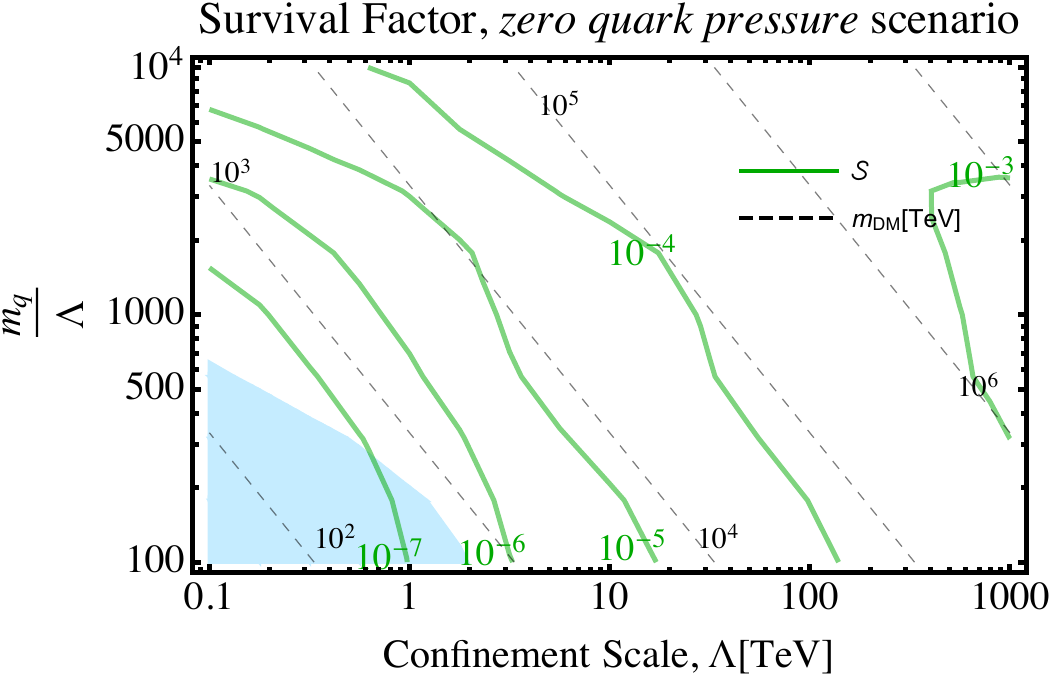}
}
\caption{Contours of constant survival factor $\mathcal{S}$ (green contours) in the two limiting scenarios that we consider: (i) assuming the asymmetry bound on $\Sfactor$ is saturated (on the \textbf{left}), or (ii) neglecting the quark pressure effect on the pocket wall velocity (on the \textbf{right}). 
The contours of constant DM mass in TeV are shown as well (black dashed). To obtain the \textbf{right} plot, we solve the full set of Boltzmann equations in Eq.~\eqref{eq:fullboltz} using the values for the initial pocket radius $R_i$ and the pocket wall velocity $v_w$ based on our simulation results discussed in Sec.~\ref{sec:overview} and App.~\ref{appx:thermo}. In the light blue region on the \textbf{right} plot we find that the suppression is so severe that only the accidental asymmetric abundance of quarks in each pockets survives after the phase transition. For the \textbf{left} plot we simply assume $\Sfactor=\eta_{\rm rms}$ for every point in the parameter space. We observe orders of magnitude suppression in the DM abundance due to the second stage of annihilation during the phase transition in either scenarios. 
Note that the small difference in the $10^{-7}$ contours in the overlap region is a plotting artifact.}
\label{fig:resultSplot}
\end{figure}

In Fig.~\ref{fig:resultSplot} we show contours of constant survival factor for both these scenarios and for different values of $\Lambda$ and $m_q/\Lambda$. The \textit{asymmetry} scenario plot shows the smallest survival factor possible while the \textit{zero quark pressure} scenario gives an upper bound on the survival factor for every point in the parameter space, see the discussion in Sec.~\ref{subsec:analytic}. 
In the \textit{asymmetry} scenario the only sources of uncertainty are those affecting the pre-confinement calculation and the initial pocket size, while in the \textit{zero quark pressure} scenario the uncertainty in determining the wall velocity $v_w$ should also be included.

\begin{figure}
\resizebox{0.7\columnwidth}{!}{
\includegraphics[scale=1]{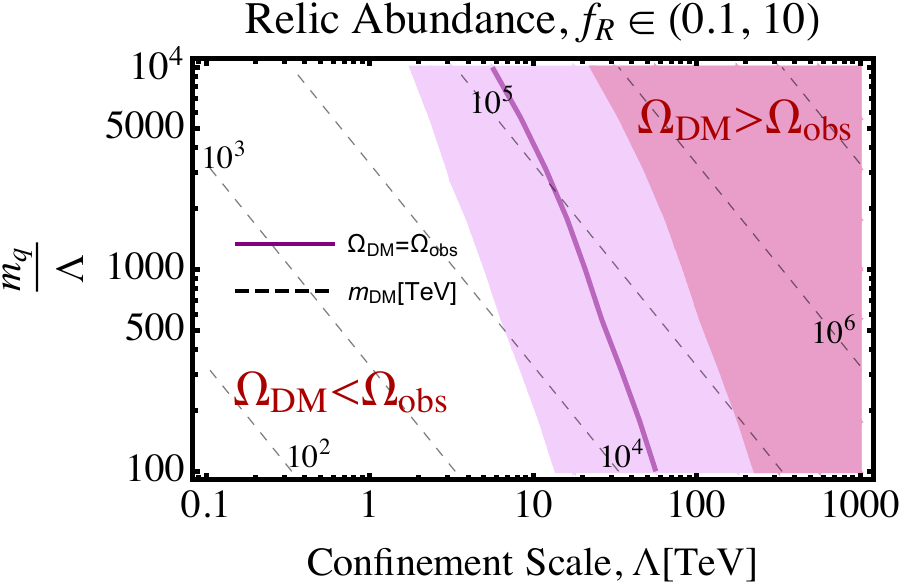}
}
\caption{
The produced abundance of dark baryons, the DM candidate, in the \textit{asymmetry} abundance scenario.
The black dashed lines are contours of constant DM mass in TeV. 
The relic abundance line (with the initial radius fixed to its central value, i.e. $f_{R}=1$) is plotted (purple line) along with its uncertainty (light purple shades) corresponding to an order of magnitude variation of the initial radius. 
The shaded red region is excluded, as it produces too much DM, while the unshaded region produces too little DM. 
The baryons can therefore constitute a sub-component of the DM within the unshaded regions of parameter space. 
The survival factor is determined by the accidental asymmetry of the pocket, which is independent of the wall velocity as long as the asymmetry bound is saturated. Thus, the major source of uncertainty in the location of the relic abundance line is the initial pocket radius. We also find that the uncertainty from \textit{microscopic} quantities is sub-dominant to those of initial pocket radius. 
This figure clearly shows that the baryon masses accounting for the observed DM abundance can be much heavier than the unitarity bound \cite{Griest:1989wd}. 
}
\label{fig:resultQP}
\end{figure}

The available parameter space in the \textit{asymmetry} limit scenario is shown in Fig.~\ref{fig:resultQP}. Equation~\eqref{eq:Rinitial} shows that as $\Lambda$ increases, $R_1$, and so the number of trapped quarks inside the pocket, decreases. Thus, as expected from Eq.~\eqref{eq:Slimit}, we find that the larger the initial radius, the smaller the survival factor. 

We should keep in mind that many simplifying approximations were made about the dynamics of the phase transition in App.~\ref{appx:thermo} in order to obtain Eq.~\eqref{eq:Ri} for the bubble radius. 
This, inevitably, introduces some uncertainty in our calculation. To characterize this uncertainty, in Fig.~\ref{fig:resultQP} we introduce a fudge factor for the bubble radius denoted by $f_R$, to be multiplied against the values from Eq.~\eqref{eq:Ri}. The observed relic abundance line moves within the light purple band as we vary $f_R$ between $0.1$ and $10$. Any point above and to the right of the relic abundance line, including the entire red region, is ruled out.

Since the \textit{asymmetry} limit scenario was the lowest attainable $\Sfactor$ in our setup, the relic abundance line in this scenario is an upper bound on the possible masses in our model. 

\begin{figure}
\resizebox{0.7\columnwidth}{!}{
\includegraphics[scale=1]{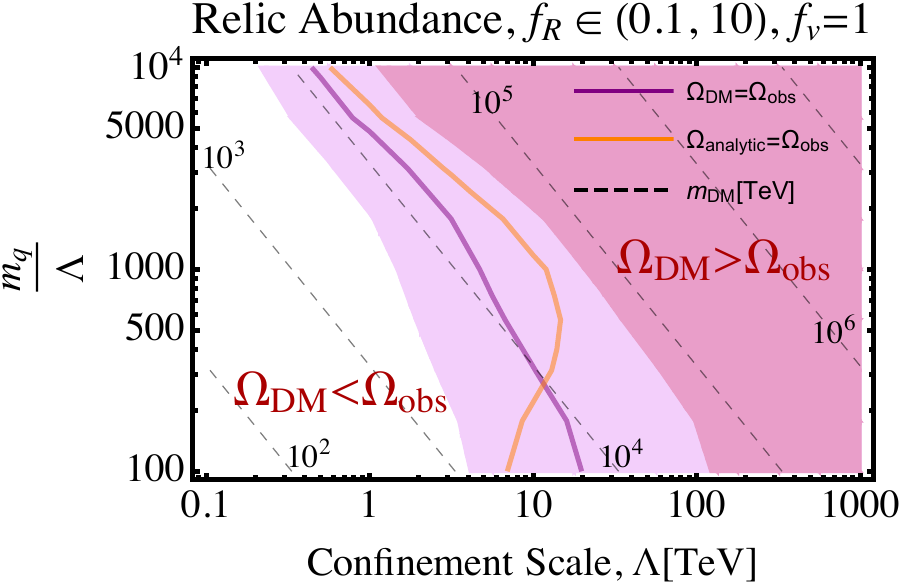}
}\\
\resizebox{0.7\columnwidth}{!}{
\includegraphics[scale=1]{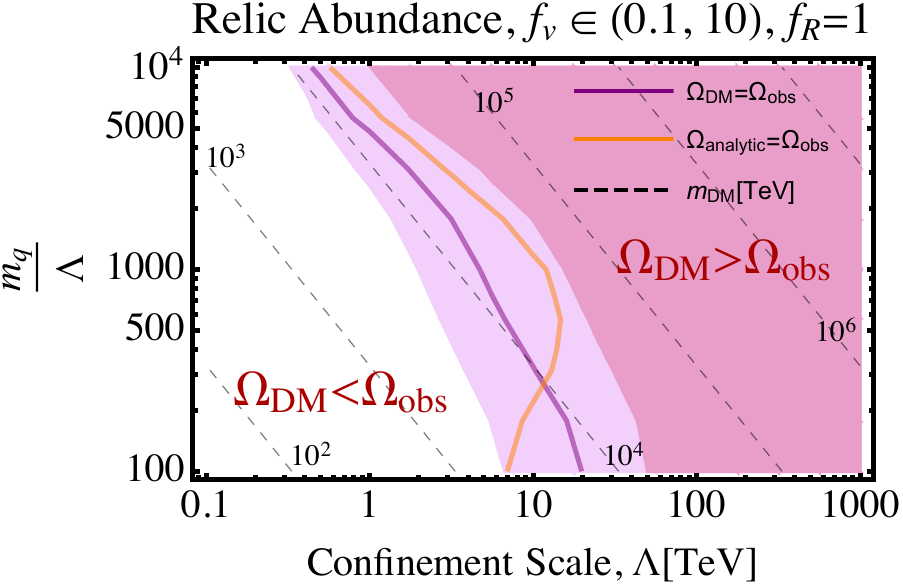}
}
\caption{Similar to Fig.~\ref{fig:resultQP} but now with \textit{zero quark pressure}. We vary the pocket initial radius (\textbf{top}) or the wall velocity (\textbf{bottom}) within one order of magnitude of the central values in Eqs.~\eqref{eq:Ri} and \eqref{eq:vw} to characterize the uncertainty in the final relic abundance calculation stemming from these quantities. The relic abundance line using the analytic approximation of Eq.~\eqref{eq:Sanalitic} for the survival factor is denoted by the orange curve too.
Even with the quark pressure neglected, we still find a substantial suppression in the DM abundance during the phase transition. We still find that the baryon masses accounting for the observed DM abundance can be much heavier than the unitarity bound \cite{Griest:1989wd}. 
}
\label{fig:result-noQP}
\end{figure}

In the other limit, the \textit{zero quark pressure} scenario provides us with a lower bound on the range of DM masses in this setup that can explain the observed DM abundance. In Fig.~\ref{fig:result-noQP} we show the available parameter space in this scenario. The calculation can now be affected by a change in both the initial pocket radius $R_1$ and its wall velocity $v_w$. To characterize this uncertainty, in Fig.~\ref{fig:result-noQP} we introduce a fudge factor for both the bubble radius 
and the wall velocity, denoted by $f_R$ and $f_v$, respectively. As expected, for any fixed $\Lambda$ the observed DM relic abundance in this scenario is obtained by smaller DM masses than that of the \textit{asymmetry} scenario in Fig.~\ref{fig:resultQP}.

In this figure we also show the relic abundance line when we use the analytic approximation of Eq.~\eqref{eq:Sanalitic} to calculate the survival factor. We find reasonable agreement between our analytic approximation (the orange curve) and the full numerical result (the purple curve).

In both these limiting scenarios studied in Figs.~\ref{fig:resultQP}-\ref{fig:result-noQP}, we find a similar range of DM masses and $\Lambda$ that can account for the present-day DM abundance. 
We expect that these two scenarios bracket the true location of the relic abundance line when the effect of the quark pressure on the pocket wall velocity is appropriately included.
The figures indicate that, depending on the \textit{macroscopic} parameters, the region of parameter space that produces the observed DM abundance predicts $m_{\mathrm{DM}} \sim \mathcal{O}(1)-\mathcal{O}(100)$~PeV, well above the thermal relic unitarity bound of $m_\text{DM} \lesssim 300$~TeV~\cite{Griest:1989wd}. 
Even with various sources of uncertainty, our results predict a confinement scale roughly in the $\mathcal{O}(1)-\mathcal{O}(100)$~TeV range, in contrast to \cite{Mitridate:2017oky}, which predicts a much wider range of confinement scales in such models.
The parameter space above this range is ruled out, while the remaining parameter space is allowed, producing a sub-component of DM.

The values of the cross sections and the binding energies entering the Boltzmann equations can be found in appendices \ref{appx:xsec} and \ref{appx:binding}, respectively. We find that the uncertainty in our results due to these \textit{microscopic} quantities is sub-dominant to the uncertainty from the \textit{macroscopic} bubble dynamics parameters discussed above. For further details about these parameters and the uncertainties in determining them see the aforementioned appendices and the references therein.

Determining the exact position of the relic abundance line requires more precise calculations of both \textit{macroscopic} and \textit{microscopic} quantities. Nonetheless, such calculations will not change our qualitative conclusions: that the phase transition gives rise to a new stage of annihilation that reduces the relic abundance by orders of magnitude and shifts the DM mass to well above the unitarity bound.

We can also understand the expected results for the parts of parameter space not plotted. 
For larger $\Lambda$s than were plotted, Figs.~\ref{fig:resultQP}-\ref{fig:result-noQP} suggests that this model always overproduces DM and is ruled out. For smaller $\Lambda$s than were plotted, our assumption that the pre-confinement abundances of bound states are negligible breaks down.
Since so many baryons are produced before the start of the phase transition, the survival factor becomes comparable to 1. For low enough $\Lambda$, we should use the combinatoric calculation of the relic abundance described in \cite{Mitridate:2017oky}. Further investigation of this region is left for future works.

As we go to larger values of $m_q/\Lambda$, our assumption that $v_q \gg v_w$ breaks down. In this case, local inhomogeneities appear in the distribution of particles in the pockets and the entire homogeneous system of equations in Eq.~\eqref{eq:fullboltz} must be modified. 
Furthermore, we find that, for higher $m_q/\Lambda$ than is shown in Figs.~\ref{fig:resultQP}-\ref{fig:result-noQP}, the quark separations during the contraction epoch can become as low as $\sim 1/\Lambda$ (due to the small cross sections allowing for a greater degree of compression). 
In this case, the picture of well-separated quarks that rebound off the stiff bubble wall (before they run into other colored particles) must be modified. Non-perturbative effects become more relevant in this case. It is also possible that at such high densities quarks bind into more stable and massive dark nuclear states such as nuggets, see Ref.~\cite{Bai:2018dxf} for a study of dark quark nuggets in the light dark quark limit and Ref.~\cite{Hong:2020est} for similar macroscopic objects. 

Finally, as we go to lower values of $m_q/\Lambda$, eventually the first-order phase transition turns into a second order one and then a cross over, see e.g. \cite{Saito:2011fs}. In this regime there will be no bubble walls to compress quarks into a second stage of annihilation. Even for lower values of $m_q/\Lambda$ for which there still exists a first-order phase transition we run the risk of breaking our assumption that the string breaking rate is negligible, so that quarks and diquarks can escape from the pocket before significant annihilation takes place.

All in all, outside of the window shown in Figs.~\ref{fig:resultQP}-\ref{fig:result-noQP}, either the parameter space is already ruled out, or at least one of the simplifying assumptions we made fails and our analysis becomes unreliable.

\subsection{Extensions of our analysis}
\label{subsec:extensions}

So far we have focused on a confining $SU(3)$ gauge group with a single generation of heavy fermions in the fundamental representation. 
Nonetheless, it is conceptually straightforward to repeat our analysis for slightly different setups. In this section we comment on the differences that we expect would have arisen had we varied the number of colors, $N_c$, or the quark representation.

Had we chosen a gauge group with a larger number of colors, $SU(N_c \geqslant 4)$, we expect that we would have found a smaller $\Sfactor_{\mathrm{symm.}}$ since the stable DM candidate in such a theory (the analogue of the baryon) requires more constituent quarks to bind together in more steps. 
(Notice that in Fig.~\ref{fig:abundances-R} as the quark number of a state increases, its abundance decreases within a pocket.) However, if even with $N_c=3$ we find $\Sfactor_{\mathrm{symm.}} \leqslant \eta_{\mathrm{rms}}$, we expect to saturate the asymmetry bound for larger gauge groups as well.
The additional $\Sfactor_{\mathrm{symm.}}$ suppression would not change the final survival factor $\Sfactor$.

The quark representation under the dark gauge group has a slightly more complicated effect on our results. For any quark representation, one must first identify the list of all possible bound states and then write down the Boltzmann equations with all possible interactions. As explained in Sec.~\ref{subsec:analytic}, the binding energies of these bound states can also have a significant effect on the solutions of the Boltzmann equations.
As an example, consider the case in which quarks are in the adjoint representation of the group. These quarks can bind with gluons to form color-neutral gluequarks \cite{Contino:2018crt}. Since the gluons can be found abundantly, we expect that quarks can easily pass through pocket walls by binding with a nearby gluon. 
Thus, pocket walls will not compress the quarks, and there will be no second stage of annihilation due to the phase transition.

Besides changing the model under consideration, our work would also benefit from improving our order of magnitude estimates and simplifying assumptions. 
Dedicated numerical simulations that more carefully model the bubble dynamics and non-perturbative physics could reduce the uncertainties in both \textit{macroscopic} and \textit{microscopic} quantities listed in Tab.~\ref{tab:quantities}, narrowing down the uncertainty on the relic abundance line in Figs.~\ref{fig:resultQP}-\ref{fig:result-noQP}. 

Finally, the glueball lifetime can have profound effects on our results and should be studied in more details in a concrete model. If they live long enough to dominate the energy budget of the universe, after their decay they inject entropy into the bath, further diluting the dark baryon. This pushes the relic abundance line in Figs.~\ref{fig:resultQP}-\ref{fig:result-noQP} to even heavier dark baryon masses. The lifetime of the glueballs depends strongly on the dimension of the portal operator that enables their decay to SM particles~\cite{Juknevich:2009gg,Forestell:2016qhc,Forestell:2017wov,Contino:2020tix}. 

When this operator is dim-8, the glueballs are long-lived and a dilution in the dark baryon abundance is expected. When the operator is dim-6 instead, the glueballs decay fast and do not affect the final baryon abundance as long as $m_q/\Lambda \lesssim 10^4$. Models that contain such dim-6 decay operators can contain a Higgs portal coupling to the dark sector quarks, see Ref.~\cite{Mitridate:2017oky} for explicit models. The dark baryon relic abundances in these models should be re-evaluated taking into account the new effects discussed in this work.

\section{Potential Experimental Signals}
\label{sec:pheno}

Our study so far only relied on fairly general properties of a dark sector. 
We only assumed the dark sector under study is a confining $SU(3)$ gauge theory with a single generation of heavy fermions; we also assumed a portal exists between the sectors that keeps them in kinetic equilibrium and allows the glueballs and mesons to decay to the SM. All the conclusions drawn in the previous sections were independent of further details of the portal and the origin of the heavy dark quark mass.

A detailed study of all the phenomenological signals of such a sector has to be carried out in a model-dependent way with a specified portal. 
As a result, here we merely list the signals and constraints that should be expected from this broad class of models.

\begin{itemize}

\item The main feature of our setup is a first-order phase transition in the early universe. Such a phase transition can also give rise to a stochastic gravitational wave (GW) background that can be detected in a host of different future experiments, e.g. see Refs.~\cite{Schwaller:2015tja,Halverson:2020xpg,Huang:2020mso,Wang:2020jrd} for a recent study of the GW signals of confining dark sectors. The characteristics of the resulting GW, such as the frequency and the strength, depend on a handful of thermodynamical parameters, see \cite{Weir:2017wfa} for a brief review. This GW signal is independent of the portal to the SM. A naive estimate\footnote{We use the formulas in \cite{Caprini:2015zlo} to estimate the GW signal produced during the phase transition. We use the interface introduced in \cite{Moore:2014lga} to compare the result to the reach of various experiments.} 
shows that different parts of our parameter space could potentially be probed in future experiments like 
DECIGO \cite{Seto:2001qf,Kawamura_2006} and BBO \cite{Harry:2006fi}. Early universe phase transitions can also give rise to anisotropies in the GW spectrum, which can potentially be detected at future facilities (see e.g.~\cite{Geller:2018mwu}). Given the extremely high mass range of the DM candidates in our model, the GW signals could have the highest discovery potential in such sectors. We leave the further study of GW signals in this class of models for future work.

\item The glueballs and the mesons are unstable due to the portal to the SM. Stringent bounds from BBN require that these relics have a short lifetime. See for instance \cite{Cyburt:2009pg,Kawasaki:2017bqm,Forestell:2018txr} for recent studies. 
As a rule of thumb, one can avoid various constraints by assuming all these bound states decay before the BBN, i.e. their lifetime is $\tau \leqslant 1$s. This bound on the lifetime introduces a lower bound on the strength of the portal. This lower bound can vary substantially depending on the details of the portal. Our requirement that both sectors are in kinetic equilibrium also imposes a lower bound, though we expect the BBN bound to be more stringent.

\item The portal to the SM introduces possible direct and indirect detection signals. However, the DM number density in the universe and in our galaxy is very suppressed due to this model's heavy DM mass. A naive estimation suggests that our model's indirect signal from DM annihilation within the Milky Way is severely suppressed and undetectable. The direct detection signal, however, depends on the details of the portal and should be studied model-dependently. We note that in this heavy mass range even very large DM-SM elastic cross sections are allowed, but within the reach of upcoming and ongoing experiments~\cite{Cappiello:2020lbk}.

\item A separate indirect detection signal comes from the observation that our composite DM model admits excited states. 
De-excitations from these excited states might lead to radiation that could be detected. 
Excitations could be produced in the early universe or via interactions with matter today.

\item Yet another indirect signal could come from the capture of DM in celestial bodies see for example Refs.~\cite{PhysRevD.40.3221,Baryakhtar:2017dbj}.\footnote{See also~\cite{VanTilburg:2020jvl} and \cite{Leane:2020wob} for studies of lighter DM capture in gravitational basins or exoplanets, respectively.   } As DM accumulates at the bottom of these potential wells, it can begin to annihilate at a significant rate, possibly affecting the evolution of these celestial bodies in an observable way or enhancing a potential annihilation signal~\cite{Leane:2017vag,Leane:2021ihh}.

\item For the $\mathcal{O}(\text{PeV})$ and above DM masses predicted in our model, direct production of DM at collider facilities is not possible in the foreseeable future. Yet, if the portal is substantially lighter, it can be directly observed at collider experiments. While the dark quarks are too heavy to produce at collider facilities, the glueballs of the new dark sector, whose mass is $\mathcal{O}(10\Lambda)$, could potentially be produced at future colliders.

\item Various studies suggest an upper bound on the DM self-scattering \cite{Markevitch:2003at,Feng:2009mn,Buckley:2009in,10.1111/j.1365-2966.2012.21182.x,10.1093/mnrasl/sls053,Tulin:2017ara,Bondarenko:2020mpf}. As a rough estimate 
\begin{alignat}{1}
\sigma_{\mathrm{SI}}/m_\text{DM} \lesssim 1~\text{cm}^2 \text{g}^{-1}  \sim  (60~\mathrm{ MeV})^{-3} \, .
\end{alignat}
It is straightforward to check that for the high confinement scales we are studying, this upper bound is easily satisfied. 

\item One can also search for signals coming from the inhomogeneities in the DM density that were produced during the phase transition when DM was compressed by contracting pockets, but this seems unlikely.  By performing a Jeans stability analysis we find that the internal baryon pressure easily overcomes the self-gravity of these overdensities.  Pockets therefore do not seed self-gravitating DM clumps.  One might also look for modifications to the matter power spectrum due to these overdensities, but initial estimates indicate that the matter power spectrum would only be modified at unobservably small mass scales if at all. Specifically, the total DM mass within a horizon radius soon after the phase-transition epoch (after which the comoving abundance is fixed) can be estimated as the DM density multiplied by $H^{-3}$, with a DM density crudely approximated (ignoring changes in the number of relativistic degrees of freedom over time) as $\sim (\Lambda/T_\text{CMB,0})^3 \times$ the present cosmological density of DM, where $T_\text{CMB,0} \sim 2\times 10^{-4}$ eV is the present-day temperature of the radiation bath. This gives an enclosed mass:
\begin{equation} M_\text{enc} \sim (\Lambda^2/M_\text{pl})^{-3} (\Lambda/T_\text{CMB,0})^3 \times 10^{-6} \text{GeV/cm}^3 \sim \left(\frac{1 \text{TeV}}{\Lambda}\right)^3 \times 10 \text{kg} \end{equation}
Thus for phase transitions at the TeV scale and above, we would expect phase-transition-induced inhomogeneities to affect DM clumps at the kg scale and below. Even if these clumps survived, this mass scale is vastly lower than can be probed by any possible observational constraints on the matter power spectrum, which are currently exploring halo masses of order $10^{7-8} M_\odot$ (e.g.~\cite{Nadler:2019zrb, Gilman:2019nap, Schutz:2020jox}).

\end{itemize}

Because of its low number density, the dark matter in our setup can have significant interactions with the SM particles and still have escaped detection so far.
Creative new search strategies will be needed to explore this possibility. Novel ideas for direct detection of such a scenario have been put forward in Ref.~\cite{Cappiello:2020lbk}, and interesting signals in heavy isotope searches \cite{Polikanov:1990sf} could arise if our dark baryons can bind to SM atoms and nuclei.

In addition to the above signals, which should exist for any specific realization of the DM-SM portal, there may exist additional portal-dependent signatures. We also find, using the results of Ref.~\cite{Mitridate:2017oky}, that depending on the type of the portal to the SM the glueballs lifetime could be larger than the Hubble time at $T=\Lambda$. In such a scenario, the delayed decay of the glueballs can further dilute the DM abundance \cite{Contino:2018crt,Dondi:2019olm} in the parameter space that we have studied, thus pushing the relic abundance line in Figs.~\ref{fig:resultQP}-\ref{fig:result-noQP} to even higher DM masses. A proper study of this effect, as well as other signals from any specific portal, is left for future works.

\section{Conclusion}
\label{sec:conclusion}

In this work we studied the consequences of a first-order phase transition in a confining dark sector with a single heavy quark in the fundamental representation. We assumed a portal exists between our sector and the dark sector that keeps the two sectors at kinetic equilibrium at the time of the phase transition and respects dark baryon number conservation. The arguments we presented do not depend on further details of the portal.

We argued that the bubbles of the confined phase, after nucleation, expand very slowly. Soon after the bubbles come in contact and coalesce, pockets of the deconfined phase form, and are submerged in a sea of the confined phase. 
The quarks are trapped inside these isolated and ever-contracting deconfined phase pockets. There is always an accidental asymmetry in the net dark baryon number in a given pocket, due to local stochastic fluctuations in the number of quarks and anti-quarks at the onset of pocket formation. As the pockets contract, the enclosed quarks compress until formerly frozen-out interactions recouple, giving rise to a second stage of annihilation. 

We wrote down the complete set of Boltzmann equations with all $2\rightarrow 2$ interactions between lowest lying bound states of the heavy quark. 
By solving these equations, we were able to calculate the fraction of dark quarks that survive the second annihilation event.  These surviving quarks bind into stable, color-singlet states that comprise the DM abundance we see today. 
We find that these Boltzmann equations predict a dramatic suppression in the DM relic abundance. This suppression is sensitive to the initial size of the pocket, the density of the quarks trapped within, and the pocket wall velocity. While there is a large uncertainty in determining these parameters, we showed that for virtually any reasonable values of these parameters there is a significant suppression in the DM relic abundance. 

We find the effect of quark pressure on the pocket wall velocity difficult to model. However, we do know that this effect will further slow down the pocket wall, which we showed will imply a smaller survival factor. We calculated the relic abundance of DM in this setup in two extreme scenarios: (i) the \textit{zero quark pressure} scenario, and (ii) when we assume the quark pressure is so severe that the \textit{asymmetry} bound on the survival factor is saturated. These two limiting scenarios bracket the range over which the relic abundance line can move when the quark pressure effects are properly taken into account. We found that for a fixed dark confinement scale $\Lambda$, the DM mass in this setup only changes by $\mathcal{O}(1)$ factors between these two scenarios.

After identifying the parts of the $m_q-\Lambda$ parameter space that predict the observed present-day DM abundance, we found that this large suppression opens up parts of the parameter space that were previously thought to be ruled out. In particular, we found a DM mass scale well above the often-quoted unitarity bound. 
Our calculation also suggests an upper bound on the dark confinement scale $\Lambda \sim \mathcal{O}(1)-\mathcal{O}(100)$~TeV. For any $\Lambda$ above this bound DM is overproduced, despite the dramatic suppression of its abundance during the phase transition. Depending on the value of $\Lambda$, the dark baryon mass that can explain the observed DM abundance varies roughly between $10^3$ to $10^5$ TeV.

There are many possible signals that our setup can give rise to. With the exception of gravitational waves, all the other potentially detectable signals depend on the specific form of this model's portal to the SM. It will be interesting to investigate the signatures of specific portals and their constraints, which we leave to future work.

There are numerous ways in which our analysis can be improved. To decrease the uncertainties in our results it will be important to perform more detailed numerical simulations of the \textit{macroscopic} bubble dynamics during the phase transition and the \textit{microscopic} strong dynamics that determine the particle interactions.
The most important quantities to be calculated would be the initial pocket size and its subsequent contraction rate, and the cross sections and binding energies included in our Boltzmann equations. Additionally, it will also be interesting to study the relic abundance calculation for other gauge groups and different quark representations. Furthermore, for any specific portal we should study the potential DM dilution due to a delayed glueball decay after the phase transition.

Confining sectors are natural dark sector candidates. In this paper we focused on such sectors with only a single species of heavy dark quark. We have pointed out the dramatic effect that this model's first-order phase transition has on the relic DM abundance of such a sector. The dynamics lead us to a sharp prediction about the natural mass scale of such DM candidates, $10^{3-5}$~TeV. It is of paramount importance to study variations of these minimal dark sectors in greater detail and their potential signatures in various upcoming experiments. 

\section*{Acknowledgments}

We thank Andrei Alexandru, Tom Cohen, Daniel Hackett, Julian Mu$\tilde{\mathrm{n}}$oz, Jessie Shelton, and Xiaojun Yao for useful discussions. We especially thank Benjamin Svetitsky for his collaboration in early stages of this work.
We also thank Yonit Hochberg, Graham Kribs, Rebecca Leane, Michele della Morte, Michele Redi, and Kai Schmidt-Hoberg for constructive comments on the draft. We especially thank Filippo Sala for raising the question of quark pressure effects. The work of PA, GWR, and TRS was supported by the U.S. Department of Energy, Office of Science, Office of High Energy Physics, under grant Contract Number DE-SC0012567. The work of PA is also supported by the MIT Department of Physics. GWR was also supported by an NSF GRFP and the U.S. Department of Energy, Office of Science, Office of Nuclear Physics under grant Contract Number DE-SC0011090.
The work of EK is supported by the Israel Science Foundation (grant No.1111/17), by the Binational Science Foundation
(grant No. 2016153), and by the I-CORE Program of the Planning Budgeting Committee (grant No. 1937/12). 
The work of EDK was supported by the Zuckerman STEM Leadership Program and by ISF and I-CORE grants of EK. 
JS is largely supported by a Feodor Lynen Fellowship from the Alexander von Humboldt foundation.

\appendix

\section{Thermodynamics of a First-order Phase Transition}
\label{appx:thermo}

In this Appendix we collect results that are helpful for understanding the dynamics of a first-order phase transition. 
We also describe a numerical simulation of the confining phase transition from the main text, which we perform to fix the initial pocket radius and its contraction rate. 
The code we used to perform this simulation can be found at \href{https://github.com/gridgway/ConfiningPT_Bubbles}{https://github.com/gridgway/ConfiningPT\_Bubbles}.

\subsection{Standard thermodynamics}

The defining characteristic of a first-order phase transition is that a first derivative of the free energy, or the free energy density $f$, is discontinuous at a critical point. In our case, 
the entropy density, $s = -\frac{\partial f}{\partial T}$, is discontinuous, corresponding to a non-zero latent heat release  due to phase conversion. In contrast, $f$ itself is continuous, meaning that the free energy of the two phases are the same at the critical point, $f_\text{deconf}=f_\text{conf}$. 

To better understand the latent heat, first notice that in the absence of chemical potentials the critical point is specified by a single parameter, the critical temperature $T_c$, and the free energy density is minus the pressure, $f = -p$. Since the free energy of both phases are the same at the critical point, their pressures are the same. Using the Euler relation, $\rho = T s - p$, where $\rho$ is the energy density, one can take the difference in energy densities of the two phases at the critical temperature to find 
\begin{alignat}{1}
	\Delta \rho = T_c \Delta s \equiv l \, ,
\label{eq:latent_heat}
\end{alignat}
which defines the latent heat density. 
Whereas the \textit{free} energy density is continuous, the energy density is not.
Another important quantity that characterizes first-order phase transitions is the surface tension, $\sigma$. The surface tension is the energy cost per unit area of an interface separating the two phases.

The latent heat density and surface tension are calculable via lattice simulations. 
We assume that at temperatures below the quark mass, the thermodynamics of the system are insensitive to the heavy quark field and we need only consider lattice simulations of pure $SU(3)$ Yang-Mills theory. 
In particular, the authors of \cite{largeN_lattice} calculate a latent heat density and surface tension in the infinite volume limit (see their tables 7 and 15) 
\begin{alignat}{1}
	l &= 1.413 \, T_c^4 \, , \nn \\
	\sigma &= .02 \, T_c^3 \, .
\label{eq:l_sigma}
\end{alignat}
These quantities have been computed elsewhere~\cite{Celik_LH, Christ_LH, Forcrand_ST, Giusti_2017}, 
and we find that the uncertainties on $l$ and $\sigma$ among these different lattice calculations are not large enough to qualitatively change our results.

As the universe expands, our thermal system, still in the deconfined phase, supercools.\footnote{
Much of the following discussion can be found in Refs.~\cite{LL_SP1} \S162 and \cite{LL_PK} \S99-100.
}
Once the system is supercooled, the free energy of the confined phase is lower than the free energy of the deconfined phase, meaning that it is energetically favorable for a phase transition to take place.
For any non-zero amount of supercooling, there now exists a critical radius, $R_c$, at which the energy cost of increasing a spherical bubble's surface area is exactly compensated by the free-energy decrease due to phase conversion, 
\begin{alignat}{1}
	&\left.\frac{\partial F}{\partial R}\right|_{R_c} = \left. \frac{\partial}{\partial R} \left( 4\pi R^2 \sigma - \frac{4\pi}{3}R^3 \Delta f \right)\right|_{R_c} = 0 \nn \\
	\implies & R_c = \frac{2\sigma}{\Delta f} \, ,
	\label{eq:Rc}
\end{alignat}
where $\Delta f$ is the confined phase free energy density minus that of the deconfined phase, and so is positive.

To relate $\Delta f$ to the latent heat density, 
first recall that at $T_c$ the entropy difference is given by $\Delta s = l/T_c$. 
If we then use the thermodynamic relation $\frac{\partial f}{\partial T} = -s$ we find $\left.\frac{\partial \Delta f}{\partial T}\right|_{T_c} = -l/T_c$.  
If we assume small supercooling we can use a Taylor expansion, which at leading order gives
\begin{alignat}{1}
	\Delta f & = \frac{\partial \Delta f}{\partial T} \, (T-T_c)  \nn \\
	& = l \, \frac{(T_c - T)}{T_c} \,.
	\label{eq:DeltaF}
\end{alignat}
We then find
\begin{alignat}{1}
	R_c = \frac{2\sigma \, T_c}{l \, (T_c-T)} \, .
	\label{eq:Rc_final}
\end{alignat}
The total free energy of a bubble at the critical radius is
\begin{alignat}{1}
    \label{eq:Fc}
	F_c &= 4\pi R_c^2 \sigma - \frac{4\pi}{3}R_c^3 \Delta f \nn \\
	& = \frac{16 \pi}{3} \frac{\sigma^3}{\Delta f^2} \nn \\
	& = \frac{16 \pi}{3} \left(\frac{\sigma}{T_c^3}\right)^3 \left(\frac{l}{T_c^4}\right)^{-2} \frac{T_c^3}{(T_c-T)^2}.
\end{alignat}

Thermal fluctuations will randomly convert regions of varying shapes and sizes from one phase to the other.  
When the system is supercooled, there is a chance that a converted region will be large enough that it expands rather than contracts. 
The probability per unit time and volume of converting a region with free energy $F$ is determined primarily by the Boltzmann factor $e^{-F/T}$. 
Using dimensional analysis we have \cite{Gorbunov:2011zz}, 
\begin{alignat}{1}
\label{eq:nucrate}
	\Gamma = A T^4 e^{-\frac{F}{T_c}}\, ,
\end{alignat}
where $A$ is assumed to be some $\mathcal{O}(1)$ number that is roughly constant with respect to temperature and determined by the microscopic theory.
Provided $A$ is indeed an $\mathcal{O}(1)$ number, we find that the exponential is by far the more important factor for determining the behavior of the phase transition, so we set $A$ to 1 without qualitatively changing our results. 
For an alternative though similar expression for $\Gamma$, See \cite{LL_PK} \S99.

Let us make the simplifying assumption that all bubbles can be approximated as spherical. 
Then bubbles that nucleate with radii below $R_c$ are ephemeral, quickly shrinking due to surface tension, 
while bubbles with radii well above $R_c$ are exponentially less likely to nucleate than critical bubbles by Eq.~\eqref{eq:nucrate}.
Then to a good approximation, we can assume that only bubbles at the critical radius nucleate. Combining Eqs.~\eqref{eq:Fc} and \eqref{eq:nucrate}, we find the nucleation rate of these critical bubbles is
\begin{alignat}{1}
\label{eq:nucrate2}
	\Gamma &= A T_c^4 e^{-\frac{\kappa \, T_c^2}{(T_c-T)^2}} \, ,\\
	\kappa & = \frac{16 \pi}{3} \left(\frac{\sigma}{T_c^3}\right)^3 \left(\frac{l}{T_c^4}\right)^{-2}\, , \nn \\
	& \sim 7 \times 10^{-5} \, ,
\label{eq:kappa}
\end{alignat}
where in the last line we have used the lattice results from Eq.~\eqref{eq:l_sigma}. 
Because the latent heat density is an order one number in units of $T_c$ while the surface tension is a small number in units of $T_c$, $\kappa$ turns out to be a very small number (which Ref.~\cite{largeN_lattice} indicates is generically true for $SU(N)$ theories).
As a result, very little supercooling is required before bubble nucleation becomes efficient.

\subsection{First half of the phase transition: bubble growth}

After bubble nucleation begins, the phase transition proceeds via the nucleation of new bubbles and the expansion of old bubbles. 
To keep track of the progress of the phase transition, we define the fraction of the universe that is in the confined phase~\cite{Gorbunov:2011zz},
\begin{alignat}{1}
	x(t) = \int_{t_c}^t dt' \Gamma(t') \frac{4\pi}{3} R^3(t,t') (1-x(t')) \, ,
\end{alignat}
where $t_c$ is the time at which the universe first reaches the critical temperature and $R(t,t')$ is the radius at time $t$ of a bubble nucleated at time $t'$. Applying a time derivative yields
\begin{alignat}{1}
	\dot{x}(t) &=  \Gamma(t) (1-x(t)) \frac{4\pi}{3} R_c^3(t) + \int_{t_c}^t dt' \Gamma(t') 4\pi R^2(t,t') \dot{R}(t, t') (1-x(t')), 
\label{eq:xdot}
\end{alignat}
where the first term corresponds to the nucleation of new bubbles while the second corresponds to the expansion of old bubbles.

The temperature evolution of the early universe plasma includes the usual adiabatic cooling term due to Hubble expansion, but now it also includes a new heating term due to the steady release of latent heat as the deconfined phase converts to the confined phase. 
By Eq.~\eqref{eq:latent_heat}, converting a fraction $dx$ from the deconfined to the confined phase releases $d\rho = l dx$ energy per unit volume. 
The released energy is absorbed in each phase with a temperature increase determined by each phase's respective specific heat, $d\rho/dT$.
For now, we will focus on the large scale average temperature so that we do not have to deal with small scale temperature gradients between points near and far from the sites of latent heat release.  
If we assume that the portal interaction between the SM bath and dark sector leads to frequent enough interactions between the two sectors per Hubble time, then in each phase the specific heat is dominated by the many degrees of freedom of the high temperature SM bath. 
For example, the deconfined phase of the dark sector is found to contribute 
about $5\%$ to the specific heat at the critical temperature~\cite{Giusti_2017}. 
We can then use 
\begin{alignat}{1}
	\rho(T) \approx \frac{g_*(T)\pi^2}{30}T^4 \, ,
\end{alignat} 
for both phases,
where $g_*(T)$ is the effective number of relativistic degrees of freedom and is $g_*(T) \approx 106.75$ for all temperatures of interest in our analysis. 
We therefore approximate both phases as having the same specific heats and find 
\begin{alignat}{1}
	dT &\approx l \left( \frac{d \rho(x)}{dT}(T_c)\right)^{-1} dx \nn \\
	& \approx 10^{-2} T_c dx.
\label{eq:bubble_heating}
\end{alignat}
The total temperature evolution of the universe during the phase transition is then given by 
\begin{alignat}{1}
	\dot{T} = -HT + 10^{-2} \, T_c \dot{x} \, ,
\label{eq:T_evolution}
\end{alignat} 
where the first term comes from the adiabatic cooling of relativistic species due to Hubble expansion. 
Had we considered a model in which few interactions take place between the standard model and dark sector baths per Hubble time, then the two sectors would be thermally decoupled. $T$ would refer to the dark sector's temperature, which would heat relative to the SM temperature, and we would have divided by the dark sector's specific heat, eliminating the factor of $10^{-2}$ in Eq.~\eqref{eq:T_evolution}.

There is an important distinction between the temperature evolution of a weakly first-order phase transition and a strongly first-order phase transition. 
In a weakly first-order phase transition the latent heat in units of $T_c$ is typically small (see \cite{Gorbunov:2011zz}) so that 
the heating term in Eq.~\eqref{eq:T_evolution} is negligible and the amount of supercooling does not change much due to the added latent heat. 
In a strongly first-order phase transition the latent heat in units of $T_c$ can be order one or larger 
(see Eq.~\eqref{eq:l_sigma}) so that $T$ can be driven back up to $T_c$ before the phase transition completes. Furthermore, since $\kappa$ in Eq.~\eqref{eq:kappa} scales inversely with $l^2$, the large value of the latent heat decreases the amount of supercooling needed to achieve efficient bubble nucleation as is seen in Eq.~\eqref{eq:nucrate2}, making it easier for the universe to reheat to a point where nucleation is negligible.
We will show in a simulation below that this reheating scenario is achieved in our phase transition. 

Notice also that $T$ cannot ever reheat all the way up to $T_c$. 
If it did, the critical radius would diverge and all bubbles would shrink. 
The second term in Eq.~\eqref{eq:T_evolution} would change sign and become a cooling term since latent heat would be absorbed, and hence $T$ would be driven back below $T_c$.  
Instead, $|T-T_c|$ decreases from its maximum to an equilibrium value very close to zero at which the heating and cooling terms in Eq.~\eqref{eq:T_evolution} nearly balance one another, as we will show below.
This equilibrium phase coexistence is exactly the regime described by the Maxwell construction for first-order phase transitions (see \cite{LL_SP1}, \S84).

As explained in the previous section, we assume that bubbles nucleated at time $t_0$ are of size $R_c(T(t_0))$, giving the initial condition $R(t=t_0,t_0) = R_c(T(t_0))$. 
To determine the radius at a future time, we need an expression for the bubble wall velocity, $\dot{R}(t,t')$. 
An accurate treatment of the bubble wall velocity, requires full 3+1 dimensional numerical simulations of bubble dynamics during the phase transition. 
However, even in simplified settings, various numerical simulations have not converged on a single, definitive answer~\cite{Ignatius:1993qn,Huber:2011aa,Ellis:2018mja,Balaji:2020yrx}. 
Instead, we will use a convenient, basic model of $\dot{R}(t,t')$. 
We require that critical bubbles not change their radius, and that larger bubbles expand while smaller bubbles contract. 
To capture this behavior, we use the simple functional form
\begin{alignat}{1}
    \dot{R}(t,t_0) = v_w(t) \, \text{sign}\left[ R(t,t_0) - R_c(t)\right] \, ,
\label{eq:vw_model}
\end{alignat}
where we define $\text{sign}(0) = 0$. 

To determine $v_w(t)$ would require a better understanding of the underlying strong dynamics. 
Instead, we will estimate an upper bound on $v_w(t)$ based on thermodynamic arguments. As argued in the main text, the larger $v_w$ is, the less the DM relic abundance will be suppressed. 
So we will always set $v_w$ to its upper bound, assuming it to be a conservative choice.

As a bubble expands, it releases latent heat near its wall, \textit{locally} heating the plasma at the wall to a temperature $T_\text{wall}>T$ and reducing the free energy difference at the interface.
Since the free energy density is minus the pressure, this local heating reduces the net pressure acting on the wall (See Eq.~\eqref{eq:DeltaF}).
Since the degree of supercooling is so small, the wall could potentially heat up to a temperature at which the net pressure balances against the surface tension. 
If the wall reached this temperature, bubble growth would no longer be thermodynamically favorable, and so the wall motion would come to a halt.
By an argument completely analogous to the one that lead to Eq.~\eqref{eq:Rc}, except we evaluate Eq.~\eqref{eq:DeltaF} at the wall temperature, we find this equilibrium wall temperature to be
\begin{alignat}{1}
    T^\text{eq}_\text{wall} &= T_c\left(1-\frac{2\sigma }{l \, R} \right) \, .
    \label{eq:T_wall}
\end{alignat}
We assume that as the wall temperature approaches $T^\text{eq}_\text{wall}$ its growth slows down gradually. 
Before the wall temperature reaches $T^\text{eq}_\text{wall}$ it will have slowed down to a steady state at which the rates of wall heating and cooling cancel one another. 
By estimating the rates of wall heating and cooling then setting them equal, we will determine an approximate expression for $v_w(t)$.

We start with the cooling rate. 
We will assume that $T_\text{wall}$ is very close to $T^\text{eq}_\text{wall}$, which is in turn very close to $T_c$ since the second term in Eq.~\eqref{eq:T_wall} is very small compared to $1$ for bubbles larger than $\Lambda^{-1}$. 
This assumption will lead to a faster $v_w$.
The fractional temperature difference between points near and far from the wall is then $(T_c-T)/T_c$. 
We assume that the heat loss rate is given by a diffusion equation, $\dot{T}_\text{cool} \sim - K \nabla^2T$, and that the transport coefficient at $T_c$ is of order $K \sim \Lambda^{-1}$. 
If we further assume that the length scale of the density gradient is $\Lambda^{-1}$, then we find $\dot{T}_\text{cool} \sim - \Lambda^2 \, (T_c-T)/T_c$. 

Now we move on to the heating rate. 
We start with the energy injected per unit wall area and time, $l \, v_w$. 
If we assume that this energy is injected within a typical length $\Lambda^{-1}$ of the wall, then the energy injected per unit volume is $l v_w \Lambda$. 
As before, dividing by the specific heat, $d\rho/dT$, converts the energy increase into a temperature increase. 
Rather than assume that the specific heat is dominated by the SM degrees of freedom as we did when deriving Eq.~\eqref{eq:bubble_heating}, we assume that for this process the SM degrees of freedom are irrelevant and the specific heat is dominated by the dark sector degrees of freedom, though this choice will not affect our final result. 
We do so because we anticipate that for most models the portal interaction between the SM and dark sector will generically take place on timescales much slower than $\Lambda^{-1}$, and so will be inefficient compared to the interactions within the dark sector. 
Hence, we assume that the dark degrees of freedom disperse latent heat on a fast timescale of order $\Lambda^{-1}$,
and only on much longer timescales does this heat find its way into the SM degrees of freedom.\footnote{We assume that the timescale over which interactions between the SM and dark sector baths exchange energy is much faster than the Hubble rate, justifying why the SM degrees of freedom are included in the heat capacity in Eq.~\eqref{eq:bubble_heating}.}
Then we can use $d\rho_\text{DS}/dT \sim T^3$~\cite{Giusti_2017},
which gives a heating rate of $\dot{T}_\text{heat} \sim \Lambda^2 v_w$.\footnote{
If the portal interaction is actually efficient enough to keep the two sectors in equilibrium on this short timescale, then the heating rate would be suppressed by a factor of $g_\star$. However, the cooling rate would also be suppressed by a factor of $g_\star$ since $K$ is inversely proportional to the number density of interacting degrees of freedom, \cite{kardar_2007} \S3.9. These two factors would then cancel in the expression for $v_w$. 
We therefore expect Eq.~\eqref{eq:vterm} to be independent of the specific portal interaction.
}
The wall velocity at which both rates are in balance is then 
\begin{alignat}{1}
    v_w = \left(\frac{T_c-T}{T_c}\right) \, .
\label{eq:vterm}
\end{alignat}

As argued in Ref.~\cite{Witten:1984rs}, we see that $v_w$ is suppressed by the small degree of supercooling during the phase transition. 

We apply a similar argument to the wall velocity of contracting bubbles, which are smaller than the critical radius. As bubbles contract, latent heat is absorbed near the wall, decreasing the wall temperature, increasing the net pressure on the wall, and thus opposing further contraction. 
The contraction rate is therefore limited by the rate at which the cold wall can heat due to heat flow from the hotter surrounding plasma. 
Balancing the heating and cooling rates as before leads to a wall velocity given by Eq.~\eqref{eq:vterm} up to a relative sign, justifying the symmetric functional form of Eq.~\eqref{eq:vw_model}.

Again, the above expression for $v_w$ is in reality an upper bound. An expanding bubble wall cannot move any faster than Eq.~\eqref{eq:vterm}, at least for an extended period, because then it would \textit{locally} overheat the wall.
There is a similar consideration that the {\it global} rate of increase in temperature from bubble expansion should not significantly outpace the Hubble cooling, lest the whole universe be heated above $T=T_c$. Consequently, we will see that the value of $v_w$ derived in Eq.~\eqref{eq:vterm} evolves so that it always lies below or close to this ``global threshold''.

\begin{figure}
    \caption{
    The wall velocity (blue) and ``global threshold'' (black dashed) above which phase conversion is so fast that the universe experiences net heating. 
    The right panel displays the same data as the left panel, but zooms in to the very end of the phase transition when pockets have contracted significantly. The discontinuity in the middle is an artifact of our modeling. It occurs when the spectrum of bubbles, which peaks at $R_0$, discontinuously jumps to a delta function spectrum of pockets centered at $R_1$.
    }
    \resizebox{\columnwidth}{!}{
    \includegraphics[scale=0.35]{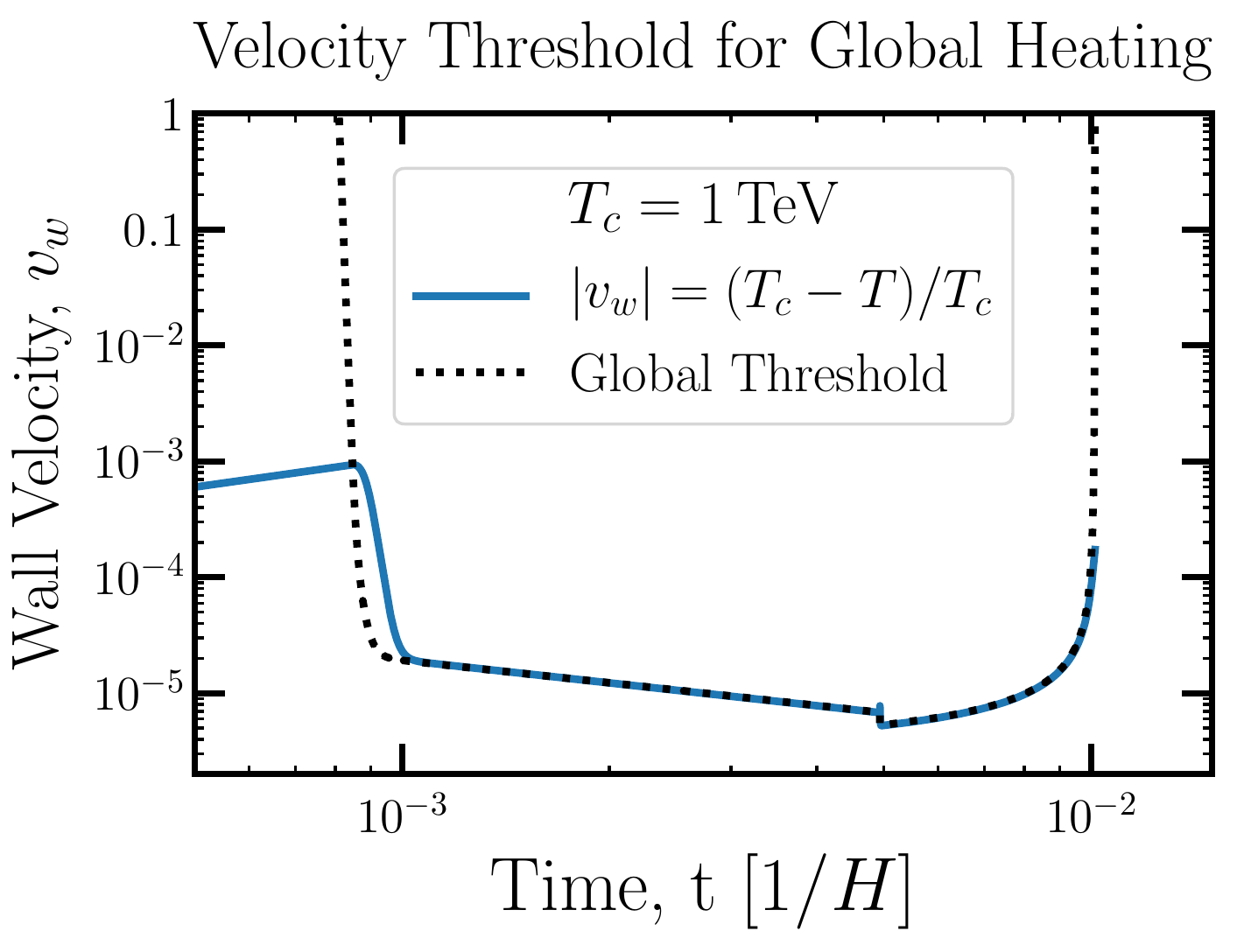}
    \includegraphics[scale=0.35]{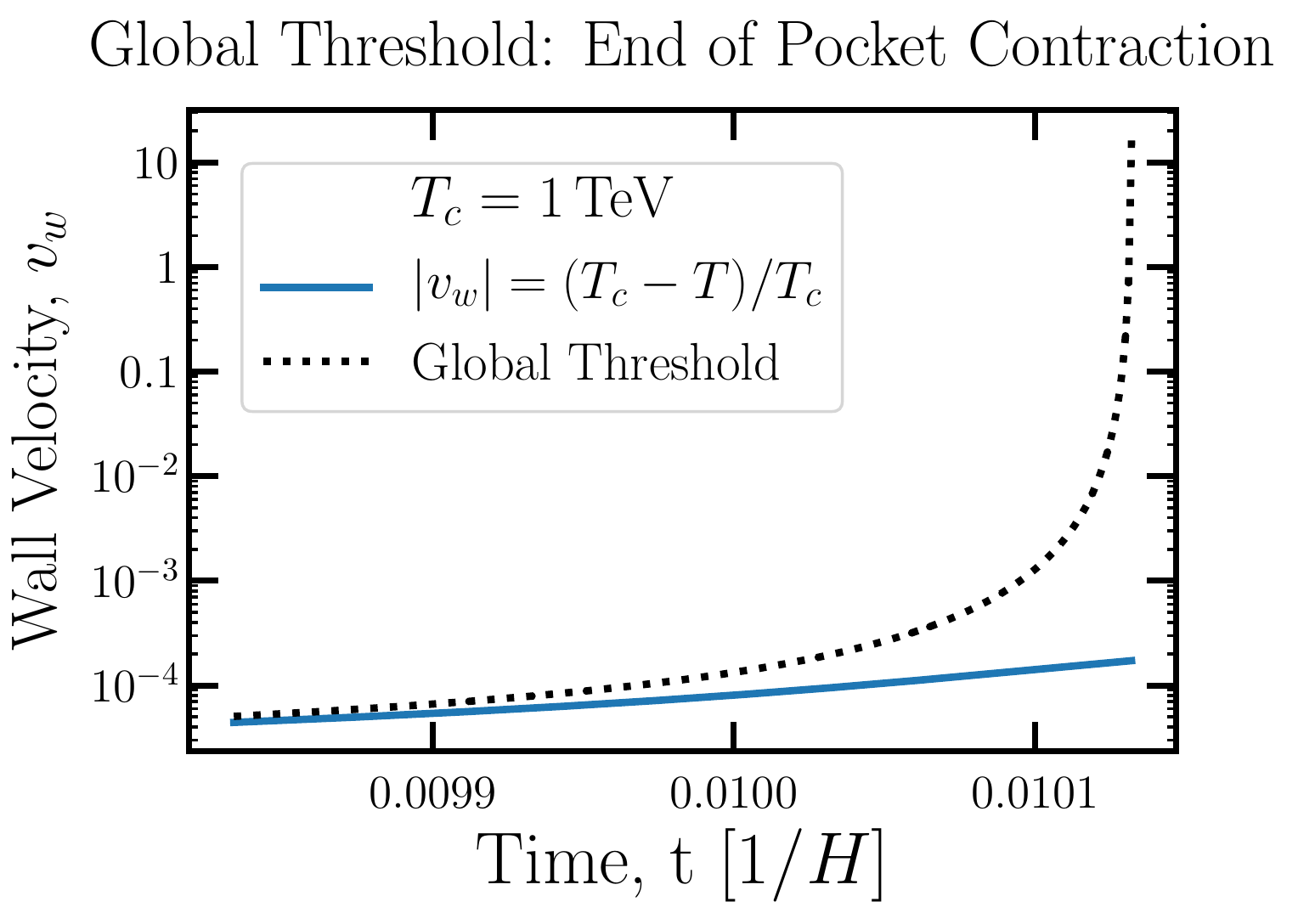}
    }
     \label{fig:glob_thresh}
\end{figure}

The wall velocity (Eq.~\eqref{eq:vterm}) and global threshold are plotted in Fig.~\ref{fig:glob_thresh}. At the start of the phase transition, when bubbles are rare and their radii are small, the global heating from bubble expansion is very small even for $v_w \rightarrow c$, and so the global threshold velocity goes to infinity. The same is true at the end of the phase transition, when there is very little volume available for phase conversion. However, during the phase transition when the rate of phase conversion is rapid, this threshold becomes relevant.

Let us consider what happens when bubble coalescence and growth causes $v_w=(T_c - T)/T_c$ to exceed this global threshold (which depends on the bubble density and typical bubble radius) for the first time. If $v_w$ overshoots the threshold, then the universe will begin to heat up on average, thus reducing $v_w$. The net effect is for the degree of supercooling to evolve such that $v_w$ tracks the global threshold velocity. In practice, we observe that there is at first an abrupt drop in both $v_w$ and the global threshold velocity, associated with a sharp increase in the bubble number density due to nucleation; in this epoch, $v_w$ slightly exceeds the global threshold velocity and this supports a fairly rapid increase in $T$. Once nucleation becomes inefficient due to the rising temperature, the global threshold velocity evolves more slowly, driven by the expansion of the largest existing bubbles. In this epoch $v_w$ tracks the global threshold closely and $\dot{T} \approx 0$, with a slow adiabatic increase in temperature toward $T_c$ driven by the slow decrease in the global threshold velocity (which requires a corresponding decrease in $v_w$ and hence in the degree of supercooling). 
We can even derive the scaling of the global threshold (and $v_w$) with $t$ during this period.  
By Eq.~\eqref{eq:T_evolution}, $\dot{T} \approx 0$ implies $\dot{x} \approx 100 H$. 
Using the fact that the spectrum of bubbles is strongly peaked at a single radius $R$, and the nucleation of bubbles is so suppressed that the number density of bubbles is constant, we have $x = 4 \pi R^3 n_\text{bub}/3$. 
Combining the two equations gives $dR/dt \propto R^{-2}$ so that $R = 3 A (t-t_0)^{1/3}$ for some constants A and $t_0$. 
Therefore, $v_w = dR/dt = A (t-t_0)^{-2/3}$.  
Indeed, we find that the slope of the line in Fig.~\ref{fig:glob_thresh} between times $.001/H$ and $.005/H$ is precisely $-2/3$.

The Hubble cooling can be relevant here, even though the phase transition takes place on timescales much smaller than a Hubble time, because the bubble expansion is so sensitive to the degree of supercooling; in contrast, we can freely drop e.g. density dilution terms corresponding to the Hubble expansion, as there is no comparably small density difference relevant to our calculation (see e.g. Eq.~\eqref{eq:L12}). After coalescence, when the heating comes from pockets of shrinking radius rather than bubbles of expanding radius, the reverse process occurs, with a slow adiabatic decrease in the equilibrium temperature due to the decreasing size of the pockets. The temperature evolution eventually switches over to the standard Hubble cooling once $v_w=(T_c-T)/T_c$ can no longer reach the global velocity threshold (and hence $\dot{T} \approx 0$ cannot be maintained). 

With Eqs.~\eqref{eq:xdot} 
and~\eqref{eq:T_evolution} through \eqref{eq:vterm} in hand we are able to simulate the first half of the phase transition. 
This system of equations models the initial bubble nucleation and accompanying latent heat release, and the proceeding equilibrium regime up until percolation when bubbles begin to overlap and new dynamics must be included.
Our initial conditions are that the universe has supercooled a little bit, $\frac{T-T_c}{T_c} = - 10^{-4}$, the universe is fully in the deconfined phase, $x=0$, and that no bubbles have nucleated yet. 
We evolve forward in small time steps, $\Delta t = 10^{-6}/H(T_c)$. 
In each time step we nucleate $\Gamma(T(t)) (1-x(t)) \Delta t$ bubbles per unit volume at radius $R_c(T(t))$ and add them to a list. 
We allow all other bubbles from previous steps to expand or contract by an amount $\dot{R}(t,t') \Delta t$, which only depends on the bubble size and temperature within that time step. 
This procedure produces bubbles with radii less than or equal to 0, so we set such bubbles' radii to zero and remove them from our list. 
Additionally, many time steps result in an additional number density of bubbles that is so exponentially small that the computer sets the number density to zero.  
We remove these bubbles from our list, too. 
Each bubble nucleation and all bubble expansions increase $x$ by $\dot{x} \Delta t$ and the temperature by $\dot{T} \Delta t$ according to Eqs.~\eqref{eq:xdot} and~\eqref{eq:T_evolution}. 
We finish our evolution once $x=\frac12$.

\begin{figure}
    \caption{
    The degree of supercooling as a function of time (left), the confined phase fraction as a function of time (middle), and the spectrum of bubble radii at percolation, when $x=1/2$ (right).
    }
    \resizebox{\columnwidth}{!}{
    \includegraphics[scale=0.35]{T_vs_t_new3.pdf}
    \includegraphics[scale=0.35]{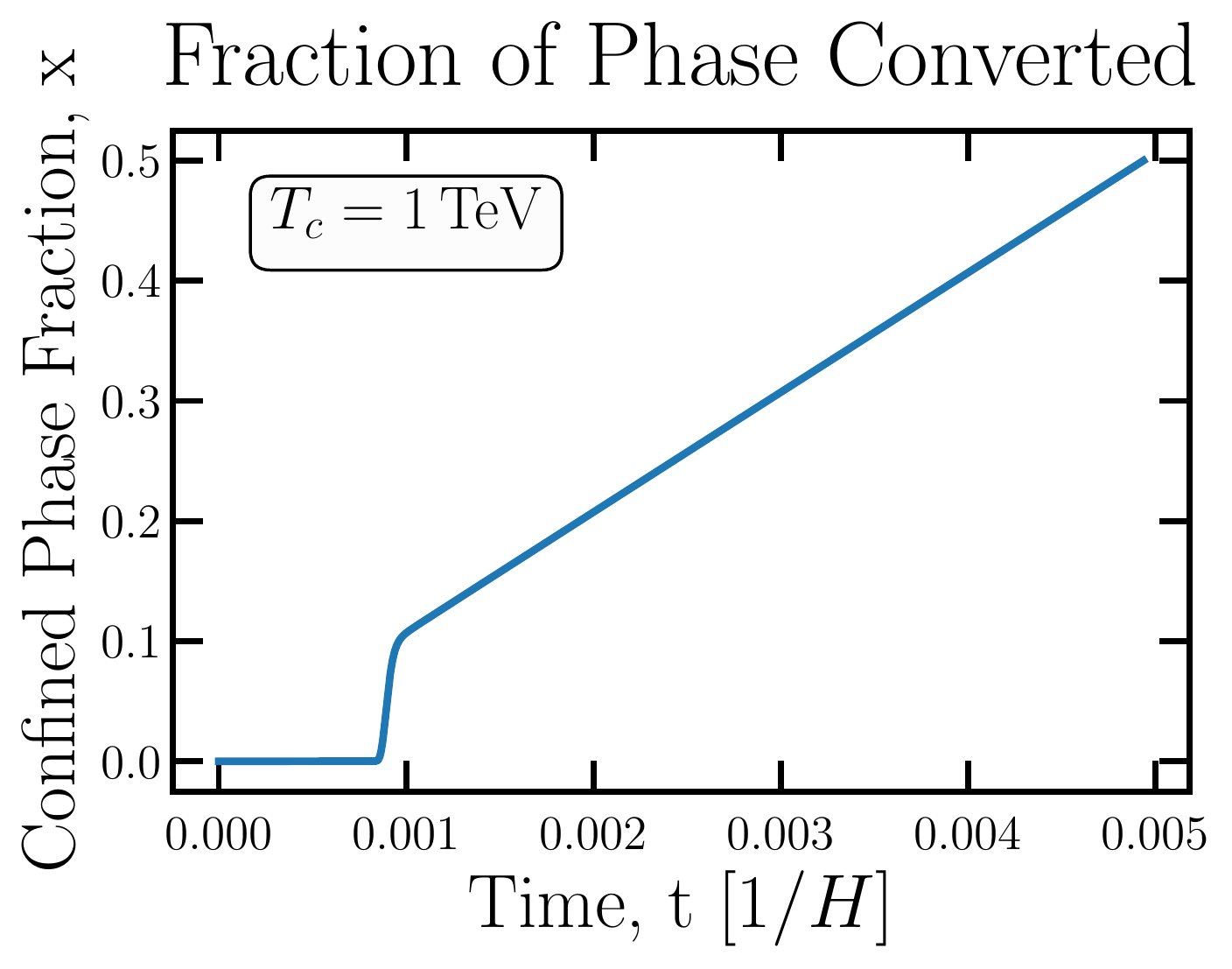}
     \includegraphics[scale=0.35]{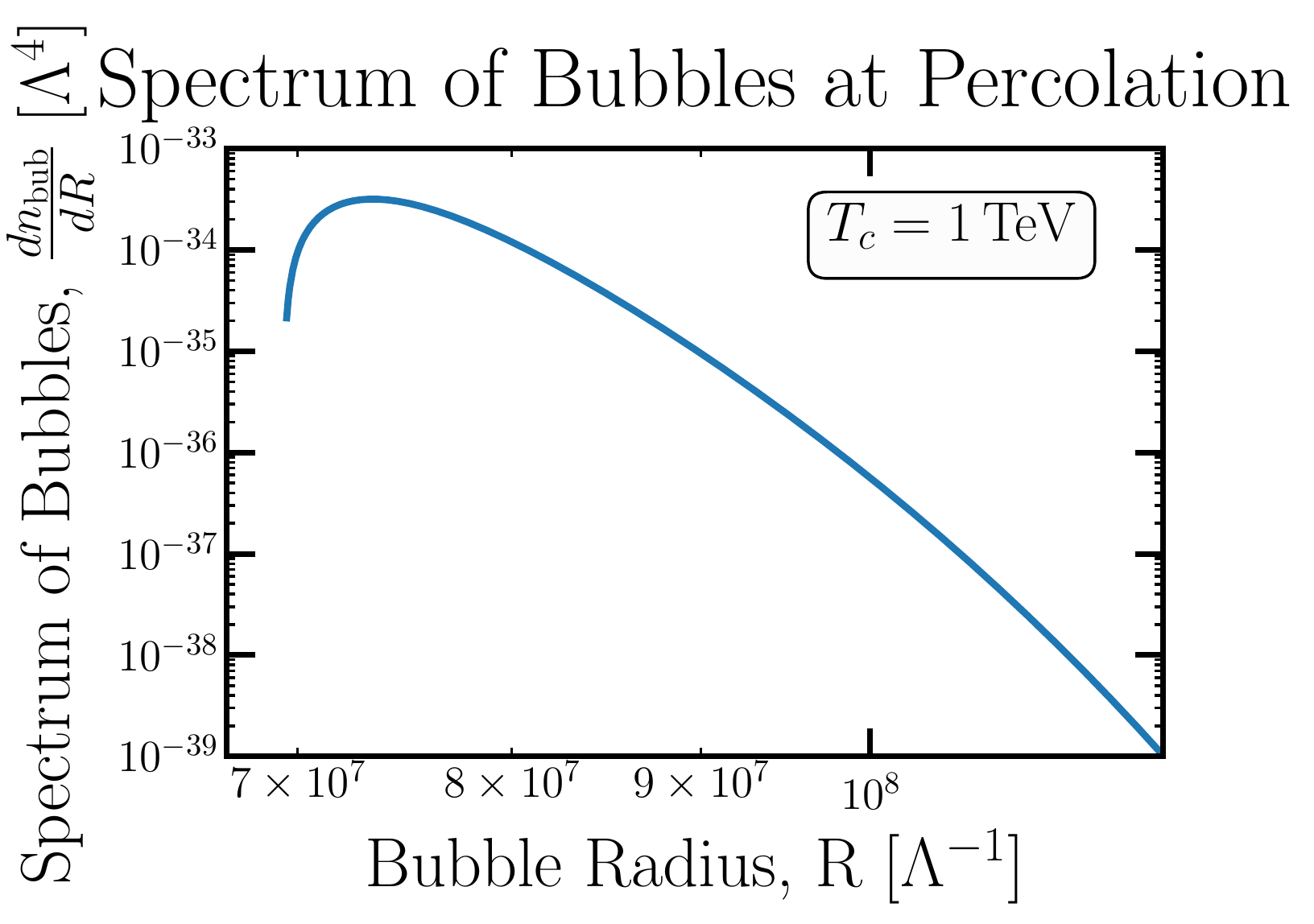}
     }
     \label{fig:sim}
\end{figure}

Outputs of our simulation are shown in Fig.~\ref{fig:sim}. 
The left plot shows the degree of supercooling before percolation and the middle plot shows the fraction of phase converted, $x$.
These plots make clear that the first half of the phase transition can be divided into three distinct stages. 
In the first stage, the degree of supercooling is so small that the bubble nucleation rate is too suppressed to have a significant effect on the simulation. 
During this period $x=0$ and the universe cools through Hubble expansion.
In the second stage, the supercooling reaches a point at which nucleation becomes efficient. 
These nucleated bubbles quickly grow and inject heat, corresponding to the sudden jump in the temperature and $x$ a little before $t = 0.001/H$. The temperature reaches a point very close to $T_c$ at which nucleation of new bubbles becomes inefficient again, leading to the third stage. 
This stage is exactly the equilibrium phase coexistence regime described above. 
There are a fixed number density of large bubbles that grow and inject latent heat at such a rate as to cancel Hubble cooling. 
The net effect is that $\dot{T} \approx 0$. 
By Eq.~\eqref{eq:T_evolution}, we have
\begin{alignat}{1}
    \dot{x} \approx 10^2 H \, .
\end{alignat}
Since the temperature is constant during this stage, the Hubble rate is as well, meaning that $x$ grows linearly in time, which can be seen in our middle plot. 
This equation explains why the phase transition occurs over a time scale of $10^{-2}/H$. Had we assumed instead that the portal interaction between the SM bath and dark sector was very weak and led to few scatters per Hubble time, then the dark sector would not have access to the SM heat capacity and the above factor of $10^2$ would be replaced by an $\mathcal{O}(1)$ factor instead.

The rightmost plot in Fig.~\ref{fig:sim} shows the spectrum of bubble radii at the time of percolation, $x=1/2$.
The shape of this spectrum is a product of the preceding stages. 
The earliest bubbles were nucleated during the first stage. They had the longest time to grow, but were nucleated at a time of relatively small supercooling, meaning that their number density was exponentially suppressed. 
So as $R$ increases away to the right of the peak of the spectrum, $dn_\text{bub}/dR$ decreases.
Just as the second stage begins the universe is maximally supercooled. 
The bubbles produced at this point are the most numerous and constitute the peak of the distribution. 
In the rest of the second stage, the supercooling quickly diminishes, producing an exponentially suppressed population of bubbles that are smaller than the peak radius since they have less time to grow, leading to the sharp decrease of the spectrum to the left of the peak.

We define the peak in $dn_\text{bub}/dR$ to occur at $R_0$ and empirically find that it is very well fit by the function
\begin{alignat}{1}
    R_0 \, \Lambda = 10^{-6} \times \left(\frac{\Lambda}{M_\text{pl}}\right)^{-0.9} = 6.7 \times 10^7 \times \left(\frac{\Lambda}{\text{TeV}}\right)^{-0.9} \, .
    \label{eq:R0}
\end{alignat}
$R_0$, however, is not the only relevant length scale for the bubbles. 
An additional length scale, $R_1$, emerges from the dynamics of bubble coalescence.

At percolation, bubbles frequently come into contact with one another and begin to coalesce. 
To model the coalescence dynamics, we borrow another argument from \cite{Witten:1984rs}. 
When two spherical bubbles of radius $R$ coalesce into one larger-radius bubble, they decrease their surface area and therefore reach a more energetically favorable configuration due to surface tension. 
The energy difference between the two configurations is $\Delta E \sim 4\pi R^2 (2-2^{\frac23}) \sigma = 4\pi R^2 (2-2^{\frac23}) \times .02 T_c^3$. 
This change in energy is achieved by applying a force to the mass $M$ in the bubbles over a characteristic distance $R$. 
So $F \sim \Delta E / R \sim M a$, where $a$ is the acceleration of the material in the bubbles. 
This acceleration can also be estimated as $a \sim R/t_\text{coalesce}^2$, where $t_\text{coalesce}$ is the coalescence timescale. 
If we then use that the total mass in the two bubbles is $M = 2\times \frac43 \pi R^3 \rho_\text{deconf} \approx \frac83 \pi R^3 T_c^4$, where we used that $\rho_\text{deconf}(T_c) \approx T_c^4$~\cite{Giusti_2017}, then we find
\begin{alignat}{1}
	t_\text{coalesce} & \sim \left( \frac{M R^2}{\Delta E} \right)^{\frac12} \\ 
	& \sim 10 T_c^{\frac12} R^{\frac32} \, .
\end{alignat}

The above equation shows that small bubbles coalesce quicker than large ones. 
Therefore, at percolation small bubbles will quickly coalesce until they reach a size $R_1$ past which $t_\text{coalesce}$ takes longer than the timescale of percolation. 
From Fig.~\ref{fig:sim} we can estimate the timescale of percolation as the time it takes $x$ to change by a couple percent at around $x=\frac12$. 
We find $t_\text{perc} \sim 10^{-3} H^{-1} \sim 10^{-3} \frac{M_\text{pl}}{T_c^2}$ where  $M_\text{pl} = \left(8\pi G\right)^{-1/2} \sim 2.4 \times 10^{18}$GeV is the reduced Planck mass and $G$ is Newton's  constant. 
Setting the percolation timescale equal to $t_\text{coalesce}$ then yields the critical bubble size of, 
\begin{alignat}{1}
	R_1 \, \Lambda & \approx 10^{-8/3} \left(\frac{M_\text{pl}}{\Lambda} \right)^{2/3} \approx  4 \times 10^{7} \left(\frac{ \text{TeV}}{\Lambda}\right)^{2/3} \, ,
\label{eqn:R1}
\end{alignat}
in units of $T_c=\Lambda$.

In Fig.~\ref{fig:R0-R1} we plot $R_0$ and $R_1$ as a function of the confinement scale. 
We find that for $\Lambda \gtrsim 1\,\text{TeV}$ , $R_0$ is less than or equal to $R_1$. Therefore, for this range of $\Lambda$, once our simulation finishes at $x=\frac12$ we assume that all bubbles begin coalescing and quickly grow to radius $R_1$.
Since $x$ does not change during this process, we assume that $T$ remains fixed, too.  
For $ \Lambda < 1\,\text{TeV}$, we assume that coalescence is inefficient and bubbles remain at radius $R_0$.

\subsection{Second half of the phase transition: pocket contraction}

After percolation, most bubbles are in contact with one another. The confined regions form a web and the deconfined regions form isolated pockets. 
We assume these pockets quickly attain spherical symmetry due to surface tension, and also that the typical size of a bubble before percolation is equal to the typical size of a pocket after percolation, $R_1$.

Since all pockets are at the same initial radius, we can solve for the initial density of pockets. 
The number density of pockets that are all of radius $R$ satisfies 
\begin{alignat}{1}
    1-x = \frac{4 \pi}3 R^3 n_\text{pocket} \, .
\label{eqn:pocket_x}
\end{alignat}
Since $x=\frac12$ at percolation, we have
\begin{alignat}{1}
n_\text{pocket} = \frac3{8 \pi R_1^3} \, .
\end{alignat}
We will find that the degree of supercooling continues to be so small during pocket contraction that the nucleation of more bubbles is completely suppressed.
Therefore $n_\text{pocket}$ remains constant and we find
\begin{alignat}{1}
x = 1 - \frac{R^3}{2 R_1^3} \, .
\label{eq:x_pt2}
\end{alignat}

As before the contraction rate of the pockets is limited due to the latent heat release near the pocket wall. However, the wall velocity estimate for pockets is slightly different than it was for bubbles.
Whereas supercooling results in a net pressure outward for bubbles, supercooling results in a net pressure inward for pockets, since the two phases are on opposite sides of the wall in either case. 
With this caveat in mind we can repeat our argument leading to Eq.~\eqref{eq:vterm}.

As before, we argue that as it expands, the wall quickly heats up, approaching a threshold temperature at which pressure and surface tension are in equilibrium. 
This threshold now corresponds to slight superheating at the temperature $T_c\left(1+\frac{2\sigma }{l \, R} \right)$. 
Before $T_\text{wall}$ reaches this threshold it achieves a steady state at which heating from latent heat injection cancels against cooling from heat diffusion from the wall. 
Using the same arguments as before, we find that this steady state corresponds to a velocity of
\begin{alignat}{1}
\frac{dR}{dt} = -(T_c-T)/T_c \, .
\label{eq:dRdt}
\end{alignat}
Again, we used that $\frac{2\sigma}{l R} = \frac{.03}{R\,\Lambda}\ll 1$, so we approximate the temperature difference between the wall and its surroundings appearing in the above equation as $T_c-T$, which leads to an overestimate of the wall velocity.

There is another much more important new effect modifying the contraction rate of a pocket: quark pressure. 
The density of quarks trapped within pockets increases as they are forced within ever-shrinking volumes. 
This pressure opposes contraction, slowing down the wall velocity relative to Eq.~\eqref{eq:dRdt}. 
For now, to build intuition, we will ignore the effect of quark pressure, and we will consider it in the next section.

\begin{figure}
\caption{
    The degree of supercooling as a function of time (\textbf{left}), the confined phase fraction as a function of time (\textbf{middle}), and the pocket wall velocity as a function of bubble radius during the second half of the phase transition (\textbf{right}).
}
    \resizebox{\columnwidth}{!}{
    \includegraphics[scale=0.35]{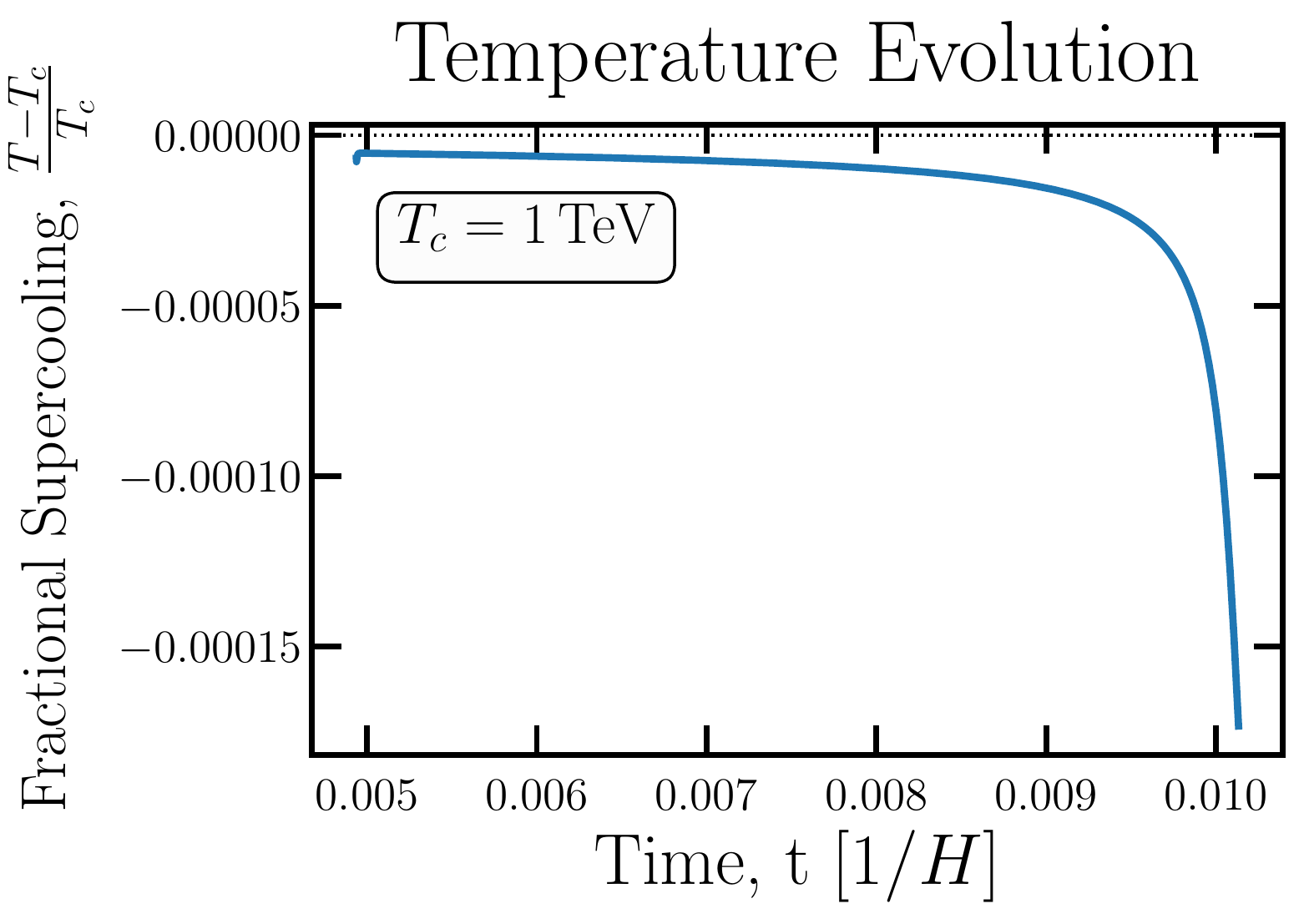}
    \includegraphics[scale=0.35]{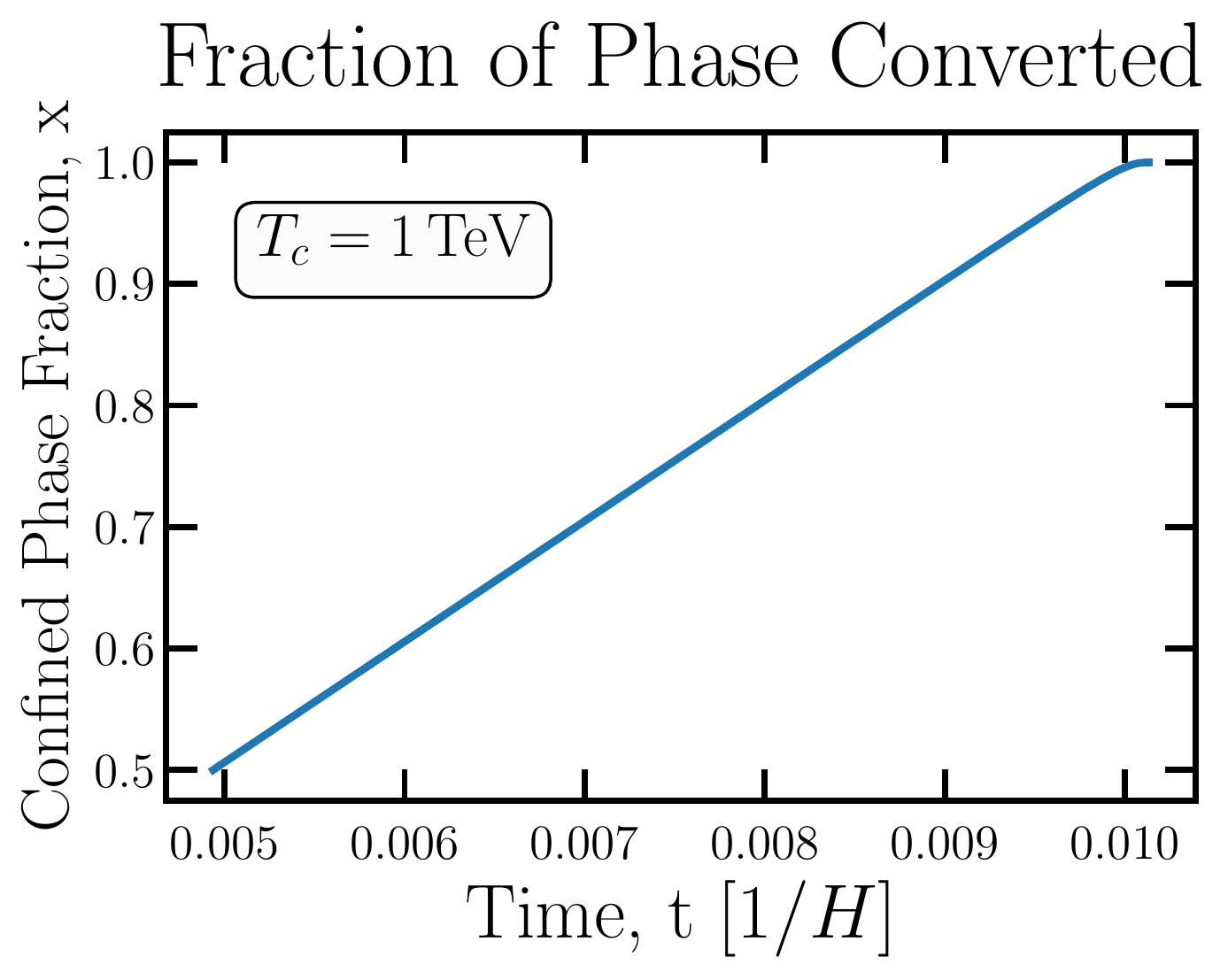}
    \includegraphics[scale=0.35]{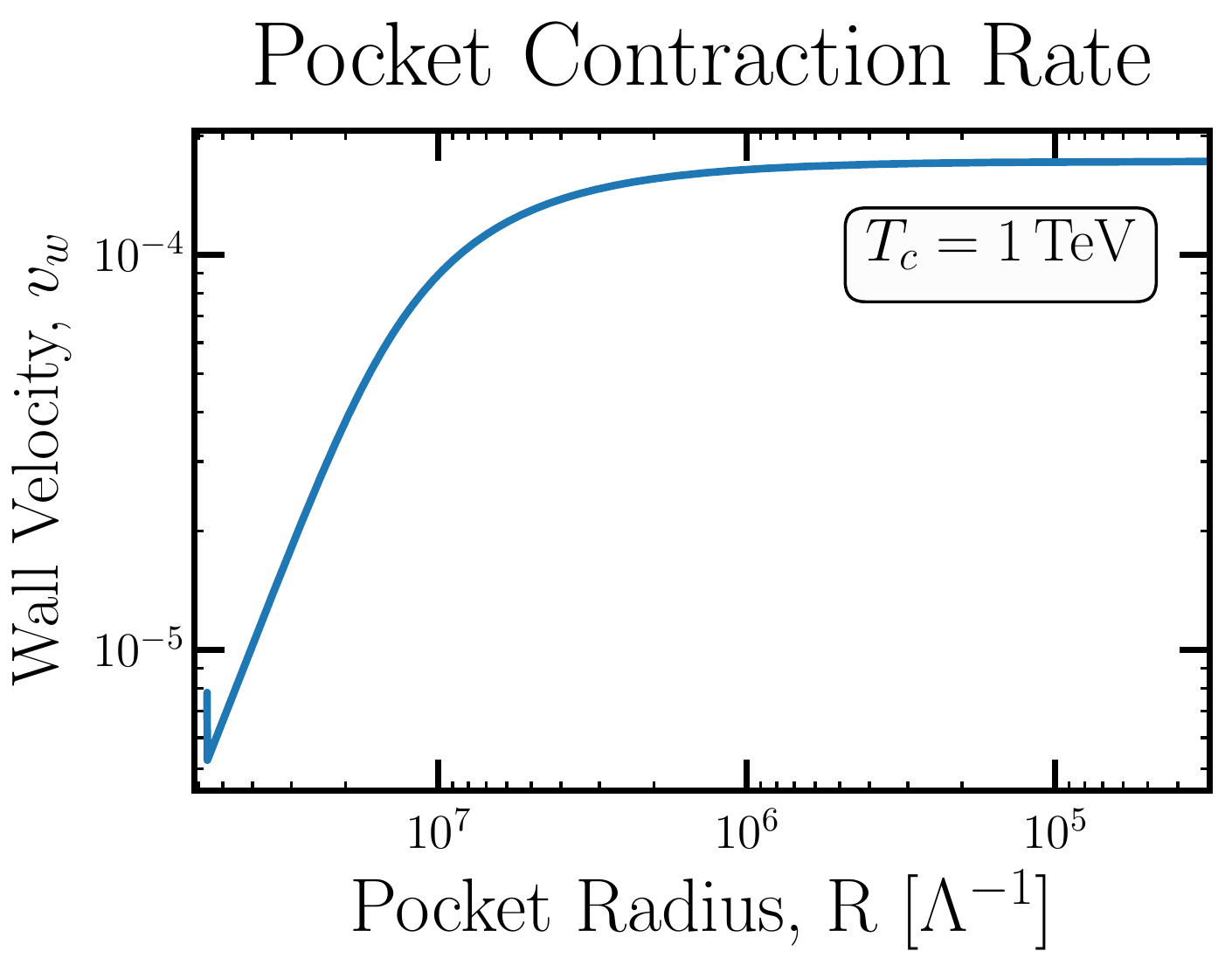}
    }
     \label{fig:sim_pt2}
\end{figure}

With Eqs.~\eqref{eq:T_evolution}, \eqref{eq:x_pt2}, and \eqref{eq:dRdt} we have all we need to simulate the second half of the phase transition.  
We use the same method and parameters as we did for the first half of the phase transition. In Fig.~\ref{fig:sim_pt2} we show the results of our simulation. 
The first instant of the simulation features a discontinuity in the temperature (and thus $v_w$). 
This discontinuity results from our discontinuous change of the spectrum of bubbles from the smooth form of $dn/dR$ in the right panel of Fig.~\ref{fig:sim} to a delta function at $R_1$. 
As the pockets get smaller, their heating rate diminishes. 
Hubble cooling becomes relatively more important over time leading to the steady decrease in $T$ over time. 
By the very end of the phase transition, the temperature evolution is purely determined by Hubble cooling. 

The right panel shows the pocket wall velocity as a function of the pocket radius. 
As a function of $R$, $v_w$ asymptotes to a well-defined value near $2 \times 10^{-4}$ for the choice of $\Lambda$ shown in the figure. 
Naively, this plot seems to be at odds with the left panel, since the two plots are equivalent up to a minus sign. 
Whereas the velocity seems to asymptote to a constant value at late times, the degree of supercooling seems to vary greatly at late times.
The apparent discrepancy is a result of the different x-axis scales.  
Whereas the x-axis in the left panel is linearly scaled in $t$, the x-axis in the right panel is log scaled in $R$.
The majority of the simulation takes place when the pocket radii are very large. 
The pocket radii are much smaller than their initial value only for a very short time at the end of the simulation, at time scales much shorter than the $1/H$. 
In fact, the small timestep $10^{-6}/H$ was chosen to resolve this smaller timescale. 
At such small timescales and pocket radii, very little Hubble cooling or latent heat injection takes place, leading to the plateau in $v_w$. 
Since most of the quark interactions recouple at pocket radii much smaller than $R_1$
(see e.g. Fig.~\ref{fig:abundances-R}), we are justified in treating $v_w$ as constant within the Boltzmann equations of Sec.~\ref{sec:boltzeqs}.\footnote{Recall that by ``recoupling'' we mean the point when the rates of quark interactions become large compared to the contraction rate of the pocket, so the quark density evolution is dominated by interactions.}
The asymptotic $v_w$ value as a function of $\Lambda$ is plotted in Fig.~\ref{fig:R0-R1} and is well fit by 
\begin{alignat}{1}
    v_w(R \ll R_1) = 0.2 \left(\frac{\Lambda}{M_\text{pl}}\right)^{0.2} =  1.7 \times 10^{-4} \times \left(\frac{\Lambda}{\text{TeV}}\right)^{0.2}.  
\end{alignat}

\subsection{The effect of quark pressure}
\label{appx:pq}
Up to this point we have neglected the effect of quark pressure on the phase transition,
\begin{alignat}{1}
    p_q = n_q T \, .
\end{alignat}
For the first half of the phase transition, this approximate treatment was justified.  
At the start of the phase transition, $p_q$ is suppressed compared to the gluon pressure. 
During the phase transition $n_q$ grows, and so the quark pressure could potentially become large enough to affect the bubble dynamics.
During the first half of the phase transition, however, the quark density, and hence $p_q$, grows by only a factor of two. 
Including this factor of two enhancement, we find that the quark pressure is sub-dominant compared to the net gluon pressure during the bubble growth stage of the phase transition, and hence can be neglected.

On the other hand, during the second half of the phase transition, the quarks are compressed much more. 
We find that for every point in the DM parameter space we consider, the quark pressure eventually becomes comparable to the other forces acting on the wall. 
Most likely, the increased quark pressure would oppose further contraction and slow $v_w$ down. 
Unfortunately, the process by which $p_q$ gradually grows and, in response, $v_w$ gradually shrinks, is a non-equilibrium, strong physics problem for which we have no solution.
Nevertheless, we can still use thermodynamic arguments as before to understand the possible limiting behavior of $v_w$. 

Consider the scenario in which the enhanced quark pressure forces the pocket into a state of mechanical equilibrium. 
Mechanical equilibrium is achieved when the four forces on the wall -- gluon pressure inside the pocket, glueball pressure outside the pocket, surface tension, and quark pressure -- are in balance.
At this equilibrium point, we must have
\begin{alignat}{1}
    0 &= dA \, \sigma + dV \left.\left(\sum p\right)\right|_{\text{wall}}  \nonumber \\
	&= 8 \pi \sigma R + 4\pi \Delta f(T_\text{wall}) R^2 - 4\pi p_q R^2 \nonumber \\
	&= 8 \pi \sigma R + 4\pi \frac{(T_c-T_\text{wall})}{T_c} l \, R^2 - 3 N_q T_\text{wall} R^{-1}\, ,
	\label{eq:MechEq}
\end{alignat}
where $\Delta f$ still refers to the confined phase minus the deconfined phase gluonic pressures. 
The temperatures are all evaluated at the local wall temperature and 
we have used $N_q = \frac{4\pi}{3}R^3 \, n_q$.
If the system ever reached this equilibrium point, it would be a stable equilibrium: 
if $R$ were to contract the increased quark density would oppose it; 
if $R$ were to expand the surface tension would increase, the quark pressure would decrease, and the wall would absorb latent heat and increase the net gluon pressure difference, all of which oppose further expansion. 
After achieving mechanical equilibrium, the wall would proceed to contract adiabatically as quark annihilation decreases the quark pressure and wall cooling increases the net gluon pressure compressing the wall. 
By differentiating Eq.~\eqref{eq:MechEq} with respect to time and defining $v_w = -\dot{R}$, we find
\begin{alignat}{1}
	 v_w = \frac{-3\dot{N_q} T_\text{wall} - \left(N_q + \frac{4\pi l R^3}{T_c}\right) \dot{T}_\text{wall}}{\left(8\pi\sigma R +  8\pi\frac{(T_c-T_\text{wall})}{T_c} l R^2 + 3N_q T_\text{wall} R^{-1}\right)}
	  \, .
	\label{eq:vw_pq}
\end{alignat}
\begin{figure}
\caption{
    The pocket wall velocity (left) and particle abundances (right) within the pocket using a model that crudely incorporates the effects of quark pressure on $v_w$. 
    We choose a confinement scale of $1\,$TeV and dark matter mass of $10^3\,$TeV.
    We begin the simulation neglecting the quark pressure, allowing us to use Eq.~\eqref{eq:dRdt} to determine the contraction rate (red-dotted line). 
    Eventually the quark pressure is so large that it is able to come into mechanical equilibrium with the other forces acting on the wall, which happens at the discontinuity near $R\,\Lambda=10^5$.
    At this point, we switch over to a contraction rate given by Eq.~\eqref{eq:vw_pq}, leading to the discontinuous drop in $v_w$ in the left panel and the sudden depletion of all particle abundances in the right panel.
}
    \resizebox{\columnwidth}{!}{
    \includegraphics[scale=0.5]{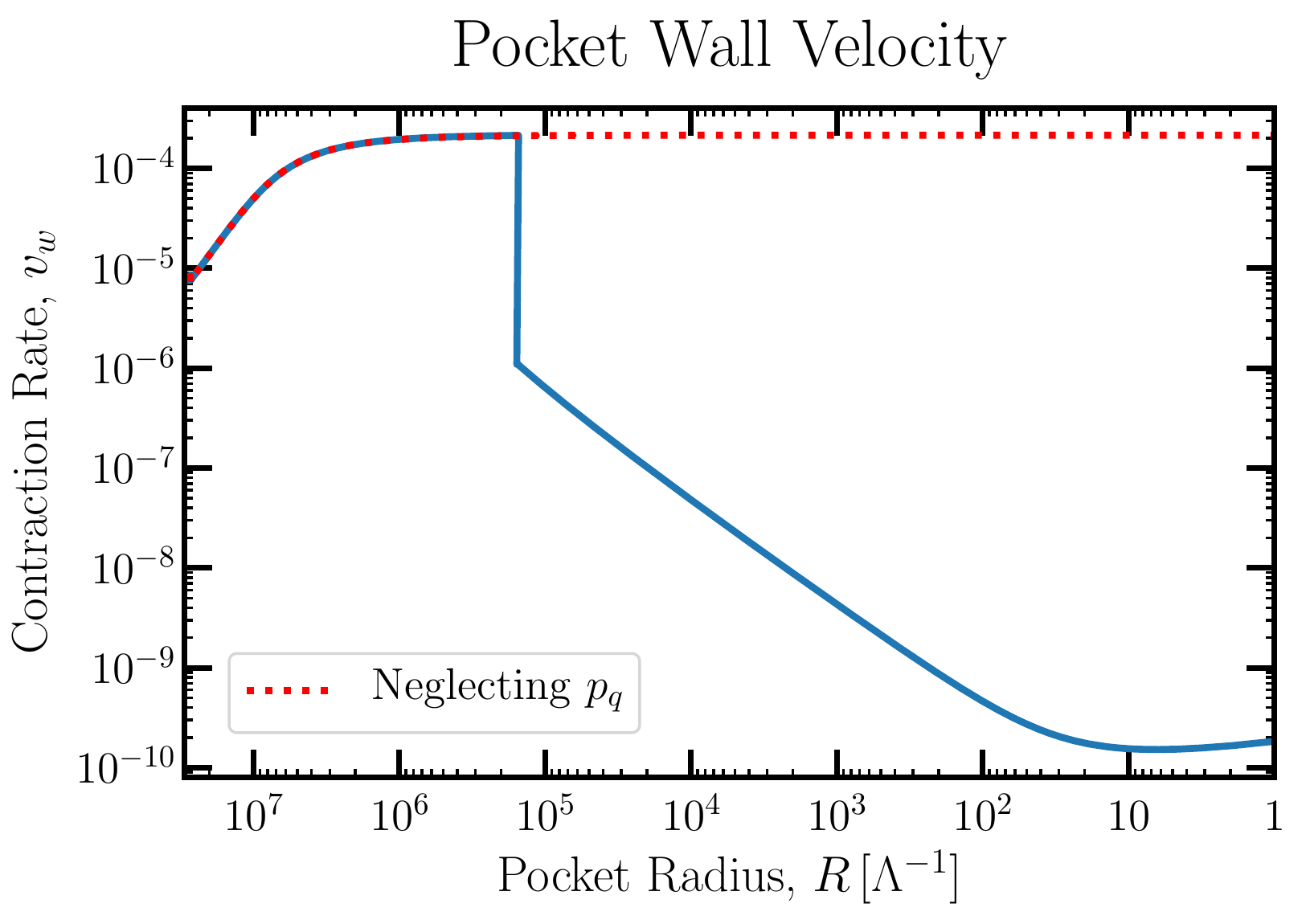}
    \includegraphics[scale=0.5]{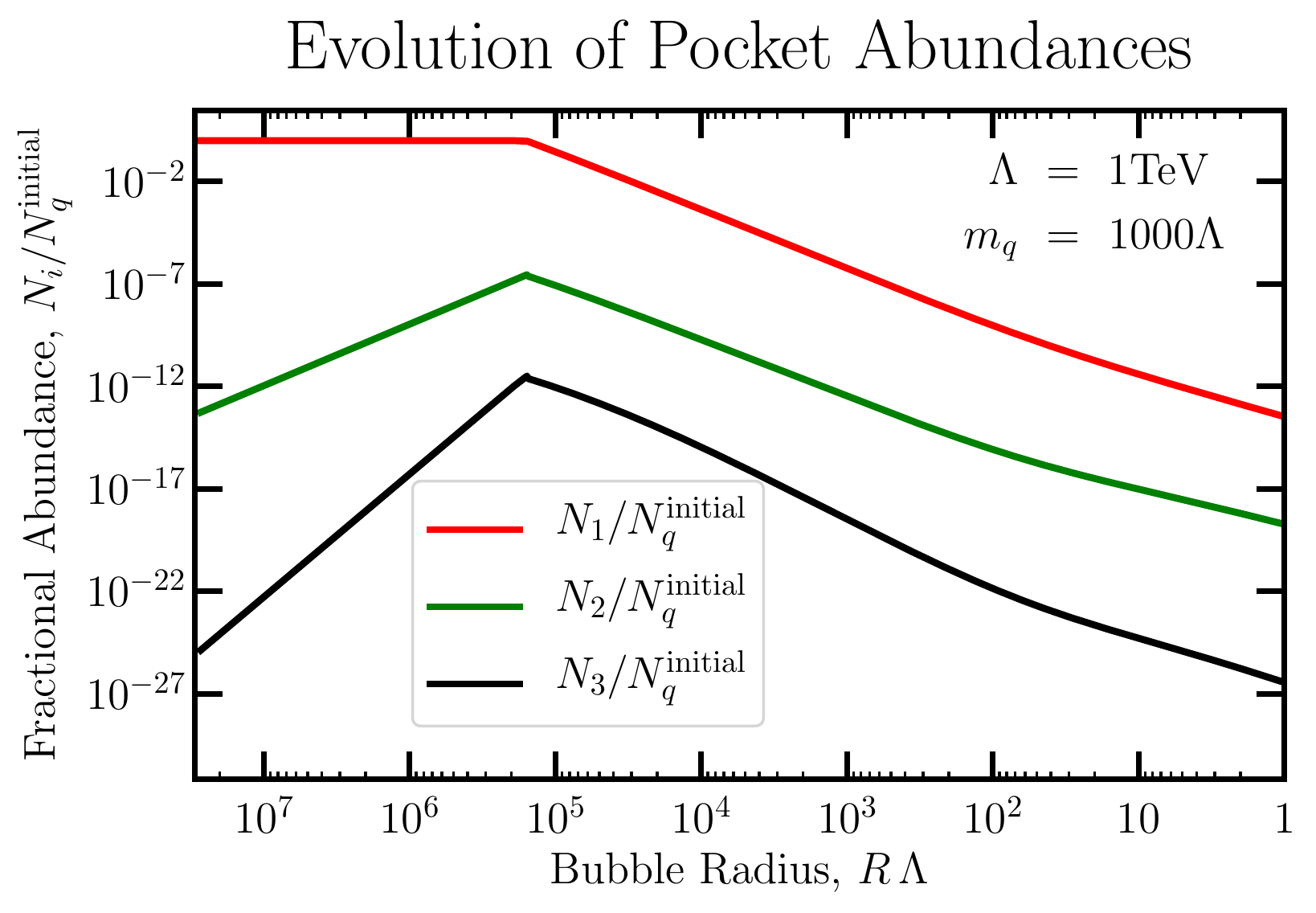}
    }
     \label{fig:pq_sim}
\end{figure}

Since we do not keep track of $T_\text{wall}$, nor can we calculate $v_w$ as a function of $T_\text{wall}$ when the wall is out of mechanical equilibrium, we have no way of knowing when or if mechanical equilibrium is achieved. 

However, to develop some intuition for the possible effects of quark pressure, we can perform a crude approximate calculation.
For this calculation, we perform our pocket contraction simulation while simultaneously keeping track of the quark abundance within each step using the Boltzmann equations of Sec.~\ref{sec:boltzeqs}. Since the quark pressure is sub-dominant initially, we start with $v_w$ given by Eq.~\eqref{eq:dRdt}.
Eventually there comes a radius when the quark pressure is so large that it is able to oppose the combined forces of surface tension and the net gluonic pressure, even when the latter pressure is at its maximum (which is attained when the wall temperature is at its minimum, $T_\text{wall}=T$). 
At this radius, we say that the wall has suddenly attained mechanical equilibrium. We then switch over to a wall velocity of Eq.~\eqref{eq:vw_pq} for the rest of the simulation.  
We assume that the system maintains mechanical equilibrium and $T_\text{wall}=T$ until the end of the simulation.

In Fig.~\ref{fig:pq_sim} we plot the velocity and particle abundances within the pocket as functions of $R$ for this new simulation. 
One can see that the velocity discontinuously drops at a radius of $R\sim10^5 \Lambda^{-1}$ when the pocket abruptly achieves mechanical equilibrium. 
When this happens, further contraction is allowed only by subsequent quark annihilations; the quark depletion processes immediately recouple and dominate the density evolution due to the sharp drop in the contraction rate $v_w/R$.
At this same radius, the baryon abundance is at a maximum and we find that the DM relic abundance is set. 
Afterwards, the pocket slowly contracts and all particle numbers deplete until the pocket vanishes.
Importantly, we find that the pocket asymmetry is saturated in this simulation (see Eq.~\eqref{eq:Slimit}), and is also saturated for every other point in the DM parameter space that we consider. Of course, this is a crude approximation -- a realistic scenario could have a smoother evolution of $v_w$, or if the wall velocity falls sharply, this could induce plasma shock waves which modify the pocket evolution. However, this explicit example supports our intuition that turning on quark pressure will drive the system rapidly into the regime where the asymmetry is saturated, and once in this regime the details of the evolution do not affect the final relic abundance.

\section{Cross Sections}
\label{appx:xsec}

For the computation of the baryon survival factor, multiple processes are relevant. We distinguish between three classes of interactions: 
\begin{itemize}
\item Annihilation process, i.e. a direct annihilation of free quarks into dark gluons: $1 + (-1) \rightarrow 0 + 0$. 
\item Capture processes, where a dark gluon is emitted for momentum conservation, for example: $1 + 1 \rightarrow 2 + 0$.
\item Rearrangement processes, where no gluon is emitted and only quark constituents exchanged, for example: $2 + 2 \rightarrow 3+1$.
\end{itemize}
Similar to Tab.~\ref{tab:relics}, in writing these equations we use each relic's quark number.

\begin{figure}
\resizebox{0.7\columnwidth}{!}{
\includegraphics[scale=0.6]{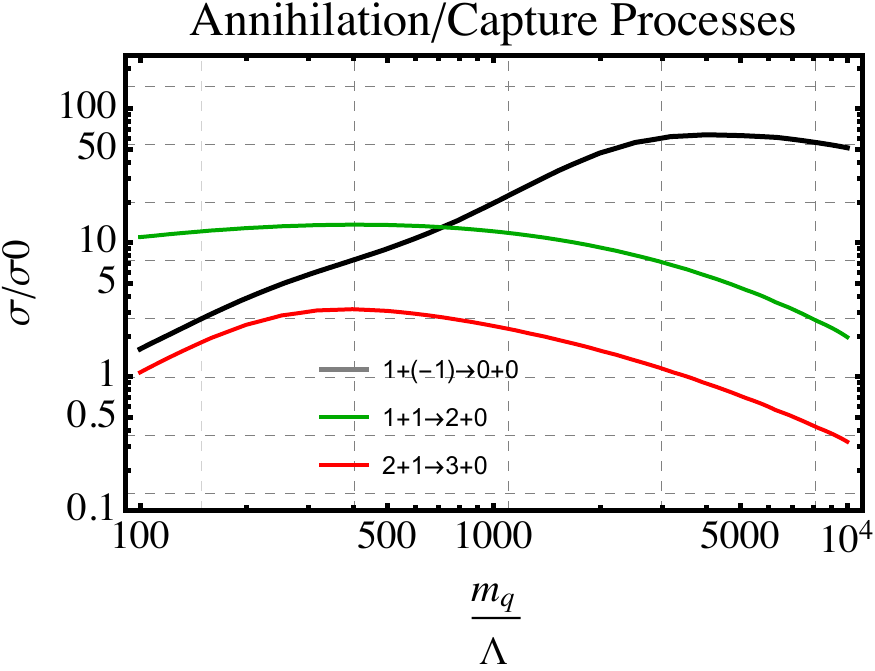}
}
\caption{Reproducing the results of \cite{Mitridate:2017izz,Mitridate:2017oky,DeLuca:2018mzn} for $\zeta$ factor in Eq.~\eqref{eqn:ann_xsec} for various annihilation or capture processes entering our full set of Boltzmann equations. }
\label{fig:crosssections}
\end{figure}

For the values of the annihilation and capture cross sections explicit calculations taking into account group theory factors have been performed in~\cite{Mitridate:2017izz,Mitridate:2017oky,DeLuca:2018mzn}.\footnote{See also Refs.~\cite{Biondini:2017ufr,Biondini:2018pwp} for a similar calculation and further details on the finite temperature effects on the cross sections.} The cross sections scale as 
\begin{align}
 \langle \sigma_{\rm ann./cap.} v \rangle =  \zeta\,\frac{\pi \alpha^2}{ m_q^2} \equiv \zeta\,\sigma_0\,,
\label{eqn:ann_xsec}
\end{align}
where $\zeta$ is a numerical factor that depends on the number of colors and flavors in the theory, and the coupling strength $\alpha$ is evaluated at the scale of the momentum transferred in the annihilation process, which is $m_q$.
In addition at low interaction energies the bound state formation and the annihilation process experience enhancement due to non-perturbative Sommerfeld corrections. The cumulative effect of those non-perturbative effects at finite temperature can be taken into account by the effective cross section $ \langle \sigma_{\text{eff}} v_{\text{rel}} \rangle$ defined in~\cite{Mitridate:2017izz,Mitridate:2017oky,DeLuca:2018mzn}. In Fig.~\ref{fig:crosssections} the factor $\zeta$ is shown for different annihilation and capture processes in our full set of Boltzmann equations.

Thermal masses of the dark gluons prevent bound state formation at large temperatures, an effect that has been confirmed by additional investigations of the process at hand in a non-equilibrium field theory treatment~\cite{Mitridate:2017izz,Binder:2019erp,Binder:2020efn}.

\begin{table}[h]
\centering
{\def\arraystretch{1.3}
\begin{tabular}{ l | l | l }
\hline\hline
Class & Process & Cross section \\
\hline
\multirow{1}{*}{Annihilation} & $1 + (-1) \rightarrow 0 + 0$ & \multirow{3}{*}{}  \\
\cline{1-2} 
\multirow{2}{*}{Capture} & $1 + 1 \rightarrow 2 + 0$ &  Fig.~\ref{fig:crosssections} $\&$ Eq.~\eqref{eqn:ann_xsec} \\
& $2 + 1 \rightarrow 3 + 0$ &  \\
\cline{1-3} 
\multirow{11}{*}{} & $(-3) + 1 \rightarrow (-1) + (-1)$ &  \multirow{11}{*}{} \\
& $(-3) + 1 \rightarrow (-2) + 0$ &  \\
& $(-2) + 1 \rightarrow (-1) + 0$ &   \\
& $3 + (-2) \rightarrow 1 + 0$ &   \\
& $2 + 2 \rightarrow 3 + 1$ &   \\
Rearrangement& $3 + (-2) \rightarrow 2 + (-1)$ &  Eq.~\eqref{eq:xsecRA} \\
& $3 + (-3) \rightarrow 2 + (-2)$ &   \\
& $2 + (-2) \rightarrow 1 + (-1)$ &   \\
& $3 + (-3) \rightarrow 1 + (-1)$ &   \\
& $3 + (-3) \rightarrow 0 + 0$ &   \\
& $2 + (-2) \rightarrow 0 + 0$ &   \\
\hline\hline
\end{tabular}}
\caption{ 
 The processes and cross section classes involved in the annihilation and baryon formation. Bound states are donated by their baryon number. Direct annihilation takes place if multiple gluons need to be emitted in the final state. If one gluon is radiated, then perturbative capture cross section calculations apply. In the case that the final states have no gluons that would be needed for momentum conservation, we use the geometric rearrangement cross sections discussed below. The 0 in the rearrangement processes denotes a pion $\bar{q}q$ that promptly decays to gluons. Generally, all processes above have equivalent reactions, where all particles are replaced by anti-particles. For those we assume the same cross sections, i.e. assuming CP is conserved. }
 \label{tab:cross}
\end{table}

The rearrangement process is more complex and requires taking into account non-perturbative effects. Here simulations and comparisons to hydrogen$-$anti-hydrogen annihilation have been performed in~\cite{Mitridate:2017oky}. 
The resulting cross section scales with the area set by the Bohr radius of the colliding bound states and contains a suppression factor which becomes effective once the kinetic energy exceeds the available binding energy:
\begin{align}
\sigma_{\rm RA} = \frac{\pi R_{\rm Bohr}^2}{\sqrt{E_{\rm kin}/E_{\rm E_B}}}.
\end{align}
This results in a constant $\sigma v$ cross section, which is expected for an exothermic reaction
\begin{align}
\langle \sigma_{\rm RA} v \rangle = \frac{1}{C_N \alpha}\frac{\pi}{m_q^2} =  \frac{\sigma_0}{C_N \alpha^3}\, ,
\label{eq:xsecRA}
\end{align}
with $C_N$ being the quadratic Casimir of the dark quark representation ($C_N = 4/3$ for quarks in the fundamental representation of $SU(N)$), that controls the interquark attraction in a non-abelian theory. 
The overall scaling is thus $\sigma_{\rm RA} v   \sim \sigma_{\rm ann./cap.} v /\alpha^3$, in agreement with the numerical results of \cite{Geller:2018biy}. This non-perturbative enhancement results from taking into account the finite size of the colliding bound states.

We summarize all the cross sections entering Eq.~\eqref{eq:fullboltz} in Tab.~\ref{tab:cross}. Notice that in this table some processes involving gluons (denoted by $0$) are listed as rearrangement. These processes, in fact, have an intermediate step in which the quarks rearrange and make pions ($\bar{q}q$), which can promptly decay into gluons, see Eq.~\eqref{eq:mesonrates}.

\section{Binding Energies}
\label{appx:binding}

The binding energies of several types of dark states enter the full set of Boltzmann equations. For the two quark states exact results are available. Since we work in the limit $m_q \gg \Lambda_{\rm DC}$ the Coulomb potential approximation is valid. For the baryon binding energy variational methods are needed, and numerical evaluations were performed in~\cite{Mitridate:2017oky}. We focus on the case that $N = 3$. The resulting binding energies are:
\begin{itemize}
\item Binding energy of a singlet diquark, or meson $\bar{q}q$: $E_B^{\bar{q}q} = \frac{1}{4} \alpha^2 C_{N}^2 m_q$.
\item Binding energy of a bound non-singlet diquark state $qq$ in a binding configuration is: $E_B^{qq} = \frac{1}{4} E_B^{\bar{q}q}$.
\item Binding energy of a baron $qqq$ singlet sate is:  $E_B^{qqq} = 0.26 \alpha^2 C_{N}^2 m_q$.
\end{itemize}

Here $\alpha$ is the gauge coupling of the confining group given by Eq.~\eqref{eq:alpha}. The relevant scale at which the coupling should be evaluated is the Bohr momentum $\alpha m_q$, which can be determined iteratively, starting from the value of $\alpha(m_q)$.  
It has been shown that the corrections due to the linear (confining) part of the potential between quarks has negligible effect on these binding energies (see Eq.~(4) of \cite{DeLuca:2018mzn}).

\afterpage{\clearpage}
\bibliographystyle{utphys_modified}
\bibliography{bib}

\providecommand{\href}[2]{#2}\begingroup\raggedright\begin{thebibliography}{100}

\bibitem{Aghanim:2018eyx}
{\bfseries Planck} Collaboration, N.~Aghanim {\em et~al.}, ``{Planck 2018
  results. VI. Cosmological parameters},''
  \href{http://dx.doi.org/10.1051/0004-6361/201833910}{{\em Astron. Astrophys.}
  {\bfseries 641} (2020) A6}, \href{http://arxiv.org/abs/1807.06209}{{\ttfamily
  arXiv:1807.06209 [astro-ph.CO]}}.

\bibitem{Gudnason:2006yj}
S.~B. Gudnason, C.~Kouvaris, and F.~Sannino, ``{Dark Matter from new
  Technicolor Theories},''
  \href{http://dx.doi.org/10.1103/PhysRevD.74.095008}{{\em Phys. Rev. D}
  {\bfseries 74} (2006) 095008},
  \href{http://arxiv.org/abs/hep-ph/0608055}{{\ttfamily arXiv:hep-ph/0608055}}.

\bibitem{Alves:2009nf}
D.~S.~M. Alves, S.~R. Behbahani, P.~Schuster, and J.~G. Wacker, ``{Composite
  Inelastic Dark Matter},''
  \href{http://dx.doi.org/10.1016/j.physletb.2010.08.006}{{\em Phys. Lett. B}
  {\bfseries 692} (2010) 323--326},
  \href{http://arxiv.org/abs/0903.3945}{{\ttfamily arXiv:0903.3945 [hep-ph]}}.

\bibitem{Kilic:2009mi}
C.~Kilic, T.~Okui, and R.~Sundrum, ``{Vectorlike Confinement at the LHC},''
  \href{http://dx.doi.org/10.1007/JHEP02(2010)018}{{\em JHEP} {\bfseries 02}
  (2010) 018}, \href{http://arxiv.org/abs/0906.0577}{{\ttfamily arXiv:0906.0577
  [hep-ph]}}.

\bibitem{Hambye:2009fg}
T.~Hambye and M.~H.~G. Tytgat, ``{Confined hidden vector dark matter},''
  \href{http://dx.doi.org/10.1016/j.physletb.2009.11.050}{{\em Phys. Lett. B}
  {\bfseries 683} (2010) 39--41},
  \href{http://arxiv.org/abs/0907.1007}{{\ttfamily arXiv:0907.1007 [hep-ph]}}.

\bibitem{Kribs:2009fy}
G.~D. Kribs, T.~S. Roy, J.~Terning, and K.~M. Zurek, ``{Quirky Composite Dark
  Matter},'' \href{http://dx.doi.org/10.1103/PhysRevD.81.095001}{{\em Phys.
  Rev. D} {\bfseries 81} (2010) 095001},
  \href{http://arxiv.org/abs/0909.2034}{{\ttfamily arXiv:0909.2034 [hep-ph]}}.

\bibitem{Alves:2010dd}
D.~Spier Moreira~Alves, S.~R. Behbahani, P.~Schuster, and J.~G. Wacker, ``{The
  Cosmology of Composite Inelastic Dark Matter},''
  \href{http://dx.doi.org/10.1007/JHEP06(2010)113}{{\em JHEP} {\bfseries 06}
  (2010) 113},
\href{http://arxiv.org/abs/1003.4729}{{\ttfamily arXiv:1003.4729 [hep-ph]}}.

\bibitem{Bai:2010qg}
Y.~Bai and R.~J. Hill, ``{Weakly Interacting Stable Pions},''
  \href{http://dx.doi.org/10.1103/PhysRevD.82.111701}{{\em Phys. Rev. D}
  {\bfseries 82} (2010) 111701},
  \href{http://arxiv.org/abs/1005.0008}{{\ttfamily arXiv:1005.0008 [hep-ph]}}.

\bibitem{Feng:2011ik}
J.~L. Feng and Y.~Shadmi, ``{WIMPless Dark Matter from Non-Abelian Hidden
  Sectors with Anomaly-Mediated Supersymmetry Breaking},''
  \href{http://dx.doi.org/10.1103/PhysRevD.83.095011}{{\em Phys. Rev. D}
  {\bfseries 83} (2011) 095011},
  \href{http://arxiv.org/abs/1102.0282}{{\ttfamily arXiv:1102.0282 [hep-ph]}}.

\bibitem{Fok:2011yc}
R.~Fok and G.~D. Kribs, ``{Chiral Quirkonium Decays},''
  \href{http://dx.doi.org/10.1103/PhysRevD.84.035001}{{\em Phys. Rev. D}
  {\bfseries 84} (2011) 035001},
  \href{http://arxiv.org/abs/1106.3101}{{\ttfamily arXiv:1106.3101 [hep-ph]}}.

\bibitem{Lewis:2011zb}
R.~Lewis, C.~Pica, and F.~Sannino, ``{Light Asymmetric Dark Matter on the
  Lattice: SU(2) Technicolor with Two Fundamental Flavors},''
  \href{http://dx.doi.org/10.1103/PhysRevD.85.014504}{{\em Phys. Rev. D}
  {\bfseries 85} (2012) 014504},
  \href{http://arxiv.org/abs/1109.3513}{{\ttfamily arXiv:1109.3513 [hep-ph]}}.

\bibitem{Frigerio:2012uc}
M.~Frigerio, A.~Pomarol, F.~Riva, and A.~Urbano, ``{Composite Scalar Dark
  Matter},'' \href{http://dx.doi.org/10.1007/JHEP07(2012)015}{{\em JHEP}
  {\bfseries 07} (2012) 015}, \href{http://arxiv.org/abs/1204.2808}{{\ttfamily
  arXiv:1204.2808 [hep-ph]}}.

\bibitem{Buckley:2012ky}
M.~R. Buckley and E.~T. Neil, ``{Thermal dark matter from a confining
  sector},'' \href{http://dx.doi.org/10.1103/PhysRevD.87.043510}{{\em Phys.
  Rev. D} {\bfseries 87} no.~4, (2013) 043510},
  \href{http://arxiv.org/abs/1209.6054}{{\ttfamily arXiv:1209.6054 [hep-ph]}}.

\bibitem{Appelquist:2013ms}
{\bfseries Lattice Strong Dynamics (LSD)} Collaboration, T.~Appelquist {\em
  et~al.}, ``{Lattice Calculation of Composite Dark Matter Form Factors},''
  \href{http://dx.doi.org/10.1103/PhysRevD.88.014502}{{\em Phys. Rev. D}
  {\bfseries 88} no.~1, (2013) 014502},
  \href{http://arxiv.org/abs/1301.1693}{{\ttfamily arXiv:1301.1693 [hep-ph]}}.

\bibitem{Bai:2013xga}
Y.~Bai and P.~Schwaller, ``{Scale of dark QCD},''
  \href{http://dx.doi.org/10.1103/PhysRevD.89.063522}{{\em Phys. Rev. D}
  {\bfseries 89} no.~6, (2014) 063522},
  \href{http://arxiv.org/abs/1306.4676}{{\ttfamily arXiv:1306.4676 [hep-ph]}}.

\bibitem{Bhattacharya:2013kma}
S.~Bhattacharya, B.~Meli\'c, and J.~Wudka, ``{Pionic Dark Matter},''
  \href{http://dx.doi.org/10.1007/JHEP02(2014)115}{{\em JHEP} {\bfseries 02}
  (2014) 115}, \href{http://arxiv.org/abs/1307.2647}{{\ttfamily arXiv:1307.2647
  [hep-ph]}}.

\bibitem{Cline:2013zca}
J.~M. Cline, Z.~Liu, G.~Moore, and W.~Xue, ``{Composite strongly interacting
  dark matter},'' \href{http://dx.doi.org/10.1103/PhysRevD.90.015023}{{\em
  Phys. Rev. D} {\bfseries 90} no.~1, (2014) 015023},
  \href{http://arxiv.org/abs/1312.3325}{{\ttfamily arXiv:1312.3325 [hep-ph]}}.

\bibitem{Boddy:2014yra}
K.~K. Boddy, J.~L. Feng, M.~Kaplinghat, and T.~M.~P. Tait, ``{Self-Interacting
  Dark Matter from a Non-Abelian Hidden Sector},''
  \href{http://dx.doi.org/10.1103/PhysRevD.89.115017}{{\em Phys. Rev. D}
  {\bfseries 89} no.~11, (2014) 115017},
  \href{http://arxiv.org/abs/1402.3629}{{\ttfamily arXiv:1402.3629 [hep-ph]}}.

\bibitem{Boddy:2014qxa}
K.~K. Boddy, J.~L. Feng, M.~Kaplinghat, Y.~Shadmi, and T.~M.~P. Tait,
  ``{Strongly interacting dark matter: Self-interactions and keV lines},''
  \href{http://dx.doi.org/10.1103/PhysRevD.90.095016}{{\em Phys. Rev. D}
  {\bfseries 90} no.~9, (2014) 095016},
  \href{http://arxiv.org/abs/1408.6532}{{\ttfamily arXiv:1408.6532 [hep-ph]}}.

\bibitem{Hochberg:2014kqa}
Y.~Hochberg, E.~Kuflik, H.~Murayama, T.~Volansky, and J.~G. Wacker, ``{Model
  for Thermal Relic Dark Matter of Strongly Interacting Massive Particles},''
  \href{http://dx.doi.org/10.1103/PhysRevLett.115.021301}{{\em Phys. Rev.
  Lett.} {\bfseries 115} no.~2, (2015) 021301},
  \href{http://arxiv.org/abs/1411.3727}{{\ttfamily arXiv:1411.3727 [hep-ph]}}.

\bibitem{Appelquist:2015yfa}
T.~Appelquist {\em et~al.}, ``{Stealth Dark Matter: Dark scalar baryons through
  the Higgs portal},'' \href{http://dx.doi.org/10.1103/PhysRevD.92.075030}{{\em
  Phys. Rev. D} {\bfseries 92} no.~7, (2015) 075030},
  \href{http://arxiv.org/abs/1503.04203}{{\ttfamily arXiv:1503.04203
  [hep-ph]}}.

\bibitem{Antipin:2015xia}
O.~Antipin, M.~Redi, A.~Strumia, and E.~Vigiani, ``{Accidental Composite Dark
  Matter},'' \href{http://dx.doi.org/10.1007/JHEP07(2015)039}{{\em JHEP}
  {\bfseries 07} (2015) 039}, \href{http://arxiv.org/abs/1503.08749}{{\ttfamily
  arXiv:1503.08749 [hep-ph]}}.

\bibitem{GarciaGarcia:2015fol}
I.~Garcia~Garcia, R.~Lasenby, and J.~March-Russell, ``{Twin Higgs WIMP Dark
  Matter},'' \href{http://dx.doi.org/10.1103/PhysRevD.92.055034}{{\em Phys.
  Rev. D} {\bfseries 92} no.~5, (2015) 055034},
  \href{http://arxiv.org/abs/1505.07109}{{\ttfamily arXiv:1505.07109
  [hep-ph]}}.

\bibitem{Soni:2016gzf}
A.~Soni and Y.~Zhang, ``{Hidden SU(N) Glueball Dark Matter},''
  \href{http://dx.doi.org/10.1103/PhysRevD.93.115025}{{\em Phys. Rev. D}
  {\bfseries 93} no.~11, (2016) 115025},
  \href{http://arxiv.org/abs/1602.00714}{{\ttfamily arXiv:1602.00714
  [hep-ph]}}.

\bibitem{Kribs:2016cew}
G.~D. Kribs and E.~T. Neil, ``{Review of strongly-coupled composite dark matter
  models and lattice simulations},''
  \href{http://dx.doi.org/10.1142/S0217751X16430041}{{\em Int. J. Mod. Phys. A}
  {\bfseries 31} no.~22, (2016) 1643004},
  \href{http://arxiv.org/abs/1604.04627}{{\ttfamily arXiv:1604.04627
  [hep-ph]}}.

\bibitem{Harigaya:2016nlg}
K.~Harigaya, M.~Ibe, K.~Kaneta, W.~Nakano, and M.~Suzuki, ``{Thermal Relic Dark
  Matter Beyond the Unitarity Limit},''
  \href{http://dx.doi.org/10.1007/JHEP08(2016)151}{{\em JHEP} {\bfseries 08}
  (2016) 151}, \href{http://arxiv.org/abs/1606.00159}{{\ttfamily
  arXiv:1606.00159 [hep-ph]}}.

\bibitem{Mitridate:2017oky}
A.~Mitridate, M.~Redi, J.~Smirnov, and A.~Strumia, ``{Dark Matter as a weakly
  coupled Dark Baryon},'' \href{http://dx.doi.org/10.1007/JHEP10(2017)210}{{\em
  JHEP} {\bfseries 10} (2017) 210},
  \href{http://arxiv.org/abs/1707.05380}{{\ttfamily arXiv:1707.05380
  [hep-ph]}}.

\bibitem{DeLuca:2018mzn}
V.~De~Luca, A.~Mitridate, M.~Redi, J.~Smirnov, and A.~Strumia, ``{Colored Dark
  Matter},'' \href{http://dx.doi.org/10.1103/PhysRevD.97.115024}{{\em Phys.
  Rev. D} {\bfseries 97} no.~11, (2018) 115024},
  \href{http://arxiv.org/abs/1801.01135}{{\ttfamily arXiv:1801.01135
  [hep-ph]}}.

\bibitem{Contino:2018crt}
R.~Contino, A.~Mitridate, A.~Podo, and M.~Redi, ``{Gluequark Dark Matter},''
  \href{http://dx.doi.org/10.1007/JHEP02(2019)187}{{\em JHEP} {\bfseries 02}
  (2019) 187}, \href{http://arxiv.org/abs/1811.06975}{{\ttfamily
  arXiv:1811.06975 [hep-ph]}}.

\bibitem{Gross:2018zha}
C.~Gross, A.~Mitridate, M.~Redi, J.~Smirnov, and A.~Strumia, ``{Cosmological
  Abundance of Colored Relics},''
  \href{http://dx.doi.org/10.1103/PhysRevD.99.016024}{{\em Phys. Rev. D}
  {\bfseries 99} no.~1, (2019) 016024},
  \href{http://arxiv.org/abs/1811.08418}{{\ttfamily arXiv:1811.08418
  [hep-ph]}}.

\bibitem{Beylin:2019gtw}
V.~Beylin, M.~Y. Khlopov, V.~Kuksa, and N.~Volchanskiy, ``{Hadronic and
  Hadron-Like Physics of Dark Matter},''
  \href{http://dx.doi.org/10.3390/sym11040587}{{\em Symmetry} {\bfseries 11}
  no.~4, (2019) 587}, \href{http://arxiv.org/abs/1904.12013}{{\ttfamily
  arXiv:1904.12013 [hep-ph]}}.

\bibitem{Dondi:2019olm}
N.~A. Dondi, F.~Sannino, and J.~Smirnov, ``{Thermal history of composite dark
  matter},'' \href{http://dx.doi.org/10.1103/PhysRevD.101.103010}{{\em Phys.
  Rev. D} {\bfseries 101} no.~10, (2020) 103010},
  \href{http://arxiv.org/abs/1905.08810}{{\ttfamily arXiv:1905.08810
  [hep-ph]}}.

\bibitem{Buttazzo:2019mvl}
D.~Buttazzo, L.~Di~Luzio, P.~Ghorbani, C.~Gross, G.~Landini, A.~Strumia,
  D.~Teresi, and J.-W. Wang, ``{Scalar gauge dynamics and Dark Matter},''
  \href{http://dx.doi.org/10.1007/JHEP01(2020)130}{{\em JHEP} {\bfseries 01}
  (2020) 130}, \href{http://arxiv.org/abs/1911.04502}{{\ttfamily
  arXiv:1911.04502 [hep-ph]}}.

\bibitem{Landini:2020daq}
G.~Landini and J.-W. Wang, ``{Dark Matter in scalar Sp($ \mathcal{N} $) gauge
  dynamics},'' \href{http://dx.doi.org/10.1007/JHEP06(2020)167}{{\em JHEP}
  {\bfseries 06} (2020) 167}, \href{http://arxiv.org/abs/2004.03299}{{\ttfamily
  arXiv:2004.03299 [hep-ph]}}.

\bibitem{Brower:2020mab}
{\bfseries Lattice Strong Dynamics} Collaboration, R.~C. Brower {\em et~al.},
  ``{Stealth dark matter confinement transition and gravitational waves},''
  \href{http://dx.doi.org/10.1103/PhysRevD.103.014505}{{\em Phys. Rev. D}
  {\bfseries 103} no.~1, (2021) 014505},
  \href{http://arxiv.org/abs/2006.16429}{{\ttfamily arXiv:2006.16429
  [hep-lat]}}.

\bibitem{Contino:2020god}
R.~Contino, A.~Podo, and F.~Revello, ``{Composite Dark Matter from
  Strongly-Interacting Chiral Dynamics},''
  \href{http://arxiv.org/abs/2008.10607}{{\ttfamily arXiv:2008.10607
  [hep-ph]}}.

\bibitem{Beylin:2020bsz}
V.~Beylin, M.~Khlopov, V.~Kuksa, and N.~Volchanskiy, ``{New physics of strong
  interaction and Dark Universe},''
  \href{http://dx.doi.org/10.3390/universe6110196}{{\em Universe} {\bfseries 6}
  no.~11, (2020) 196}, \href{http://arxiv.org/abs/2010.13678}{{\ttfamily
  arXiv:2010.13678 [hep-ph]}}.

\bibitem{Barr:1991}
S.~M. Barr, ``Baryogenesis, sphalerons, and the cogeneration of dark matter,''
  \href{http://dx.doi.org/10.1103/PhysRevD.44.3062}{{\em Phys. Rev. D}
  {\bfseries 44} (Nov, 1991) 3062--3066}.
  \url{https://link.aps.org/doi/10.1103/PhysRevD.44.3062}.

\bibitem{Huang:2017kzu}
F.~P. Huang and C.~S. Li, ``{Probing the baryogenesis and dark matter relaxed
  in phase transition by gravitational waves and colliders},''
  \href{http://dx.doi.org/10.1103/PhysRevD.96.095028}{{\em Phys. Rev. D}
  {\bfseries 96} no.~9, (2017) 095028},
  \href{http://arxiv.org/abs/1709.09691}{{\ttfamily arXiv:1709.09691
  [hep-ph]}}.

\bibitem{Ben_PT}
B.~Svetitsky and L.~G. Yaffe, ``Critical behavior at finite-temperature
  confinement transitions,''
  \href{http://dx.doi.org/https://doi.org/10.1016/0550-3213(82)90172-9}{{\em
  Nuclear Physics B} {\bfseries 210} no.~4, (1982) 423 -- 447}.
  \url{http://www.sciencedirect.com/science/article/pii/0550321382901729}.

\bibitem{Kaczmarek:1999mm}
O.~Kaczmarek, F.~Karsch, E.~Laermann, and M.~Lutgemeier, ``{Heavy quark
  potentials in quenched QCD at high temperature},''
  \href{http://dx.doi.org/10.1103/PhysRevD.62.034021}{{\em Phys. Rev. D}
  {\bfseries 62} (2000) 034021},
  \href{http://arxiv.org/abs/hep-lat/9908010}{{\ttfamily
  arXiv:hep-lat/9908010}}.

\bibitem{PhysRevD.60.034504}
C.~Alexandrou, A.~Bori\ifmmode~\mbox{\c{c}}\else \c{c}\fi{}i, A.~Feo,
  P.~de~Forcrand, A.~Galli, F.~Jegerlehner, and T.~Takaishi, ``Deconfinement
  phase transition in one-flavor qcd,''
  \href{http://dx.doi.org/10.1103/PhysRevD.60.034504}{{\em Phys. Rev. D}
  {\bfseries 60} (Jun, 1999) 034504}.
  \url{https://link.aps.org/doi/10.1103/PhysRevD.60.034504}.

\bibitem{Aoki:2006we}
Y.~Aoki, G.~Endrodi, Z.~Fodor, S.~Katz, and K.~Szabo, ``{The Order of the
  quantum chromodynamics transition predicted by the standard model of particle
  physics},'' \href{http://dx.doi.org/10.1038/nature05120}{{\em Nature}
  {\bfseries 443} (2006) 675--678},
  \href{http://arxiv.org/abs/hep-lat/0611014}{{\ttfamily
  arXiv:hep-lat/0611014}}.

\bibitem{Saito:2011fs}
{\bfseries WHOT-QCD} Collaboration, H.~Saito, S.~Ejiri, S.~Aoki, T.~Hatsuda,
  K.~Kanaya, Y.~Maezawa, H.~Ohno, and T.~Umeda, ``{Phase structure of finite
  temperature QCD in the heavy quark region},''
  \href{http://dx.doi.org/10.1103/PhysRevD.85.079902}{{\em Phys. Rev. D}
  {\bfseries 84} (2011) 054502},
  \href{http://arxiv.org/abs/1106.0974}{{\ttfamily arXiv:1106.0974 [hep-lat]}}.
  [Erratum: Phys.Rev.D 85, 079902 (2012)].

\bibitem{Lucini:2003zr}
B.~Lucini, M.~Teper, and U.~Wenger, ``{The High temperature phase transition in
  SU(N) gauge theories},''
  \href{http://dx.doi.org/10.1088/1126-6708/2004/01/061}{{\em JHEP} {\bfseries
  01} (2004) 061}, \href{http://arxiv.org/abs/hep-lat/0307017}{{\ttfamily
  arXiv:hep-lat/0307017}}.

\bibitem{largeN_lattice}
B.~Lucini, M.~Teper, and U.~Wenger, ``{Properties of the deconfining phase
  transition in SU(N) gauge theories},''
  \href{http://dx.doi.org/10.1088/1126-6708/2005/02/033}{{\em JHEP} {\bfseries
  02} (2005) 033}, \href{http://arxiv.org/abs/hep-lat/0502003}{{\ttfamily
  arXiv:hep-lat/0502003}}.

\bibitem{Baker:2019ndr}
M.~J. Baker, J.~Kopp, and A.~J. Long, ``{Filtered Dark Matter at a First Order
  Phase Transition},''
  \href{http://dx.doi.org/10.1103/PhysRevLett.125.151102}{{\em Phys. Rev.
  Lett.} {\bfseries 125} no.~15, (2020) 151102},
  \href{http://arxiv.org/abs/1912.02830}{{\ttfamily arXiv:1912.02830
  [hep-ph]}}.

\bibitem{Chway:2019kft}
D.~Chway, T.~H. Jung, and C.~S. Shin, ``{Dark matter filtering-out effect
  during a first-order phase transition},''
  \href{http://dx.doi.org/10.1103/PhysRevD.101.095019}{{\em Phys. Rev. D}
  {\bfseries 101} no.~9, (2020) 095019},
  \href{http://arxiv.org/abs/1912.04238}{{\ttfamily arXiv:1912.04238
  [hep-ph]}}.

\bibitem{Davoudiasl:2019xeb}
H.~Davoudiasl and G.~Mohlabeng, ``{Getting a THUMP from a WIMP},''
  \href{http://dx.doi.org/10.1007/JHEP04(2020)177}{{\em JHEP} {\bfseries 04}
  (2020) 177}, \href{http://arxiv.org/abs/1912.05572}{{\ttfamily
  arXiv:1912.05572 [hep-ph]}}.

\bibitem{Chao:2020adk}
W.~Chao, X.-F. Li, and L.~Wang, ``{Filtered pseudo-scalar dark matter and
  gravitational waves from first order phase transition},''
  \href{http://arxiv.org/abs/2012.15113}{{\ttfamily arXiv:2012.15113
  [hep-ph]}}.

\bibitem{Kang:2021epo}
Z.~Kang, S.~Matsuzaki, and J.~Zhu, ``{Dark Confinement-Deconfinement Phase
  Transition: A Roadmap from Polyakov Loop Models to Gravitational Waves},''
  \href{http://arxiv.org/abs/2101.03795}{{\ttfamily arXiv:2101.03795
  [hep-ph]}}.

\bibitem{Azatov:2021ifm}
A.~Azatov, M.~Vanvlasselaer, and W.~Yin, ``{Dark Matter production from
  relativistic bubble walls},'' {\em JHEP} {\bfseries 03} (2021) 288,
  \href{http://arxiv.org/abs/2101.05721}{{\ttfamily arXiv:2101.05721
  [hep-ph]}}.

\bibitem{Konstandin:2011dr}
T.~Konstandin and G.~Servant, ``{Cosmological Consequences of Nearly Conformal
  Dynamics at the TeV scale},''
  \href{http://dx.doi.org/10.1088/1475-7516/2011/12/009}{{\em JCAP} {\bfseries
  12} (2011) 009}, \href{http://arxiv.org/abs/1104.4791}{{\ttfamily
  arXiv:1104.4791 [hep-ph]}}.

\bibitem{Hambye:2018qjv}
T.~Hambye, A.~Strumia, and D.~Teresi, ``{Super-cool Dark Matter},''
  \href{http://dx.doi.org/10.1007/JHEP08(2018)188}{{\em JHEP} {\bfseries 08}
  (2018) 188}, \href{http://arxiv.org/abs/1805.01473}{{\ttfamily
  arXiv:1805.01473 [hep-ph]}}.

\bibitem{Baratella:2018pxi}
P.~Baratella, A.~Pomarol, and F.~Rompineve, ``{The Supercooled Universe},''
  \href{http://dx.doi.org/10.1007/JHEP03(2019)100}{{\em JHEP} {\bfseries 03}
  (2019) 100}, \href{http://arxiv.org/abs/1812.06996}{{\ttfamily
  arXiv:1812.06996 [hep-ph]}}.

\bibitem{Baldes:2020kam}
I.~Baldes, Y.~Gouttenoire, and F.~Sala, ``{String Fragmentation in Supercooled
  Confinement and implications for Dark Matter},''
  \href{http://arxiv.org/abs/2007.08440}{{\ttfamily arXiv:2007.08440
  [hep-ph]}}.

\bibitem{Witten:1984rs}
E.~Witten, ``{Cosmic Separation of Phases},''
  \href{http://dx.doi.org/10.1103/PhysRevD.30.272}{{\em Phys. Rev. D}
  {\bfseries 30} (1984) 272--285}.

\bibitem{Griest:1989wd}
K.~Griest and M.~Kamionkowski, ``{Unitarity Limits on the Mass and Radius of
  Dark Matter Particles},''
  \href{http://dx.doi.org/10.1103/PhysRevLett.64.615}{{\em Phys. Rev. Lett.}
  {\bfseries 64} (1990) 615}.

\bibitem{vonHarling:2014kha}
B.~von Harling and K.~Petraki, ``{Bound-state formation for thermal relic dark
  matter and unitarity},''
  \href{http://dx.doi.org/10.1088/1475-7516/2014/12/033}{{\em JCAP} {\bfseries
  12} (2014) 033}, \href{http://arxiv.org/abs/1407.7874}{{\ttfamily
  arXiv:1407.7874 [hep-ph]}}.

\bibitem{Smirnov:2019ngs}
J.~Smirnov and J.~F. Beacom, ``{TeV-Scale Thermal WIMPs: Unitarity and its
  Consequences},'' \href{http://dx.doi.org/10.1103/PhysRevD.100.043029}{{\em
  Phys. Rev. D} {\bfseries 100} no.~4, (2019) 043029},
  \href{http://arxiv.org/abs/1904.11503}{{\ttfamily arXiv:1904.11503
  [hep-ph]}}.

\bibitem{Moore:2014lga}
C.~J. Moore, R.~H. Cole, and C.~P.~L. Berry, ``{Gravitational-wave sensitivity
  curves},'' \href{http://dx.doi.org/10.1088/0264-9381/32/1/015014}{{\em Class.
  Quant. Grav.} {\bfseries 32} no.~1, (2015) 015014},
\href{http://arxiv.org/abs/1408.0740}{{\ttfamily arXiv:1408.0740 [gr-qc]}}.

\bibitem{Bonati:2012pe}
C.~Bonati, P.~de~Forcrand, M.~D'Elia, O.~Philipsen, and F.~Sanfilippo,
  ``{Constraints on the two-flavor QCD phase diagram from imaginary chemical
  potential},'' \href{http://dx.doi.org/10.22323/1.139.0189}{{\em PoS}
  {\bfseries LATTICE2011} (2011) 189},
  \href{http://arxiv.org/abs/1201.2769}{{\ttfamily arXiv:1201.2769 [hep-lat]}}.

\bibitem{Pelaggi:2017abg}
G.~M. Pelaggi, A.~D. Plascencia, A.~Salvio, F.~Sannino, J.~Smirnov, and
  A.~Strumia, ``{Asymptotically Safe Standard Model Extensions?},''
  \href{http://dx.doi.org/10.1103/PhysRevD.97.095013}{{\em Phys. Rev. D}
  {\bfseries 97} no.~9, (2018) 095013},
  \href{http://arxiv.org/abs/1708.00437}{{\ttfamily arXiv:1708.00437
  [hep-ph]}}.

\bibitem{Patel:1983sc}
A.~Patel, ``{A Flux Tube Model of the Finite Temperature Deconfining Transition
  in {QCD}},'' \href{http://dx.doi.org/10.1016/0550-3213(84)90484-X}{{\em Nucl.
  Phys. B} {\bfseries 243} (1984) 411--422}.

\bibitem{Mitridate:2017izz}
A.~Mitridate, M.~Redi, J.~Smirnov, and A.~Strumia, ``{Cosmological Implications
  of Dark Matter Bound States},''
  \href{http://dx.doi.org/10.1088/1475-7516/2017/05/006}{{\em JCAP} {\bfseries
  05} (2017) 006}, \href{http://arxiv.org/abs/1702.01141}{{\ttfamily
  arXiv:1702.01141 [hep-ph]}}.

\bibitem{PDG}
{Particle Data Group}, ``{Review of Particle Physics},''
  \href{http://dx.doi.org/10.1093/ptep/ptaa104}{{\em Progress of Theoretical
  and Experimental Physics} {\bfseries 2020} no.~8, (08, 2020) }.
  \url{https://doi.org/10.1093/ptep/ptaa104}.

\bibitem{Balaji:2020yrx}
S.~Balaji, M.~Spannowsky, and C.~Tamarit, ``{Cosmological bubble friction in
  local equilibrium},'' \href{http://arxiv.org/abs/2010.08013}{{\ttfamily
  arXiv:2010.08013 [hep-ph]}}.

\bibitem{Kang:2008ea}
J.~Kang and M.~A. Luty, ``{Macroscopic Strings and 'Quirks' at Colliders},''
  \href{http://dx.doi.org/10.1088/1126-6708/2009/11/065}{{\em JHEP} {\bfseries
  11} (2009) 065}, \href{http://arxiv.org/abs/0805.4642}{{\ttfamily
  arXiv:0805.4642 [hep-ph]}}.

\bibitem{PhysRev.82.664}
J.~Schwinger, ``On gauge invariance and vacuum polarization,''
  \href{http://dx.doi.org/10.1103/PhysRev.82.664}{{\em Phys. Rev.} {\bfseries
  82} (Jun, 1951) 664--679}.
  \url{https://link.aps.org/doi/10.1103/PhysRev.82.664}.

\bibitem{Karsch:2001vs}
F.~Karsch, ``{Lattice results on QCD thermodynamics},''
  \href{http://dx.doi.org/10.1016/S0375-9474(01)01365-3}{{\em Nucl. Phys. A}
  {\bfseries 698} (2002) 199--208},
  \href{http://arxiv.org/abs/hep-ph/0103314}{{\ttfamily arXiv:hep-ph/0103314}}.

\bibitem{Ignatius:1993qn}
J.~Ignatius, K.~Kajantie, H.~Kurki-Suonio, and M.~Laine, ``{The growth of
  bubbles in cosmological phase transitions},''
  \href{http://dx.doi.org/10.1103/PhysRevD.49.3854}{{\em Phys. Rev. D}
  {\bfseries 49} (1994) 3854--3868},
  \href{http://arxiv.org/abs/astro-ph/9309059}{{\ttfamily
  arXiv:astro-ph/9309059}}.

\bibitem{Huber:2011aa}
S.~J. Huber and M.~Sopena, ``{The bubble wall velocity in the minimal
  supersymmetric light stop scenario},''
  \href{http://dx.doi.org/10.1103/PhysRevD.85.103507}{{\em Phys. Rev. D}
  {\bfseries 85} (2012) 103507},
  \href{http://arxiv.org/abs/1112.1888}{{\ttfamily arXiv:1112.1888 [hep-ph]}}.

\bibitem{Ellis:2018mja}
J.~Ellis, M.~Lewicki, and J.~M. No, ``{On the Maximal Strength of a First-Order
  Electroweak Phase Transition and its Gravitational Wave Signal},''
  \href{http://dx.doi.org/10.1088/1475-7516/2019/04/003}{{\em JCAP} {\bfseries
  04} (2019) 003}, \href{http://arxiv.org/abs/1809.08242}{{\ttfamily
  arXiv:1809.08242 [hep-ph]}}.

\bibitem{Bai:2018dxf}
Y.~Bai, A.~J. Long, and S.~Lu, ``{Dark Quark Nuggets},''
  \href{http://dx.doi.org/10.1103/PhysRevD.99.055047}{{\em Phys. Rev. D}
  {\bfseries 99} no.~5, (2019) 055047},
  \href{http://arxiv.org/abs/1810.04360}{{\ttfamily arXiv:1810.04360
  [hep-ph]}}.

\bibitem{Hong:2020est}
J.-P. Hong, S.~Jung, and K.-P. Xie, ``{Fermi-ball dark matter from a
  first-order phase transition},''
  \href{http://dx.doi.org/10.1103/PhysRevD.102.075028}{{\em Phys. Rev. D}
  {\bfseries 102} no.~7, (2020) 075028},
  \href{http://arxiv.org/abs/2008.04430}{{\ttfamily arXiv:2008.04430
  [hep-ph]}}.

\bibitem{Juknevich:2009gg}
J.~E. Juknevich, ``{Pure-glue hidden valleys through the Higgs portal},''
  \href{http://dx.doi.org/10.1007/JHEP08(2010)121}{{\em JHEP} {\bfseries 08}
  (2010) 121}, \href{http://arxiv.org/abs/0911.5616}{{\ttfamily arXiv:0911.5616
  [hep-ph]}}.

\bibitem{Forestell:2016qhc}
L.~Forestell, D.~E. Morrissey, and K.~Sigurdson, ``{Non-Abelian Dark Forces and
  the Relic Densities of Dark Glueballs},''
  \href{http://dx.doi.org/10.1103/PhysRevD.95.015032}{{\em Phys. Rev. D}
  {\bfseries 95} no.~1, (2017) 015032},
  \href{http://arxiv.org/abs/1605.08048}{{\ttfamily arXiv:1605.08048
  [hep-ph]}}.

\bibitem{Forestell:2017wov}
L.~Forestell, D.~E. Morrissey, and K.~Sigurdson, ``{Cosmological Bounds on
  Non-Abelian Dark Forces},''
  \href{http://dx.doi.org/10.1103/PhysRevD.97.075029}{{\em Phys. Rev. D}
  {\bfseries 97} no.~7, (2018) 075029},
  \href{http://arxiv.org/abs/1710.06447}{{\ttfamily arXiv:1710.06447
  [hep-ph]}}.

\bibitem{Contino:2020tix}
R.~Contino, K.~Max, and R.~K. Mishra, ``{Searching for Elusive Dark Sectors
  with Terrestrial and Celestial Observations},''
  \href{http://arxiv.org/abs/2012.08537}{{\ttfamily arXiv:2012.08537
  [hep-ph]}}.

\bibitem{Schwaller:2015tja}
P.~Schwaller, ``{Gravitational Waves from a Dark Phase Transition},''
  \href{http://dx.doi.org/10.1103/PhysRevLett.115.181101}{{\em Phys. Rev.
  Lett.} {\bfseries 115} no.~18, (2015) 181101},
  \href{http://arxiv.org/abs/1504.07263}{{\ttfamily arXiv:1504.07263
  [hep-ph]}}.

\bibitem{Halverson:2020xpg}
J.~Halverson, C.~Long, A.~Maiti, B.~Nelson, and G.~Salinas, ``{Gravitational
  waves from dark Yang-Mills sectors},''
  \href{http://arxiv.org/abs/2012.04071}{{\ttfamily arXiv:2012.04071
  [hep-ph]}}.

\bibitem{Huang:2020mso}
W.-C. Huang, M.~Reichert, F.~Sannino, and Z.-W. Wang, ``{Testing the Dark
  Confined Landscape: From Lattice to Gravitational Waves},''
  \href{http://arxiv.org/abs/2012.11614}{{\ttfamily arXiv:2012.11614
  [hep-ph]}}.

\bibitem{Wang:2020jrd}
X.~Wang, F.~P. Huang, and X.~Zhang, ``{Phase transition dynamics and
  gravitational wave spectra of strong first-order phase transition in
  supercooled universe},''
  \href{http://dx.doi.org/10.1088/1475-7516/2020/05/045}{{\em JCAP} {\bfseries
  05} (2020) 045}, \href{http://arxiv.org/abs/2003.08892}{{\ttfamily
  arXiv:2003.08892 [hep-ph]}}.

\bibitem{Weir:2017wfa}
D.~J. Weir, ``{Gravitational waves from a first order electroweak phase
  transition: a brief review},''
  \href{http://dx.doi.org/10.1098/rsta.2017.0126}{{\em Phil. Trans. Roy. Soc.
  Lond.} {\bfseries A376} no.~2114, (2018) 20170126},
\href{http://arxiv.org/abs/1705.01783}{{\ttfamily arXiv:1705.01783 [hep-ph]}}.

\bibitem{Caprini:2015zlo}
C.~Caprini {\em et~al.}, ``{Science with the space-based interferometer eLISA.
  II: Gravitational waves from cosmological phase transitions},''
  \href{http://dx.doi.org/10.1088/1475-7516/2016/04/001}{{\em JCAP} {\bfseries
  1604} no.~04, (2016) 001},
\href{http://arxiv.org/abs/1512.06239}{{\ttfamily arXiv:1512.06239
  [astro-ph.CO]}}.

\bibitem{Seto:2001qf}
N.~Seto, S.~Kawamura, and T.~Nakamura, ``{Possibility of direct measurement of
  the acceleration of the universe using 0.1-Hz band laser interferometer
  gravitational wave antenna in space},''
  \href{http://dx.doi.org/10.1103/PhysRevLett.87.221103}{{\em Phys. Rev. Lett.}
  {\bfseries 87} (2001) 221103},
  \href{http://arxiv.org/abs/astro-ph/0108011}{{\ttfamily
  arXiv:astro-ph/0108011}}.

\bibitem{Kawamura_2006}
S.~K. et~al., ``The japanese space gravitational wave
  antenna{\textemdash}{DECIGO},''
  \href{http://dx.doi.org/10.1088/0264-9381/23/8/s17}{{\em Classical and
  Quantum Gravity} {\bfseries 23} no.~8, (Mar, 2006) S125--S131}.
  \url{https://doi.org/10.1088%2F0264-9381%2F23%2F8%2Fs17}.

\bibitem{Harry:2006fi}
G.~Harry, P.~Fritschel, D.~Shaddock, W.~Folkner, and E.~Phinney, ``{Laser
  interferometry for the big bang observer},''
  \href{http://dx.doi.org/10.1088/0264-9381/23/15/008}{{\em Class. Quant.
  Grav.} {\bfseries 23} (2006) 4887--4894}. [Erratum: Class.Quant.Grav. 23,
  7361 (2006)].

\bibitem{Geller:2018mwu}
M.~Geller, A.~Hook, R.~Sundrum, and Y.~Tsai, ``{Primordial Anisotropies in the
  Gravitational Wave Background from Cosmological Phase Transitions},''
  \href{http://dx.doi.org/10.1103/PhysRevLett.121.201303}{{\em Phys. Rev.
  Lett.} {\bfseries 121} no.~20, (2018) 201303},
  \href{http://arxiv.org/abs/1803.10780}{{\ttfamily arXiv:1803.10780
  [hep-ph]}}.

\bibitem{Cyburt:2009pg}
R.~H. Cyburt, J.~Ellis, B.~D. Fields, F.~Luo, K.~A. Olive, and V.~C. Spanos,
  ``{Nucleosynthesis Constraints on a Massive Gravitino in Neutralino Dark
  Matter Scenarios},''
  \href{http://dx.doi.org/10.1088/1475-7516/2009/10/021}{{\em JCAP} {\bfseries
  10} (2009) 021}, \href{http://arxiv.org/abs/0907.5003}{{\ttfamily
  arXiv:0907.5003 [astro-ph.CO]}}.

\bibitem{Kawasaki:2017bqm}
M.~Kawasaki, K.~Kohri, T.~Moroi, and Y.~Takaesu, ``{Revisiting Big-Bang
  Nucleosynthesis Constraints on Long-Lived Decaying Particles},''
  \href{http://dx.doi.org/10.1103/PhysRevD.97.023502}{{\em Phys. Rev. D}
  {\bfseries 97} no.~2, (2018) 023502},
  \href{http://arxiv.org/abs/1709.01211}{{\ttfamily arXiv:1709.01211
  [hep-ph]}}.

\bibitem{Forestell:2018txr}
L.~Forestell, D.~E. Morrissey, and G.~White, ``{Limits from BBN on Light
  Electromagnetic Decays},''
  \href{http://dx.doi.org/10.1007/JHEP01(2019)074}{{\em JHEP} {\bfseries 01}
  (2019) 074}, \href{http://arxiv.org/abs/1809.01179}{{\ttfamily
  arXiv:1809.01179 [hep-ph]}}.

\bibitem{Cappiello:2020lbk}
C.~V. Cappiello, J.~I. Collar, and J.~F. Beacom, ``{New experimental
  constraints in a new landscape for composite dark matter},''
  \href{http://dx.doi.org/10.1103/PhysRevD.103.023019}{{\em Phys. Rev. D}
  {\bfseries 103} no.~2, (2021) 023019},
  \href{http://arxiv.org/abs/2008.10646}{{\ttfamily arXiv:2008.10646
  [hep-ex]}}.

\bibitem{PhysRevD.40.3221}
I.~Goldman and S.~Nussinov, ``Weakly interacting massive particles and neutron
  stars,'' \href{http://dx.doi.org/10.1103/PhysRevD.40.3221}{{\em Phys. Rev. D}
  {\bfseries 40} (Nov, 1989) 3221--3230}.
  \url{https://link.aps.org/doi/10.1103/PhysRevD.40.3221}.

\bibitem{Baryakhtar:2017dbj}
M.~Baryakhtar, J.~Bramante, S.~W. Li, T.~Linden, and N.~Raj, ``{Dark Kinetic
  Heating of Neutron Stars and An Infrared Window On WIMPs, SIMPs, and Pure
  Higgsinos},'' \href{http://dx.doi.org/10.1103/PhysRevLett.119.131801}{{\em
  Phys. Rev. Lett.} {\bfseries 119} no.~13, (2017) 131801},
  \href{http://arxiv.org/abs/1704.01577}{{\ttfamily arXiv:1704.01577
  [hep-ph]}}.

\bibitem{VanTilburg:2020jvl}
K.~Van~Tilburg, ``{Stellar Basins of Gravitationally Bound Particles},''
  \href{http://arxiv.org/abs/2006.12431}{{\ttfamily arXiv:2006.12431
  [hep-ph]}}.

\bibitem{Leane:2020wob}
R.~K. Leane and J.~Smirnov, ``{Exoplanets as New Sub-GeV Dark Matter
  Detectors},'' \href{http://arxiv.org/abs/2010.00015}{{\ttfamily
  arXiv:2010.00015 [hep-ph]}}.

\bibitem{Leane:2017vag}
R.~K. Leane, K.~C.~Y. Ng, and J.~F. Beacom, ``{Powerful Solar Signatures of
  Long-Lived Dark Mediators},''
  \href{http://dx.doi.org/10.1103/PhysRevD.95.123016}{{\em Phys. Rev. D}
  {\bfseries 95} no.~12, (2017) 123016},
  \href{http://arxiv.org/abs/1703.04629}{{\ttfamily arXiv:1703.04629
  [astro-ph.HE]}}.

\bibitem{Leane:2021ihh}
R.~K. Leane, T.~Linden, P.~Mukhopadhyay, and N.~Toro, ``{Celestial-Body Focused
  Dark Matter Annihilation Throughout the Galaxy},''
  \href{http://arxiv.org/abs/2101.12213}{{\ttfamily arXiv:2101.12213
  [astro-ph.HE]}}.

\bibitem{Markevitch:2003at}
M.~Markevitch, A.~Gonzalez, D.~Clowe, A.~Vikhlinin, L.~David, W.~Forman,
  C.~Jones, S.~Murray, and W.~Tucker, ``{Direct constraints on the dark matter
  self-interaction cross-section from the merging galaxy cluster 1E0657-56},''
  \href{http://dx.doi.org/10.1086/383178}{{\em Astrophys. J.} {\bfseries 606}
  (2004) 819--824}, \href{http://arxiv.org/abs/astro-ph/0309303}{{\ttfamily
  arXiv:astro-ph/0309303}}.

\bibitem{Feng:2009mn}
J.~L. Feng, M.~Kaplinghat, H.~Tu, and H.-B. Yu, ``{Hidden Charged Dark
  Matter},'' \href{http://dx.doi.org/10.1088/1475-7516/2009/07/004}{{\em JCAP}
  {\bfseries 07} (2009) 004}, \href{http://arxiv.org/abs/0905.3039}{{\ttfamily
  arXiv:0905.3039 [hep-ph]}}.

\bibitem{Buckley:2009in}
M.~R. Buckley and P.~J. Fox, ``{Dark Matter Self-Interactions and Light Force
  Carriers},'' \href{http://dx.doi.org/10.1103/PhysRevD.81.083522}{{\em Phys.
  Rev. D} {\bfseries 81} (2010) 083522},
  \href{http://arxiv.org/abs/0911.3898}{{\ttfamily arXiv:0911.3898 [hep-ph]}}.

\bibitem{10.1111/j.1365-2966.2012.21182.x}
M.~Vogelsberger, J.~Zavala, and A.~Loeb, ``{Subhaloes in self-interacting
  galactic dark matter haloes},''
  \href{http://dx.doi.org/10.1111/j.1365-2966.2012.21182.x}{{\em Monthly
  Notices of the Royal Astronomical Society} {\bfseries 423} no.~4, (07, 2012)
  3740--3752}. \url{https://doi.org/10.1111/j.1365-2966.2012.21182.x}.

\bibitem{10.1093/mnrasl/sls053}
J.~Zavala, M.~Vogelsberger, and M.~G. Walker, ``{Constraining self-interacting
  dark matter with the Milky Way’s dwarf spheroidals},''
  \href{http://dx.doi.org/10.1093/mnrasl/sls053}{{\em Monthly Notices of the
  Royal Astronomical Society: Letters} {\bfseries 431} no.~1, (02, 2013)
  L20--L24}. \url{https://doi.org/10.1093/mnrasl/sls053}.

\bibitem{Tulin:2017ara}
S.~Tulin and H.-B. Yu, ``{Dark Matter Self-interactions and Small Scale
  Structure},'' \href{http://dx.doi.org/10.1016/j.physrep.2017.11.004}{{\em
  Phys. Rept.} {\bfseries 730} (2018) 1--57},
  \href{http://arxiv.org/abs/1705.02358}{{\ttfamily arXiv:1705.02358
  [hep-ph]}}.

\bibitem{Bondarenko:2020mpf}
K.~Bondarenko, A.~Sokolenko, A.~Boyarsky, A.~Robertson, D.~Harvey, and
  Y.~Revaz, ``{From dwarf galaxies to galaxy clusters: Self-Interacting Dark
  Matter over 7 orders of magnitude in halo mass},''
  \href{http://arxiv.org/abs/2006.06623}{{\ttfamily arXiv:2006.06623
  [astro-ph.CO]}}.

\bibitem{Nadler:2019zrb}
E.~O. Nadler, V.~Gluscevic, K.~K. Boddy, and R.~H. Wechsler, ``{Constraints on
  Dark Matter Microphysics from the Milky Way Satellite Population},''
  \href{http://dx.doi.org/10.3847/2041-8213/ab1eb2}{{\em Astrophys. J. Lett.}
  {\bfseries 878} no.~2, (2019) 32},
  \href{http://arxiv.org/abs/1904.10000}{{\ttfamily arXiv:1904.10000
  [astro-ph.CO]}}. [Erratum: Astrophys.J.Lett. 897, L46 (2020), Erratum:
  Astrophys.J. 897, L46 (2020)].

\bibitem{Gilman:2019nap}
D.~Gilman, S.~Birrer, A.~Nierenberg, T.~Treu, X.~Du, and A.~Benson, ``{Warm
  dark matter chills out: constraints on the halo mass function and the
  free-streaming length of dark matter with eight quadruple-image strong
  gravitational lenses},'' \href{http://dx.doi.org/10.1093/mnras/stz3480}{{\em
  Mon. Not. Roy. Astron. Soc.} {\bfseries 491} no.~4, (2020) 6077--6101},
  \href{http://arxiv.org/abs/1908.06983}{{\ttfamily arXiv:1908.06983
  [astro-ph.CO]}}.

\bibitem{Schutz:2020jox}
K.~Schutz, ``{Subhalo mass function and ultralight bosonic dark matter},''
  \href{http://dx.doi.org/10.1103/PhysRevD.101.123026}{{\em Phys. Rev. D}
  {\bfseries 101} no.~12, (2020) 123026},
  \href{http://arxiv.org/abs/2001.05503}{{\ttfamily arXiv:2001.05503
  [astro-ph.CO]}}.

\bibitem{Polikanov:1990sf}
S.~Polikanov, C.~S. Sastri, G.~Herrmann, K.~Lutzenkirchen, M.~Overbeck,
  N.~Trautmann, A.~Breskin, R.~Chechik, and Z.~Frankel, ``{Search for
  supermassive nuclei in nature},''
  \href{http://dx.doi.org/10.1007/BF01288200}{{\em Z. Phys. A} {\bfseries 338}
  no.~3, (11, 1990) 357--361}.

\bibitem{Celik_LH}
T.~Celik, J.~Engels, and H.~Satz, ``{The Order of the Deconfinement Transition
  in SU(3) Yang-Mills Theory},''
\href{http://dx.doi.org/10.1016/0370-2693(83)91314-X}{{\em Phys. Lett.}
  {\bfseries 125B} (1983) 411--414}.

\bibitem{Christ_LH}
N.~H. Christ and H.-Q. Ding, ``Latent heat and critical temperature of the
  color-deconfining phase transition,''
  \href{http://dx.doi.org/10.1103/PhysRevLett.60.1367}{{\em Phys. Rev. Lett.}
  {\bfseries 60} (Apr, 1988) 1367--1370}.
  \url{https://link.aps.org/doi/10.1103/PhysRevLett.60.1367}.

\bibitem{Forcrand_ST}
B.~Lucini, P.~de~Forcrand, and M.~Vettorazzo, ``Measuring interface tensions in
  4d su(n) lattice gauge theories,'' 2004.

\bibitem{Giusti_2017}
L.~Giusti and M.~Pepe, ``Equation of state of the su(3) yang–mills theory: A
  precise determination from a moving frame,''
  \href{http://dx.doi.org/10.1016/j.physletb.2017.04.001}{{\em Physics Letters
  B} {\bfseries 769} (Jun, 2017) 385–390},
  \href{http://arxiv.org/abs/1612.00265}{{\ttfamily arXiv:1612.00265
  [hep-lat]}}. \url{http://dx.doi.org/10.1016/j.physletb.2017.04.001}.

\bibitem{LL_SP1}
L.~Landau and E.~Lifshitz, {\em {Statistical Physics, Part 1}}, vol.~5 of {\em
  Course of Theoretical Physics}.
\newblock Pergamon Press, Oxford, 1980.

\bibitem{LL_PK}
L.~Landau and E.~Lifshitz, {\em {Physical Kinetics}}, vol.~10 of {\em Course of
  Theoretical Physics}.
\newblock Pergamon Press, Oxford, 1981.

\bibitem{Gorbunov:2011zz}
V.~A. Rubakov and D.~S. Gorbunov, \href{http://dx.doi.org/10.1142/10447}{{\em
  {Introduction to the Theory of the Early Universe}: {Hot big bang theory}}}.
\newblock World Scientific, Singapore, 2017.

\bibitem{kardar_2007}
M.~Kardar, \href{http://dx.doi.org/10.1017/CBO9780511815898}{{\em Statistical
  Physics of Particles}}.
\newblock Cambridge University Press, 2007.

\bibitem{Biondini:2017ufr}
S.~Biondini and M.~Laine, ``{Re-derived overclosure bound for the inert doublet
  model},'' \href{http://dx.doi.org/10.1007/JHEP08(2017)047}{{\em JHEP}
  {\bfseries 08} (2017) 047}, \href{http://arxiv.org/abs/1706.01894}{{\ttfamily
  arXiv:1706.01894 [hep-ph]}}.

\bibitem{Biondini:2018pwp}
S.~Biondini and M.~Laine, ``{Thermal dark matter co-annihilating with a
  strongly interacting scalar},''
  \href{http://dx.doi.org/10.1007/JHEP04(2018)072}{{\em JHEP} {\bfseries 04}
  (2018) 072}, \href{http://arxiv.org/abs/1801.05821}{{\ttfamily
  arXiv:1801.05821 [hep-ph]}}.

\bibitem{Binder:2019erp}
T.~Binder, K.~Mukaida, and K.~Petraki, ``{Rapid bound-state formation of Dark
  Matter in the Early Universe},''
  \href{http://dx.doi.org/10.1103/PhysRevLett.124.161102}{{\em Phys. Rev.
  Lett.} {\bfseries 124} no.~16, (2020) 161102},
  \href{http://arxiv.org/abs/1910.11288}{{\ttfamily arXiv:1910.11288
  [hep-ph]}}.

\bibitem{Binder:2020efn}
T.~Binder, B.~Blobel, J.~Harz, and K.~Mukaida, ``{Dark matter bound-state
  formation at higher order: a non-equilibrium quantum field theory
  approach},'' \href{http://dx.doi.org/10.1007/JHEP09(2020)086}{{\em JHEP}
  {\bfseries 09} (2020) 086}, \href{http://arxiv.org/abs/2002.07145}{{\ttfamily
  arXiv:2002.07145 [hep-ph]}}.

\bibitem{Geller:2018biy}
M.~Geller, S.~Iwamoto, G.~Lee, Y.~Shadmi, and O.~Telem, ``{Dark quarkonium
  formation in the early universe},''
  \href{http://dx.doi.org/10.1007/JHEP06(2018)135}{{\em JHEP} {\bfseries 06}
  (2018) 135}, \href{http://arxiv.org/abs/1802.07720}{{\ttfamily
  arXiv:1802.07720 [hep-ph]}}.

\end{thebibliography}\endgroup

\end{document}